\documentclass[numberwithinsect,thm-restate]{lipics-v2021}

\nolinenumbers
\usepackage{hyperref}
\usepackage{array,enumerate}
\usepackage{xcolor,xspace}
\usepackage{amsmath}
\usepackage{amssymb} 
\usepackage{tikz}
\usetikzlibrary{automata, arrows.meta, positioning}
\usepackage{yhmath} 
\usepackage{minted} 

\usepackage[textsize=scriptsize,textwidth=4.3cm]{todonotes}
\usepackage{stmaryrd}

\usepackage[capitalize,nameinlink]{cleveref}

\crefname{appendix}{Appendix}{appendices}
\Crefname{appendix}{Appendix}{Appendices}

\AddToHook{cmd/appendix/before}{%
  \crefalias{section}{appendix}%
}

\usepackage{caption}
\usepackage{subcaption}
\let\no\undefined
\DeclareRobustCommand\longtwoheadrightarrow
     {\relbar\joinrel\relbar\joinrel\twoheadrightarrow}
     
          \makeatletter
\newcommand\bigDiamond{\mathop{\mathpalette\bigDi@mond\relax}}
\newcommand\bigDi@mond[1]{%
  \vcenter{\hbox{\m@th
    \scalebox{\ifx#1\displaystyle 2\else1.2\fi}{$#1\Diamond$}%
  }}%
}
\newcommand\bigLozenge{\mathop{\mathpalette\bigL@zenge\relax}}
\newcommand\bigL@zenge[2]{%
  \vcenter{\hbox{\m@th
    \scalebox{\ifx#1\displaystyle 2\else1.2\fi}{$#1\blacklozenge$}%
  }}%
}
\makeatother

\newcommand{\extc}{\mathrel{+}}

\newcommand{\rec}[2][X]{\mathsf{rec}\ #1. #2}

\newcommand{\mends}{\mathsf{end}}
\newcommand{\yess}{\mathsf{yes}}
\newcommand{\nos}{\mathsf{no}}
\newcommand{\mend}{\mathsf{END}}
\newcommand{\yes}{\mathsf{YES}}
\newcommand{\no}{\mathsf{NO}}
\newcommand{\smend}{\mathsf{VEND}}
\newcommand{\syes}{\mathsf{VYES}}
\newcommand{\sno}{\mathsf{VNO}}
\newcommand{\instr}{\lhd}

\newcommand{\parts}[1]{\mathcal{P}({#1})}

\newcommand{\BNFsep}{\;\;|\;\;}

\newcommand{\stt}[1]{\ensuremath{\mathrel{\overset{#1} \longrightarrow}}}

\newcommand{\wt}[1]{\ensuremath{\overset{#1}{\longtwoheadrightarrow}}}
\newcommand{\wti}[1]{\ensuremath{\overset{#1}{\longtwoheadrightarrow}}_{inev}}
\newcommand{\nwt}[1]{\ensuremath{\overset{#1}{\longarrownot\longtwoheadrightarrow}}}
\newcommand{\Nst}[1]{\ensuremath{\mathrel{\overset{#1}{\longarrownot\longrightarrow}}}}

\newcommand{\pf}[1]{\mathit{Pf(#1)}}
\newcommand{\ext}[1]{\ensuremath{Ext(#1)}}

\newcommand{\rname}[1]{\textsc{#1}}

\newcommand{\sem}[1]{\llbracket #1 \rrbracket}

\newcommand{\set}[1]{\{ #1 \}}
\newcommand{\setof}[2]{\{ #1 \mid #2 \}}

\newcommand{\act}{\textsc{Act}}

\newcommand{\lab}{\textsc{Lab}}

\newcommand{\guards}[1]{\mathcal{B}(#1)}

\newcommand{\rplus}{\mathbb{R}_{\geq 0}}

\newcommand{\sver}{E}

\newcommand{\labb}{\mathcal{L}}

\newcommand{\compile}[1]{\llparenthesis #1 \rrparenthesis}

\newcommand{\true}{\mathsf{t \:\!\! t}} 
\newcommand{\false}{\mathsf{f \:\!\! f}}
\newcommand{\dmd}[1]{\langle #1 \rangle}
\newcommand{\lbox}[1]{[#1]}

\newcommand{\lin}{\mathop{\mathtt{\underline{in}}}}

\newcommand{\minfp}[2][X]{\mathsf{min }(#1, #2)}

\newcommand{\maxfp}[2][X]{\mathsf{max }(#1, #2)}

\newcommand{\conslvar}{\ensuremath{\textsc{cLVar}}}
\newcommand{\vars}{\ensuremath{\textsc{Var}}}
\newcommand{\vto}{\ensuremath{\mathbf{0}}}

\newcommand{\lcomplement}[1]{\ensuremath{\overleftarrow{#1}}}

\newcommand{\mon}{\textsc{Mon}}
\newcommand{\terms}{\logicsym[T]}
\newcommand{\tmon}{\textsc{TMon}}
\newcommand{\crmon}{\textsc{Mon}}
\newcommand{\mfv}[1]{\mathit{mfv}(#1)}
\newcommand{\mfvtime}[1]{\mathit{mfv}_{time}(#1)}

\newcommand{\semdiamond}[2]{\langle\cdot #1 \cdot\rangle(#2)}
\newcommand{\sembox}[2]{[\cdot #1 \cdot ](#2)}

\newcommand{\defeq}{\ensuremath{\stackrel{\text{def}}{=}}}

\newcommand{\tohmlSym}{\textsc{Mto\logicsym}}
\newcommand{\totlinSym}{\textsc{Pto\logicsym}}
\newcommand{\tohml}[1]{\ensuremath{\tohmlSym(#1)}}
\newcommand{\totlinn}[1]{\ensuremath{\totlinSym(#1)}}

\newcommand{\tomtlinSym}{\textsc{consMto\logic}}
\newcommand{\tomtlin}[1]{\ensuremath{\tomtlinSym(#1)}}
\newcommand{\tomtlinnSym}{\textsc{consPto\logic}}
\newcommand{\tomtlinn}[1]{\ensuremath{\tomtlinnSym(#1)}}

\newcommand{\tomax}[1]{\ensuremath{\langle\langle #1 \rangle\rangle}}
\newcommand{\tomin}[1]{\ensuremath{\Vert #1 \Vert}}
\newcommand{\setrho}{\ensuremath{\Upsilon}}

\newcommand{\vemptyset}{\ensuremath{\mathbf{0}}}

\newcommand{\iv}{\ensuremath{\mathcal{I}}}
\newcommand{\iiv}{\ensuremath{\mathbb{I}}}
\newcommand{\iem}{\ensuremath{{I_\emptyset}}}

\newcommand{\nxt}[2]{\bigcirc_{#1}\, #2}
\newcommand{\unxt}[1]{\bigcirc\, #1}
\newcommand{\until}[3]{\ensuremath{{#1}\, \, \mathcal{U}_{#2}\, {#3}}}
\newcommand{\uuntil}[2]{\ensuremath{{#1}\, \, \mathcal{U}\, {#2}}}
\newcommand{\globally}[2]{\ensuremath{{\square}_{#1}\, {#2}}}
\newcommand{\uglobally}[1]{\ensuremath{{\square}\, {#1}}}
\newcommand{\eventually}[2]{\ensuremath{{\textstyle\bigDiamond}_{#1}\, {#2}}}
\newcommand{\ueventually}[1]{\ensuremath{{\textstyle\bigDiamond}\, {#1}}}

\newcommand{\uuntilt}[2]{\ensuremath{{#1}\, \, \mathcal{U}\, {#2}}}

\newcommand{\wuntil}[3]{\ensuremath{{#1}\, \, \mathcal{U}^{\raisebox{0.5ex}{\scalebox{0.5}{\(\leq\)}}}_{#2}\, {#3}}}
\newcommand{\uwuntil}[2]{\ensuremath{{#1}\, \, \mathcal{U}^{\raisebox{0.5ex}{\scalebox{0.5}{\(\leq\)}}}\, {#2}}}
\newcommand{\wglobally}[2]{\ensuremath{{\square}^{\raisebox{0.5ex}{\scalebox{0.5}{\(\leq\)}}}_{#1}\, {#2}}}
\newcommand{\uwglobally}[1]{\ensuremath{{\square}^{\raisebox{0.5ex}{\scalebox{0.5}{\(\leq\)}}}\, {#1}}}
\newcommand{\weventually}[2]{\ensuremath{{\textstyle\bigDiamond}^{\raisebox{0.5ex}{\scalebox{0.5}{\(\leq\)}}}_{#1}\, {#2}}}
\newcommand{\uweventually}[1]{\ensuremath{{\textstyle\bigDiamond}^{\raisebox{0.5ex}{\scalebox{0.5}{\(\leq\)}}}\, {#1}}}

\newcommand{\precrI}{\ensuremath{\, \, {\prec}_{\raisebox{-0.5ex}{\scalebox{0.5}{I}}}}\, \, }
\newcommand{\preclI}{\ensuremath{\, \, {\prec}^{\raisebox{0.5ex}{\scalebox{0.5}{I}}}}\, \, }
\newcommand{\preccI}{\ensuremath{\, \, {\prec}_{\raisebox{-0.5ex}{\scalebox{0.5}{\lcomplement{I}}}}}\, \, }

\newcommand{\varphileq}{\ensuremath{{\varphi}_{\raisebox{-0.5ex}{\scalebox{0.5}{\(\leq 1\)}}}}}

\newcommand{\rb}[1]{\ensuremath{\mathcal{RB}(#1)}}
\newcommand{\lb}[1]{\ensuremath{\mathcal{LB}(#1)}}

\newcommand{\twords}{\ensuremath{\textsc{TAct}^\omega}}
\newcommand{\twordsr}{\ensuremath{\textsc{TAct}^\omega_{div}}}
\newcommand{\twordss}{\ensuremath{\textsc{TAct}^\infty}}
\newcommand{\twordstar}{\ensuremath{\textsc{TAct}^*}}
\newcommand{\twordspi}[1]{\ensuremath{\textsc{TAct}^\infty_{\vec{#1}}}}

\newcommand{\wordd}[1]{\textsc{Trn}(#1)}

\newcommand{\letin}[2]{\textsc{let}\;#1\;\textsc{in}\;#2}

\newcommand{\semdiv}[1]{\llbracket #1 \rrbracket^{div}}

\newcommand{\funof}[4]{\Lbag #1 \Rbag_{#2, #3, {#4}}}
\newcommand{\funofN}[5]{\Lbag #1 \Rbag^{#5}_{#2, #3, #4}}

\newcommand{\dom}[1]{\mathit{dom}(#1)}

\newcommand{\synsub}[2]{\ensuremath{[^{#1}/_{#2}]}}

\renewcommand{\defeq}{\mathrel{:=}}
\newcommand{\naturals}{\mathbb{N}}
\newcommand{\N}{\naturals}

\newcommand{\update}[2]{\ensuremath{[#1 := #2]}}

\newcommand{\suchthat}{: \,}
\newcommand{\wehavethat}{: \,}

\newcommand{\fvSym}{\mathit{fv}}
\newcommand{\fv}[1]{\fvSym(#1)}
\newcommand{\bvSym}[1]{\mathit{bv}}

\newcommand{\fvTstampSym}{\mathit{fv}_{\textsc{time}}}
\newcommand{\fvTstamp}[1]{\fvTstampSym(#1)}

\newcommand{\poset}[2]{\langle #1, #2 \rangle}

\newcommand{\semsubst}[2]{\ensuremath{[{#1} \mapsto {#2}]}}
\newcommand{\termF}{\phi}
\newcommand{\termG}{\psi}

\newcommand{\bigdisjunction}[1][Y]{\bigvee_{ \alpha \in \act}{\dmd{ \alpha }{#1}}}

\newcommand{\auxfun}[1][Y]{\textsc{DsjAct}(#1)}
\newcommand{\semmtl}[1]{\sem{#1}^{\textsc{MTL}}}

\newcommand{\compl}[1]{\ensuremath{\twords\setminus{#1}}}
\newcommand{\tlinnot}[1]{\ensuremath{\mathsf{not}(#1)}}
\usepackage{prooftree}
\usepackage{timestamp}

\usepackage{stmaryrd}

\newcommand{\ie}{\mbox{i.e.,\@}\xspace}%
\newcommand{\eeg}{\mbox{e.g.,\@}\xspace}%
\newcommand{\wrt}{\mbox{w.r.t.\@}\xspace}%
\newcommand{\etal}{\mbox{\emph{et~al.\@}}\xspace}%
\newcommand{\resp}{resp.\@\xspace}%
\newcommand{\nonl}{\renewcommand{\nl}{\let\nl\oldnl}}

\newcommand{\logicsym}[1][]{\ensuremath{\mathit{{#1}T^{lin}}}\xspace}
\newcommand{\logic}{\logicsym[]}%
\newcommand{\mlogic}{\logicsym[M]}%
\newcommand{\vlogic}{\logicsym[V]}%
\newcommand{\slogic}{\logicsym[S]}%
\newcommand{\clogic}{\logicsym[C]}%
\newcommand{\consrthml}{\logicsym[C]}%
\newcommand{\freeze}[2]{\ensuremath{#1.#2}}

\definecolorseries{test}{rgb}{grad}[rgb]{.95,.55,.55}{11,11,17}
\resetcolorseries[10]{test}
\newcommand{\addtodoeditor}[1]{%
    \colorlet{#1}{test!!+!50}
    \expandafter\newcommand\csname#1\endcsname [1]{%
        \todo[color=#1,size=\tiny]{\sffamily\textbf{{#1}:}
    ##1}\xspace%
    }
    \expandafter\newcommand\csname#1i\endcsname [1]{%
        \todo[inline, color=#1]{\sffamily\textbf{{#1}:} ##1}\xspace%
    }
}

\title{On the monitorability of linear-time timed logics}

\author{Mouloud Amara}{Université Paris Cité, CNRS, IRIF, F-75013 Paris, France}{}{}{}%
\author{Mohammed Aristide Foughali}{Université Paris Cité, CNRS, IRIF, F-75013 Paris, France}{foughai@irif.fr}{https://orcid.org/0000-0002-5348-5465}{}
\author{Giovanni Bernardi}{Université Paris Cité, CNRS, IRIF, F-75013 Paris, France}{}{}{} %
  \author{Adrian Francalanza}{Dept. of Computer Science, ICT, University of Malta, Msida, Malta}{}{}{}
  \author{Nicanora Ordoño Wallig}{Université Paris Cité, CNRS, IRIF, F-75013 Paris, France}{}{}{}
\authorrunning{Amara, Foughali, Bernardi, Francalanza, Ordoño Wallig}
\funding{This work has received support from the ANR project ERVoRS (ANR-25-CE48-6220).}

\begin{document}
\maketitle

\begin{abstract}
Runtime Verification (RV) is gaining popularity due to its scalability and ability to analyse block-box systems. %
\emph{Monitoring} is at the heart of RV; a logical formula $\phi$, formalising some property of interest, is typically translated into a monitor that checks whether the system under scrutiny satisfies $\phi$ during its execution. %
A logical formula $\phi$ is violation (\resp satisfaction) \emph{monitorable} iff there exists a monitor for $\phi$ that is both sound and complete \wrt its violation (\resp satisfaction) by a system; and  $\phi$ is monitorable if it is violation or satisfaction monitorable. %
The \emph{monitorability problem} is thus concerned with determining the largest subset of a logic $L$ that is monitorable. %
Although this problem has been solved for expressive untimed logics, 
 it remained open for timed logics, \ie where formulae can express both the order of events and the quantity of time separating them, until recently. %
In~\cite{DBLP:conf/ecoop/Amara0FF25}, we solved the monitorability problem for \logic, a new expressive (linear-time) timed $\mu$-calculus that we proposed. %
We showed that \logic  is strictly more expressive than MTL, the de facto timed extension of the well-known 
LTL. 
Moreover, we identified \mlogic, the largest monitorable fragment of \logic and characterised its largest subsets of formulae that are violation monitorable, satisfaction monitorable, and \emph{complete monitorable} (both satisfaction and violation monitorable). %
The maximality of these subsets was proven modulo an expressive monitoring framework called $\mon$, yet under the assumption of reactivity (\ie for monitorability purposes, a monitor was assumed able to consume any action the system might produce during its execution). %
Our work in~\cite{DBLP:conf/ecoop/Amara0FF25} was the first to solve the monitorability problem for such an expressive timed logic, and its theoretical results were accompanied with compilers (i) from MTL to \logic and (ii)  from \logic formulae to monitors. %

This paper is an extension of~\cite{DBLP:conf/ecoop/Amara0FF25} with the following new contributions. 
First, we 
generalise the maximality results to arbitrary monitors, which include both reactive and non-reactive monitors. %
Second, we extend the expressiveness results of \logic. %
We completely revise the embedding of MTL into \logic: we show that the syntax of MTL given in~\cite{DBLP:conf/ecoop/Amara0FF25} was incomplete, and provide a new compiler from MTL, under its standard syntax with the strict Until operator and negation over arbitrary formulae, into \logic. 
In passing, we demonstrate that \logic is closed under complement and provide a fully syntactic complementation procedure. %
Moreover, we prove that \logic is strictly more expressive than TPTL, a popular timed logic more expressive than MTL, and develop an automatic embedding from TPTL to \logic. %
This gives $\logic$ and the machinery around it the power of a \emph{unified framework} to reason on the monitorability of linear-time timed logics. %
Accordingly, we identify important monitorable fragments for both MTL and TPTL. 
In particular, we show that any MTL formula $\phi$ is satisfaction (\resp violation) monitorable regardless of its depth as long as every \emph{unconstrained Until}  sub-formula of $\phi$ is in the scope of an even (\resp odd) number of negations. 
We provide therefore complete monitorability results for MTL, TPTL and \logic, endowed with automatic synthesis of monitors from formulae within a unified framework.


\end{abstract}
\section{Introduction}
\label{sec:intro}

Runtime Verification (RV) is a lightweight formal verification technique~\cite{leucker2009brief,bartocci2018introduction},  %
where the system at hand is 
\emph{instrumented} with {\em monitors}~\cite{DBLP:journals/iandc/Francalanza21,FrancalanzaAAAC17}. %
Each monitor $M$ is a software component that corresponds to a logical formula $\phi$ expressing some property of interest. %
At runtime, $M$ {\em
incrementally} analyses the execution of the system until it reaches a {\em verdict} \wrt the satisfaction or violation of $\phi$~\cite{leucker2009brief,DBLP:journals/pacmpl/AcetoAFIL19}. %
RV can mitigate two major drawbacks of exhaustive, offline techniques such as model checking. 
First, by trading exhaustiveness for scalability, RV does not suffer from the well-known state-space explosion problem. %
Second, provided that actions of interest are visible for instrumentation purposes, RV is suitable for systems where no model is known beforehand~\cite{DBLP:journals/fmsd/FrancalanzaAI17}, \eeg black-box and self-adaptive systems~\cite{macias2013self}. 

Due to the lack of exhaustiveness, the delineation between the properties that can be handled by RV and those that cannot, \ie the \emph{monitorability} problem, becomes a central issue~\cite{pnueli2006psl,bartocci2018introduction,aceto2019operational,DBLP:conf/ecoop/Amara0FF25,DBLP:conf/concur/GrosenKL025}. 
Intuitively, a formula $\phi$ is monitorable iff its violation or satisfaction can be detected by a monitor. %
Consider for instance the following property: 

$$\text{ ``every \emph{request} is eventually followed by a \emph{grant}'' }$$

The above classical ``leads to'' property can be formalised in \eeg LTL, and there exists algorithms to model-check it offline~\cite{lichtenstein1985checking,clarke1994another,vardi2005automata};  
it is however \emph{not} monitorable.   %
Indeed, any satisfaction verdict after a \emph{grant} may be contradicted by a future \emph{request} not followed by a \emph{grant} (and dually, any  violation verdict after a \emph{request} may be contradicted by a future \emph{grant}). %
In technical terms, a formula $\phi$, expressing some property of interest, is said to be satisfaction (\resp violation)  monitorable iff there exists a monitor for $\phi$ that is both sound and satisfaction (\resp violation) complete~\cite{DBLP:journals/pacmpl/AcetoAFIL19,DBLP:journals/fmsd/FrancalanzaAI17} (other definitions of monitorability will be discussed in~\cref{sec:rw}). %
Formula $\phi$ is said \emph{monitorable} iff it is satisfaction or violation monitorable, and \emph{complete monitorable} iff it is both satisfaction and violation monitorable. %
The \emph{monitorability} problem is thus concerned with determining the largest monitorable subset of a given logic $L$. %

The monitorability problem has been studied for the modal $\mu$-calculus, also known as the Hennessy-Milner Logic with recursion ($\mu$-HML)~\cite{kozen1983results,larsen1990proof}. %
Notably, the \emph{maximally-expressive monitorable subset} of $\mu$-HML, \ie a subset within which any monitorable formula in $\mu$-HML is expressible, has been identified for both linear-time and branching-time semantics~\cite{DBLP:journals/pacmpl/AcetoAFIL19,DBLP:journals/fmsd/FrancalanzaAI17}. 
However, $\mu$-HML  is untimed, and thus cannot express \emph{timed properties} such as the following  \emph{bounded response} property 

$$\text{ ``every \emph{request} is eventually followed by a \emph{grant} within $5$ time units''}$$

Timed properties are extremely important in real-time systems, ubiquitous nowadays (see \eeg~\cite{maler2007synthesizing,foughali2020runtime}). %
There is therefore an urgent need to tackle the monitorability problem for timed logics. %
Even though the body of research on RV for timed systems is extensive (more in~\cref{sec:rw}), there was, until very recently, no work that studied monitorability in a timed setting. %

In 2025, the first works around timed monitorability emerged throughout two consecutive papers. %
The earlier one is ours~\cite{DBLP:conf/ecoop/Amara0FF25}, where we solved the monitorability problem for an expressive timed logic that we proposed, namely \logic. %
The present paper is an extension of~\cite{DBLP:conf/ecoop/Amara0FF25} (see below). 
The later one~\cite{DBLP:conf/concur/GrosenKL025}, providing decidability results on monitorability of $\omega$ timed regular languages, will be discussed~\cref{sec:rw}. %

\paragraph{Contributions.} %
This paper is an extension of~\cite{DBLP:conf/ecoop/Amara0FF25} which enclosed the following contributions. %
First, we propose~\logic, a new\footnote{Note that a number of timed $\mu$-calculi exist (see~\cite{cleaveland2024expressive} for an overview), yet out of this paper's scope due to their branching-time semantics.} timed $\mu$-calculus with a
linear-time interpretation, \ie the semantics of a formula in~\logic
is defined as a set of infinite timed executions that satisfy it. %
In particular, we prove that~\logic is strictly more expressive than MTL. %
Second, we develop a rich monitoring language called $\mon$, and solve the monitorability problem for \logic within $\mon$ and under the assumption of \emph{reactivity} of monitors. %
We precisely identify 
\vlogic, \slogic and \clogic, the largest fragments (\ie maximally-expressive subsets) of  \logic that are violation monitorable, satisfaction monitorable and complete monitorable, respectively, and accordingly conclude that $\mlogic = \vlogic \cup \slogic$ is the largest monitorable fragment of \logic. %
For practical reasons, we also provide automatic translators (i) from MTL to \logic and (ii) from \logic to $\mon$. %

In addition to the above contributions of~\cite{DBLP:conf/ecoop/Amara0FF25}, this extension comprises the following new results: 
\begin{itemize}
\item{} \cref{sec:clos} is new. In it, we prove that \logic is closed under complement using a fully syntactic complementation procedure. %
This is necessary to handle negation in the new translation from MTL to \logic (see below). 
\item{} \cref{sec:expressiveness} is completely revised and substantially extended: 
\begin{itemize} 
\item We completely revise our compiler from MTL to \logic (\cref{sec:mtllog}). %
We show that the syntax of MTL in~\cite{DBLP:conf/ecoop/Amara0FF25}, proposed in negative normal form was incomplete, leading to a semantics that cannot capture the negation of Until formulae (\cref{rk:wrong}). %
To demonstrate that this under-characterisation does not impair the expressiveness results, \ie the containment of MTL in $\logic$,  %
we fall back on the standard syntax of MTL with the strict Until operator and negation over arbitrary formulae (\cref{sec:mtl}) and develop a new compiler thereof from MTL to \logic that we prove correct. %
\item We show that \logic does not only embed MTL, but also the more expressive TPTL, and provide accordingly an automatic compiler from TPTL to \logic that we prove correct (\cref{sec:tptllog}). 
\item We demonstrate that \logic is strictly more expressive than TPTL and therefore strictly more expressive than MTL (\cref{sec:expre}), establishing therefore a strict expressiveness hierarchy between the three logics (\cref{venn}). %
\end{itemize}
\item{} In the new \cref{genmon}, we generalise the monitorability results of \logic to a more flexible monitoring framework, where monitors do not have to be reactive by design, and therefore extend our maximality results to arbitrary monitors in $\mon$. 
\item{} In the new \cref{sec:monmtl} and \cref{sec:montptl}, we use $\logic$ and the machinery around it (compilations from MTL and TPTL to $\logic$, and from $\logic$ to $\mon$) as a unified framework to generalise the monitorability results to both MTL and TPTL. %
In particular, we show that the depth of formulae in MTL does not impact the monitorability within the identified fragments. %
More precisely, every MTL formula $\phi$ is satisfaction (\resp violation) monitorable as long as every \emph{unconstrained Until} sub-formula of $\phi$ is in the scope of an even (\resp odd) number of negations. %
\item{} \cref{sec:rw} is substantially extended to include a comparison with~\cite{DBLP:conf/concur/GrosenKL025}, a very recent work on monitorability that appeared after the publication of~\cite{DBLP:conf/ecoop/Amara0FF25}. %
\end{itemize}
To the best of our knowledge, this is the first complete work on timed monitorability of prominent linear-time timed logics within a unified framework.

\paragraph{Outline.}%
The rest of this paper is organised as follows. %
After overviewing the preliminaries in \cref{sec:preliminaries},  we discuss our core contribution in \cref{sec:logics}
through~\cref{sec:montptl}. %
\cref{sec:logics} defines our logic \logic and proves that it is more
expressive than MTL (\resp TPTL), providing along the way an automatic translation from MTL (\resp TPTL) to \logic. %
In~\cref{sec:monitoring}, we lay out our theoretical foundations for timed monitors and instrumentation. %
We then tackle the monitorability problem of $\logic$ in~\cref{sec:monitorability}. %
In~\cref{sec:monmtl}, we identify important fragments of MTL with violation-, satisfaction- and complete-monitorable formulae. %
In~\cref{sec:montptl}, we identify analogously monitorable fragments of TPTL. 
\cref{sec:rw} compares our contributions with related work and~\cref{sec:conclusion} wraps up with concluding remarks and future work directions. %

\paragraph{Proofs.} 
In order to avoid redundancy, only the proofs related to new contributions \wrt~\cite{DBLP:conf/ecoop/Amara0FF25} are enclosed in the present paper. The locations of the remaining proofs within~\cite{DBLP:conf/ecoop/Amara0FF25} are referred to whenever needed throughout the remainder of this paper. 
\section{Preliminaries}
\label{sec:preliminaries}

In this section, we introduce preliminary notions on TLTSs and timed temporal logics (MTL and TPTL). %
All definitions are \emph{action based}, \ie rely on the actions a system under scrutiny may perform~\cite{leucker2009brief}. %
In the remainder of this paper, we denote by~$\act$ a \emph{non-empty, finite set} of such actions. %

\subsection{Timed Labelled Transition Systems}\label{sec:sum} 
The following definition is inspired from~\cite{rs_2007}. %
    \begin{definition}
  \label{def:tlts}
  A Timed Labelled Transition System (TLTS) $P$ over $\act$ is a tuple 
  $\langle S_P, \lab, \longrightarrow_P \rangle$ where $S_P$ is a set of states,
  $\lab = \act \cup \rplus \cup \{{\tau}\}$ a set of labels,
and
  $ {\longrightarrow_P} \subseteq S_P \times \lab \times S_P $ the transition relation satisfying: 
  \begin{enumerate}
    \item Time determinism: 
      for every $p,p',p'' \in S_P$ and $d \in \rplus$, if $(p,d,p')\in\longrightarrow_P$ and $(p,d,p'')\in\longrightarrow_P$ then $p' = p''$,
    \item Time additivity:
      for every $p,p'' \in S_P$ and $d_1,d_2 \in \rplus$, $(p,d_1 + d_2,p'')\in\longrightarrow_P$
        iff $(p,d_1,p')\in\longrightarrow_P$ and $(p',d_2,p'')\in\longrightarrow_P$ from some $p'\in S_P$,
    \item Zero delay: 
      for every $p,p' \in S_P$, $(p,0,p')\in\longrightarrow_P$ iff $p' = p$.
  \end{enumerate}
\end{definition}

A transition labelled $\alpha\in \act$ (\resp $\alpha\in \rplus$) is a called an \emph{action} transition (\resp delay transition). %
A $\tau$ transition is a non-observable transition labelled with the special symbol $\tau$. %
We use $a,a_i,b,c$ and $d,d_i$ to denote, respectively, an element of $\act$ and $\rplus$. %
We recall that $\act$ is finite. %
In the following, we write $p_1\stt{\alpha_1}p_2\stt{\alpha_2}p_3\ldots$ to denote a \emph{sequence of transitions} with $(p_1,\alpha_1,p_2),(p_2,\alpha_2,p_3)\ldots\in \longrightarrow_P$. %

\paragraph{Runs.}

A \emph{run} in the TLTS $P$ is a sequence of labels (in $\lab^\omega\cup \lab^* = \lab^\infty$) corresponding to a sequence of transitions. %
For example, if $p_1\stt{\alpha_1}p_2 \stt{\alpha_2} p_3$ then $\theta = \alpha_1\, \alpha_2 \in \lab^*$ is a finite run feasible from $p_1$. %
If there exists an infinite transition sequence
$p_1 \stt{\alpha_1} p_2 \stt{\alpha_2} p_3 \stt{\alpha_3}\ldots$,
we say that run~$\theta = \alpha_1\alpha_2\alpha_3\ldots \in \lab^\omega$ is \emph{infinite} from~$p_1 \in S_P$. %

\paragraph{Weak runs.}
Let $\labb = \lab \setminus{\{\tau\}}$. %
To define weak transitions, we write $p
\wt{} p'$ instead of $p \mathrel{(\stt{\tau})^*} p'$ which is the
reflexive-transitive closure of the relation $\stt{\tau}$, and following
\cite{DBLP:conf/concur/Yi90}, we write $p \wt{\alpha} p'$ for $\alpha \in \labb$
in the following cases,
  \begin{enumerate}[(a)]
  \item $p \wt{ a } p'$
    if $p \wt{} \cdot \stt{ a } \cdot \wt{} p'$,
  \item $p \wt { d } p'$ if there exists $n \geq 0$ s.t.
    $p \wt{} \cdot \stt { d_0 } \cdot \wt{} \cdot \stt { d_1 }
      \ldots \stt{d_n} \cdot \wt{} p'$ where $d = \sum_{i \leq n} d_i$,
  \end{enumerate}
with the convention that if the set ${d_1, \ldots, d_n}$ is empty, then
$\sum_{1\leq i \leq n} d_i = 0$. %
Following \cite{DBLP:journals/fmsd/FrancalanzaAI17}, for $\alpha_1, \dots,
\alpha_n \in \labb$ and $p,p' \in S_P$, we write sequences of transitions $p
\wt{ \alpha_1 } \dots \wt{ \alpha_n } p'$ as $p \wt{ \theta } p'$ where $\theta = \alpha_1
\dots \alpha_n$. %
Such $\theta$ is a finite weak run (in $\labb^*$). %
We say that a finite weak run $\theta'\in\labb^*$ \emph{corresponds to} a finite run $\theta\in\lab^*$ iff removing $\tau$ transitions then summing consecutive delays in $\theta$ gives $\theta'$. %
For instance, the  finite weak run $5\, a\, 3\, b\, 2$ corresponds to \eeg finite runs $2\, \tau\, 3\, a\, 3\, b\, \tau\, 2$ and $5\, a\, 3\, b\, 1\, \tau\, 1$. %
The \emph{empty weak run} is $\epsilon = 0$. %

\paragraph{Reachability and inevitability.}

In TLTS $P$, a state $p'\in S_P$ is reachable from state $p\in S_P$ along a finite weak run $\theta\in\labb^*$ iff $p\wt{\theta}p'$. %
We say that $p'$ is \emph{inevitably reachable} from $p$ along $\theta$, denoted  $p\wti{\theta} p'$, if $p'$ is the only reachable state from $p$ along $\theta$, that is: $$p\wt{\theta}p' \text{ and } \forall p''\in S_P : p\wt{\theta}p''  \text{ implies } p' = p''$$ %
Similarly, we write $p\nwt{\theta}p'$ if $p'$ is not reachable from $p$ along $\theta$, $p\wt{\theta}$ if there exists a state $p'$ reachable from $p$ along $\theta$, and $p\nwt{\theta}$ if there is no reachable state from $p$ along $\theta$.

\subsection{Timed Words}\label{sec:tw}

  As in \eeg~\cite{DBLP:journals/tosem/BauerLS11,DBLP:conf/formats/GrosenKLZ22},
  we interpret timed logics over infinite timed words. %
  
  \begin{definition}[Timed words]
    A timed word $\pi$ over $\act$ is a possibly infinite sequence $(a_1,t_1)(a_2,t_2)...$
    of timed events $(a_i,t_i) \in \act \times \rplus$, where
    $t_i$ are monotonically non-decreasing timestamps, \ie $\forall i \in \naturals$: $t_i \leq t_{i+1}$.
    An empty timed word contains no timed events. 
      \label{def:tw}
      \end{definition}

  We denote $\twords$ and $\twordstar$ the set of infinite (omega) and finite timed words, 
  respectively, and $\twordss$ the set of all timed words ($\twordss = \twords \cup \twordstar$). 
  We denote by $|\pi|$ the \emph{length} of a timed word $\pi\in \twordss$. %
  The length of a finite timed word $\pi = (a_1,t_1)...(a_n,t_n) \in \twordstar$ equals $n$. %
  The length of an infinite timed word $\pi\in \twords$ equals $\omega$. %
  We use the notation~$a_i(\pi)$ (resp.~$t_i(\pi)$) for
  the action (\resp the timestamp) of the timed event at position~$i$ of~$\pi\in\twordss$, with $i\in\naturals$ and $i\leq |\pi|$. %
  For a timed event $(a_i,t_i)$, we say that $t_i$ is associated with $a_i$. %
  The equality between two timed words $\pi,\pi'\in\twordss$ is defined in the standard manner:$$
  |\pi| = |\pi'| \wedge \forall i\in \naturals: i \leq |\pi| \implies (a_i(\pi) = a_i(\pi') \wedge t_i(\pi) = t_i(\pi'))$$  
  For some $\pi\in \twordss$, we write $\pi^i$ (\resp $^i\pi$), $i\leq |\pi|$,  
  for the 
  suffix 
  of~$\pi$ starting  at position~$i$ (\resp the prefix of $\pi$ ending at $i$). 
The partial function $\mathit{Concat:  \twordstar \times \twordss \rightharpoonup \twordss}$ returns, for a finite timed word $\pi$ and a timed word $\pi'$, the timed word corresponding to $\pi$ concatenated with $\pi'$ (\ie $\pi'$ appended at the tail of $\pi$). %
$\mathit{Concat}$ is defined only for couples $\mathit{(\pi,\pi')\in \twordstar \times \twordss}$ where the first timestamp of $\pi'$ is larger than the last timestamp of $\pi$ (otherwise timestamp monotonicity is violated). %
Let $\twordspi{\pi}$ be the set defined as follows, \wrt to some finite timed word $\pi$:$$\twordspi{\pi} = \{\pi' \in \twordss\, \,  |\, \,  t_1(\pi') \geq t_{|\pi|}(\pi)\}$$
Accordingly, $\mathit{Concat(\pi,\pi')}$ is defined iff $\pi' \in \twordspi{\pi}$. %
We use the shorthand notation $\pi.\pi'$ instead of $\mathit{Concat(\pi,\pi')}$. %
For any timed word $\pi'\in\twordss$, we define the set of its prefixes:$$\pf{\pi'} = \{\pi\in\twordstar \, \, |\, \,  \exists \pi''\in \twordspi{\pi} : \pi.\pi'' = \pi'\}$$
Dually, we define for any finite timed word $\pi'\in\twordstar$ its infinite \emph{extensions}, \ie the set of all infinite timed words resulting from concatenating $\pi'$ with an infinite timed word: $$\ext{\pi'} = \{\pi\in\twords \, \, |\, \,  \pi'\in\pf{\pi}\}$$

Note that a timed word corresponds necessarily to a weak run in $\labb^*$, with each timestamp recording the \emph{absolute time} of occurrence of the action associated with it. %
For example, the finite timed word $(a,0)(b,3)(a,3)(c,7.2)$ results from the finite weak run $a\, 3\, b\, a\, 4.2\, c$. %
We will formalise the function to retrieve a finite weak run from a finite timed word in~\cref{sec:framework}. %

\begin{remark}
Note that, in the timed systems literature, every execution in a TLTS corresponds to an infinite \emph{time divergent} run/timed word (see \eeg ~\cite{henzinger1992sooner,alur1997modularity}), \ie all runs/timed words diverge in time in the future. %
We do not make this assumption here since our results up to and including \cref{sec:monitoring} do not depend on it. %
Instead, we formalise, justify and enforce it in \cref{sec:framework} for monitorability purposes. %
\end{remark}

\subsection{Intervals}

We now define $\iiv$, the set of dense-time intervals with natural left bounds and natural or infinite right bounds. %
Hereafter, we refer to an element of $\iiv$ %
simply as ``interval''. %

\begin{definition}[Interval syntax]
\label{def:int-synt}
  The set $\iiv$ is given by the following grammar:
  $$
    \begin{array}{lll}
      I \in \iiv &::=& \sqsubset l,u \sqsupset \\ 
    \end{array}
    $$
    where
    ${\sqsubset} \in \set{ \,[\, , \,(\, }$, %
      ${\sqsupset} \in \set{ \, ]\, ,\, )\, }$, %
    $l\in \N$ and $u \in \N  \cup \set{ \infty }$. %
  As per standard notation, we require that ${\sqsupset}  =  {)} $ if
  $u = \infty$. %
  $l$ (resp. $u$) is called the left (\resp right) bound of $I$, we write accordingly $\lb{I} = l$ (\resp $\rb{I} = u$). %
  $I$ is said left closed iff ${\sqsubset} \, = \, [$, right closed iff ${\sqsupset} \, = \, ]$, and \emph{bounded} iff $\rb{I} \neq \infty$.
\end{definition}

\begin{definition}[Interval Semantics]
\label{def:int-sem}
  $\llbracket I \rrbracket$, the meaning of an interval %
  $I  \in \iiv$, %
  is a convex subset of $\mathit{\mathbb{R}_{\geq 0}}$: 
  $$
    \begin{array}{lll}
      \llbracket I \rrbracket &=& \setof{x \in \rplus}{\lb{I} \, \preclI\,  x \, \precrI\,  \rb{I}}\\
    \end{array}
    $$
    where  
\begin{itemize}
\item ${\preclI}  = {\leq}$ if $I$ is left closed and  ${\preclI}  =  {<}$ otherwise, 
\item ${\precrI}  = {\leq}$ if $I$ is right closed and  ${\precrI} =  {<}$ otherwise.
\end{itemize}
\end{definition}

Since we are mostly interested in the set of non-empty intervals, we let $\iv = \setof{I\in \iiv}{ \ \llbracket I \rrbracket  \neq \emptyset}$ denote such set (syntactically, $\iv$ contains all elements of $\iiv$ excluding those of the form $(l,l)$, $(l,l]$ and $[l,l)$). 
Hereafter, as by standard notation, we use $I$ to denote both the syntax and semantics of interval $I$. %

For convenience, we define the \emph{left complement} of an arbitrary non-empty interval. %
Intuitively, the left complement of some interval $I\in \iv$ covers fully and exactly the interval to the left of $I$. %
For instance, if $I = (1,5]$ then its left complement should be $[0,1]$. %

\begin{definition}[Left complement of a non-empty interval]\label{def:lcomp}
Let $\iem$ denote the empty interval 
$[0,0)$. %
The left complement of $I$, denoted $\lcomplement{I}$, is  defined over $\iv \cup \iem$ as  $\lcomplement{I} = [0,\lb{I})$ if $I$ is left closed and $\lcomplement{I} = [0,\lb{I}]$ otherwise. %
\end{definition}

\begin{example} 
 $\lcomplement{[0,2)} = \lcomplement{[0,\infty)} = \iem$, $\lcomplement{(6,8)} = [0,6]$, $\lcomplement{[3,\infty)} = [0,3)$. %
\end{example}

\begin{remark}\label{rk:lcomp}
As a consequence of~\cref{def:int-sem} and~\cref{def:lcomp}, we have for any $x\in\rplus$ and $I\in\iv$, $x\in I$ iff $x$ does \emph{not} satisfy $x\preccI \rb{\lcomplement{I}}$ and satisfies $x\precrI\rb{I}$. %
\end{remark}

\subsection{Timed Temporal Logics}
We now introduce two prominent timed extensions of the well-known Linear-time Temporal Logic (LTL), namely Metric Temporal Logic (MTL) and Timed Propositional Temporal Logic (TPTL). %

\subsubsection{Metric Temporal Logic (MTL)}\label{sec:mtl}
MTL~\cite{koymans1990specifying} is a well-known timed logic with linear-time semantics;  it is the de facto timed extension of LTL. %
Note that we consider future MTL (without past operators
), which we refer to, following \eeg~\cite{bouyer2005expressiveness,prabhakar2006expressiveness}, simply as MTL. %

\paragraph{Syntax and Semantics.}

The standard (minimal) syntax of MTL, used classically in the literature, features one timed operator, namely the \emph{strict} Until  ~\cite{bouyer2005expressiveness,ouaknine2008some,ouaknine2005decidability,DBLP:conf/time/Foughali26}. In~\cref{fig:mitl-syntax}, we provide such syntax extended with $\top$ and $\bot$. %
The latter are added for convenience, in particular when reasoning about monitorability of MTL (\cref{sec:monmtl}). %

\begin{figure}
  \hrulefill
  $$
    \begin{array}{llllllll}
      \phi,\psi &:= \top &\BNFsep \bot &\BNFsep a &\BNFsep \neg \phi &\BNFsep \phi \wedge \psi &\BNFsep \until{\phi}{I}{\psi}
    \end{array}
  $$
  \hrulefill
  \caption{MTL syntax.}
  \label{fig:mitl-syntax}
  \hrulefill
\end{figure}

The \emph{point-wise} semantics of MTL is given by the truth value of $\pi,i
\models \phi$, denoting ``$\pi\in\twords$ satisfies formula $\phi$ at
position $i$'', 
with the convention that if $\pi,1 \models \phi$, then $\pi$ is said to satisfy~$\phi$. %
Here, we define the semantics of a formula as the set of all
infinite timed words that satisfy it (\cref{fig:mitl-semantics}), relying on the definitions
given in~\cite{bouyer2005expressiveness,ouaknine2008some,ouaknine2005decidability,DBLP:conf/time/Foughali26}. 
Note that $\bot$ and $\top$ are equivalent to, respectively, $a\wedge \neg a$ and $\neg \bot$. %
Syntactic sugar is defined as usual: $\phi \vee \psi \equiv \neg(\neg \phi \wedge \neg \psi)$ ($\phi$ or $\psi$), $\nxt{I}{\phi} \equiv \until{\bot}{I}{\phi}$ (\emph{Next} $\phi$), 
$\eventually{I}{\phi} \equiv \until{\top}{I}{\phi}$ (\emph{Finally} $\phi$) and  
$\globally{I}{\phi} \equiv \neg \eventually{I}{\neg \phi}$ (\emph{Globally} $\phi$). %

The \emph{non-strict} Until operator $\wuntil{}{I}{}$, which has arguably a more ``natural'' semantics 
$$\sem{\wuntil{\phi}{I}{\psi}} = \setof{\pi}
            {\exists i\geq 1 \suchthat t_i(\pi) - t_1(\pi) \in I \text{ and } \pi^i \in \sem{\psi}
              \text{ and } \forall 1 \leq j < i \suchthat \pi^j \in \sem{\phi}}$$
              
 is expressible in terms of the strict Until: $\wuntil{\phi}{I}{\psi}\equiv \psi\vee(\phi\wedge (\until{\phi}{I}{\psi}))$ if $I$ is left closed at $0$ and $\wuntil{\phi}{I}{\psi}\equiv \phi\wedge (\until{\phi}{I}{\psi})$ otherwise~\cite{furia2007expressiveness}. %
The non-strict globally and eventually operators are then defined analogously to the strict ones, respectively $\weventually{I}{\phi} \equiv \wuntil{\top}{I}{\phi}$ and $\wglobally{I}{\phi} \equiv \neg \weventually{I}{\neg\phi}$. %

\begin{figure}
  \hrulefill
  $$
    \begin{array}{lllllllll}
      \sem{\top} &=& \twords & \quad\quad\quad\sem{\bot} &=& \emptyset
      &\quad\quad\quad
      \sem{a} &=& \setof{\pi}{a_1(\pi) = a}
      \\[2pt]
      \sem{\neg \phi} &=& \multicolumn{4}{c}{\compl{\sem{\phi}}}
      &
      \sem{\phi \wedge \psi} &=& \multicolumn{1}{c}{\sem{\phi} \cap \sem{\psi}}
      \\[2pt]
      \sem{\until{\phi}{I}{\psi}} &=& 
      \multicolumn{7}{c}{\setof{\pi}
            {\exists i>1 \suchthat t_i(\pi) - t_1(\pi) \in I \text{ and } \pi^i \in \sem{\psi}
              \text{ and } \forall 1 < j < i \suchthat \pi^j \in \sem{\phi}}}
    \end{array}
  $$
  \hrulefill
  \caption{MTL semantics.}
  \label{fig:mitl-semantics}
  \hrulefill
\end{figure}

In the remainder of this paper, we omit the subscript $I$ of temporal
operators if~$I = [0,\infty)$. %
Accordingly, if all operators appearing in a formula $\phi$ are
subscript free, then $\phi$ is said to be \emph{untimed}. 
Note that untimed non-strict Until formulae correspond to standard Until formulae in LTL. %

\begin{example}
\label{ex:until}
Let $b,c\in\act$. %
$\sem{\until{b}{I}{c}}$ contains all infinite
timed words where
$b$ is satisfied in every position (excluding the first one), until $c$
becomes satisfied within interval $I$ (relative to the first timestamp in $\pi$). %
For instance, for $I = [2,5]$, any $\pi\in\twords$ starting with the prefix $(b,3)(a,4)(c,7)$, $\pi'\in\twords$ starting with $(a,3)(b,4)(c,7)$ and $\pi''\in\twords$ starting with $(b,2)(c,3)$, we have:
\begin{itemize}
\item{} $\pi\notin\sem{\until{b}{I}{c}}$ because even though $c$ is satisfied at position $3$, $b$ is not satisfied at every position strictly between $1$ and $3$ ($a_2(\pi) = a \neq b$), 
\item{} $\pi'\in\sem{\until{b}{I}{c}}$ because $c$ is satisfied at position $3$ within $I$ ($7-3 \in [2,5]$) and $b$ is satisfied at the only position strictly between $1$ and $3$ (notice that here, however, $\pi'\notin\sem{\wuntil{b}{I}{c}}$), 
\item{} $\pi''\notin\sem{\until{b}{I}{c}}$ because the first $c$ occurs ``too early'' with timestamp $3$ and $3-2 \notin [2,5]$. %
\end{itemize}
\end{example}

\begin{theorem}\label{thm:mtl}
Satisfiability and validity of MTL are undecidable~\cite[Theorem 3]{ouaknine2006metric}. 
\end{theorem}

\begin{remark}\label{rk:wrong} The syntax given in~\cite{DBLP:conf/ecoop/Amara0FF25} was the following, using the non-strict Until and Globally operators as well as the next operator, and negation only allowed at the level of propositions $$\phi := \BNFsep a \BNFsep \neg a \BNFsep \phi \vee \phi \BNFsep \phi\wedge\psi  \BNFsep \nxt{I}{\phi}\BNFsep \wuntil{\phi}{I}{\psi}\BNFsep \wglobally{I}{\phi}$$ %
We claimed that every formula following the standard syntax in~\cref{fig:mitl-syntax} can be rewritten in the above negative normal form, and in particular for the negation of the non-strict Until we used the equivalence $\neg(\wuntil{\phi}{I}{\psi}) = (\wglobally{I}{\neg\psi}) \vee \wuntil{\neg\psi}{I}{(\neg\psi \wedge \neg\phi)}$. %
Such equivalence, classical in the untimed context (\ie in LTL), is inaccurate in the timed case. %
Take for example the formula $\wuntil{(a\vee b)}{[1,2]}{b}$ and any timed word $\pi$ starting with the prefix $(c,0)(b,1)$. %
Then we have $\pi\notin\sem{\wuntil{(a\vee b)}{[1,2]}{b}}$, but also neither in $\sem{\wglobally{[1,2]}{\neg b}}$ nor in $\sem{\wuntil{\neg b}{I}{(\neg b \wedge \neg a)}}$. %
Put succinctly, the definition of MTL syntax and semantics that we provided in~\cite{DBLP:conf/ecoop/Amara0FF25} cannot capture the negation of Until formulae. %
The correct way to deal with negative normal form is to explicitly include both the strict Until and its dual  operator~\cite{bouyer2026simple}, which is an unnecessary complication for our translation from MTL to $\logic$: we instead provide a syntactic complementation over $\logic$ to handle the case $\neg\phi$ (\cref{sec:clos}) and operate directly on the syntax given in~\cref{fig:mitl-syntax} (\cref{sec:expressiveness}). %
\end{remark}

\subsubsection{Timed Propositional Temporal Logic (TPTL)}
TPTL~\cite{alur1994really} is an expressive timed logic that uses clock-like variables. %
As for MTL, we consider TPTL without past operators, 
and adapt the original pointwise semantics (see \eeg~\cite{bouyer2005expressiveness}) to 
refer to sets of infinite timed words without loss of generality. %
Such syntax and semantics are given in~\cref{fig:tptl}. %

\label{sec:tptl}
\begin{figure}[!ht]
  \hrulefill
    $$
    \begin{array}{llllllllll}
      \phi &  ::=   & \top &\BNFsep \bot &\BNFsep a  
                    &\BNFsep \phi \wedge \phi
                    &\BNFsep \neg \phi
                    &\BNFsep \uuntilt{\phi}{\phi} 
                    &\BNFsep \freeze{x}{\phi} 
                    &\BNFsep x \sim c \\
    \end{array}
    $$
  \hrulefill
    $$
    \begin{array}{llllll}
    \sem{\top}_v &=&\twords &
      \sem{\bot}_v &=& \emptyset\\[0.1em]
      \sem{a}_v &=& \setof{\pi}{a_1(\pi) = a} &
      \sem{\neg \phi}_v &=& \compl{\sem{\phi}_v}\\[0.1em]
      \sem{\phi \wedge \psi}_v &=& \sem{\phi}_v \cap \sem{\psi}_v &
      \sem{\freeze{x}{\phi}}_v &=& \setof{\pi}{\pi \in \sem{\phi}_{v\update{x}{t_1(\pi)}}} \\[0.1em]
      \sem{x \sim n}_v &=& \setof{\pi}{t_1(\pi) - v(x) \sim n} \\
      \sem{\uuntilt{\phi}{\psi}}_v &=&
      \multicolumn{4}{l}{ \setof{\pi}
          {\exists i > 1: \pi^i \in \sem{\psi}_v \text{ and } \forall 1 < j < i: \pi^j \in \sem{\phi}_v}} \\
    \end{array}
    $$
  \hrulefill
  \caption{TPTL syntax and semantics}
  \hrulefill
  \label{fig:tptl}
\end{figure}

TPTL formulae resemble those of MTL in the use of the strict Until  operator, which however does not carry a subscripts in TPTL and is  therefore untimed by definition. %
To reason about time, TPTL uses instead clock-like variables ($x$ in~\cref{fig:tptl}). %
Each such variable can be used with the \emph{freeze} quantifier ($\freeze{x}{\phi}$) which allows to record the timestamp of some action. %
Propositions of the form $x \sim n$ with ${\sim \in \set{<,\leq,>,\geq}}$ and $n\in \naturals$ are then useful to reason on the quantity of time since $x$ has been frozen last (example below). %
For simplicity, we write $x \in I$ and $x\notin I$ for some interval $I\in \mathcal{I}$ as they can be expressed as conjunctions and disjunctions of $x \sim n$ propositions. %
Note that the semantics of TPTL is given over \emph{terms}, \ie where clock-like variables can appear free. %
A TPTL formula is however necessarily closed, and therefore we drop the subscript $v$ hereafter when referring to the semantics of a formula. %
$\unxt{\phi}$, $\uglobally{\phi}$ and $\ueventually{\phi}$ are derived using the same classical equivalences as for MTL, and the same notation is used for non-strict operators ($\uwuntil{}{}$, $\uwglobally{}$ and $\uweventually{}$), with the difference that neither of these operators carry an interval in TPTL. %

The decorrelation between temporal operators and clock-like variables gives TPTL a greater expressive power than MTL~\cite{bouyer2005expressiveness}:  
any MTL formula can be written equivalently in TPTL, for instance $\until{b}{I}{c}$ can be written in TPTL as $\freeze{x}({\uuntilt{b}{(c \wedge x \in I)}})$, whereas not all TPTL formulae are expressible in MTL as shows the following example.   

\begin{example}\label{ex:tptl}
Consider the following TPTL formula with $\act=\{a,b,c\}$:

$$\varphileq = \freeze{x}{\ueventually{(b \wedge x\leq 1 \wedge (\uglobally{(x\leq 1 \Rightarrow \neg b)})})}$$

The semantics of the above formula equals the set of all infinite timed words satisfying that, within less than $1$ time unit \wrt the timestamp of the first action, an action $b$ occurs at a position $i>1$ and no other $b$ occurs afterwards (we recall that $\uglobally{}$ and $\ueventually{}$ are strict). %
More formally, an infinite timed word $\pi$ belongs to the semantics of the above formula iff $$\exists i > 1 : t_i(\pi) - t_1(\pi) \leq 1 \wedge a_i(\pi) = b \wedge \forall j > i : t_j(\pi) - t_1(\pi) \leq 1 \Rightarrow \neg (a_j(\pi) = b)$$. 

As shown in~\cite{bouyer2005expressiveness}, this formula cannot be expressed in MTL. %
The closest MTL formula would be $\eventually{[0,1]}{(b \wedge \globally{[0,1]}{\neg b)}}$. Yet, this formula is more restrictive as the $[0,1]$ subscript of the \emph{Globally} operator is relative to the occurrence of $b$ and not the that of the first action. %
\end{example}

\begin{theorem}\label{thm:tptl}
Satisfiability and validity of TPTL are undecidable~\cite[Theorem 7]{alur1994really}. 
\end{theorem}

\section{The logic~\logic}
\label{sec:logics}

We present the syntax and semantics of~\logic (\cref{sec:ltrt-HML}), 
our novel extension of (untimed) modal $\mu$-calculus with linear-time %
semantics, and %
prove that~\logic is strictly more expressive than TPTL and MTL (\cref{sec:expressiveness}). 
The results stated in \cref{sec:ltrt-HML} are proven in~\cite[Appendix B]{DBLP:conf/ecoop/Amara0FF25}.

\subsection{\logic Syntax and Semantics}
\label{sec:ltrt-HML}

We first introduce a syntactic characterisation of timing constraints which we call  \emph{guards}. %
In the sequel, let $\mathcal{D}$ be an infinite set of 
variables taking values in~$\rplus$, that we call \emph{timestamp variables}, and let $\mathcal{V}$ be the set
of valuation functions $v\in \mathcal{D}\rightarrow \rplus$. %

\begin{definition}[Guards]
  \label{def:guards}
  \label{def:time-constraints}
  The set $\guards{\mathcal{D}}$ of guards over~$\mathcal{D}$ is given by the syntax 
  $$
  \begin{array}{llllll}
    g & := & x \prec n &  \BNFsep (x - y) \prec n & \BNFsep g \wedge g & \BNFsep  \overline{g}
  \end{array}
  $$
  where $n \in \naturals$, ${\prec} \in \set{ {<}, {\leq} }$,  
  $x,y \in \mathcal{D}$ and $\overline{g}$ is a compact notation for $\neg g$. 
  \end{definition}
  A valuation $v$ \emph{satisfies} a guard $g$, denoted $v \models g$, iff replacing each variable $x$ by  $v(x)$ in $g$ gives an expression that evaluates to true. %
  For example,  $v \models (x - y) < 2$ iff $v(x) - v(y) < 2$. %
  We write $v \not\models g$ interchangeably with $v \models \overline{g}$. %
If $I$ is an interval, then we write $(x-y) \in I$ as an abbreviation 
for the guard $\overline{(x-y) \preccI \rb{\lcomplement{I}}}  \wedge (x-y) \precrI \rb{I}$ (\cref{rk:lcomp}), and $(x-y) \notin I$ for $\overline{((x-y) \in I)}$.  
We write $v\update{x}{m}$, $x \in \mathcal{D}$ and $m \in \rplus$, to denote  
the valuation that maps $x$ to $m$ and every $y\in\mathcal{D}\setminus{\{x\}}$ to $v(y)$. %
We denote by $[x_1 = m_1,\ldots,x_n = m_n]$ the current valuations of some variables of interest $x_1,\ldots,x_n$ when valuation $v$ is understood from the context. %

Our $\logic$ consists of a set of \emph{formulae} with linear-time timed semantics. %
Thus, before defining $\logic$, we first define $\terms$, the set of \emph{terms} that can be used to write a formula ($\logic \subset \terms$). %
The set~$\terms$ 
is generated by the grammar in \cref{fig:lhml-syntax}, with $a\in\act$, $X$ an element of a countable set of
recursion variables~$\vars$,~$x\in\mathcal{D}$, and~$g\in \guards{\mathcal{D}}$.
Syntactically $\minfp{\termF}$ and $\maxfp{\termF}$ are binders of $X$
in $\termF$. The free recursion variables of a term $\termF$ are given by a function
$\fvSym : \terms \longrightarrow \parts{ \vars }$, defined
in the expected way, \eeg $\fv{ \minfp{ X \wedge x \lin Y }} = \set{ Y }$.
A term may contain timestamp variables in $\mathcal{D}$, for instance $\lbox{ x - y  < 0.7} X$
contains the two timestamp variables~$x,y\in \mathcal{D}$. In our presentation, $x \lin \termF$
binds~$x\in\mathcal{D}$ in~$\termF$, and we define accordingly the function
$\fvTstampSym : \terms \longrightarrow \parts{ \mathcal{D} }$ which gives the free timestamp
variables contained in its argument. For instance $\fvTstamp{ \lbox{ x  < 1} X } = \set{ x }$
and $\fvTstamp{ x \lin y \lin \dmd{ x - y  \in [2,4]} X} = \emptyset$.
A term in which no variable is free is {\em closed}.

\begin{figure}[th!]
    \hrulefill
    $$
    \begin{array}{lllllll}
      \phi,\psi & ::= & \true & true 
      & \BNFsep & \false & false\\ 
      & \BNFsep& X & variable
      & \BNFsep& x \lin \phi & time stamp \\
      & \BNFsep& \phi \wedge \psi & conjunction
      & \BNFsep & \phi \vee \psi & disjunction \\
      & \BNFsep & \dmd{a}\phi & diamond
      &\BNFsep & \dmd{g} \phi & diamond\ on\ time \\
      & \BNFsep& \minfp{\phi} & min.\ fixpoint
      & \BNFsep& \maxfp{\phi} & max.\ fixpoint\\
    \end{array}
    $$
    \hrulefill
    \caption{Syntax of $\terms$.}
    \label{fig:lhml-syntax}
    \hrulefill
\end{figure}

Compared to the untimed version of linear-time $\mu$-calculi (in particular linear-time HML in~\cite{DBLP:journals/pacmpl/AcetoAFIL19}), $\terms$ provides two new syntactic constructs to handle dense time: $x \lin \phi$ and $\dmd{g}\phi$. %
These are best discussed when introducing formulae later in this subsection. %

   \newcommand{\semmu}[3]{\sem{#1, #2}_{#3}}

  \begin{figure}[ht!]
    \hrulefill
    $$
      \begin{array}{llllll}
        \semmu{\true}{ \rho}{ v} &=& \twords 
        & \semmu{\false}{ \rho}{ v} &=& \emptyset\\
        \semmu{\phi \vee \psi}{ \rho}{ v} &=& \semmu{\phi}{ \rho}{v} \cup \semmu{\psi}{ \rho}{v} 
        & \semmu{\phi \wedge \psi}{ \rho}{ v} &=& \semmu{\phi}{ \rho}{v} \cap \semmu{\psi}{ \rho}{v}\\
        \semmu{ x \lin{\phi} }{ \rho}{ v} &=& \setof{\pi}{\pi \in \semmu{\phi}{ \rho }{v[x := t_1(\pi)]}}
        &\semmu{\dmd{a}\phi}{ \rho}{ v} &=& \semdiamond{ a }{\semmu{\phi}{ \rho}{ v}} \\
        \semmu{\minfp{\phi}}{ \rho}{ v} &=& \bigcap{\setof{S}{ \semmu{\phi}{ \rho[X \rightarrow S]}{ v } \subseteq S }} 
        &\semmu{X}{ \rho}{ v} &=& \rho(X) \\
        \semmu{\maxfp{\phi}}{ \rho}{ v} &=& \bigcup{\setof{S}{ S \subseteq \semmu{\phi}{ \rho[X \rightarrow S]}{ v}}} 
        &\semmu{\dmd{g}\phi}{ \rho}{ v} &=& \semdiamond{ g }{v, \semmu{\phi}{ \rho}{ v}}\\
      \end{array}
    $$
    \hrulefill
    \caption{Linear time semantics of $\terms$}
    \label{fig:lhml-semantics}
    \hrulefill
  \end{figure}
  
  The semantics of terms over infinite timed words is given in \cref{fig:lhml-semantics},
  where we have used the auxiliary functions %

  $$
  \semdiamond{ a }{-}, \,  \semdiamond{ g }{v, -} : \parts{\twords} \longrightarrow \parts{\twords}
  $$
  defined by the following equalities:

  \begin{equation}
    \label{eq:semantic-functions}
    \begin{array}{l@{\hskip 3pt}l@{\hskip 3pt}ll@{\hskip 3pt}l@{\hskip 3pt}l}
    \semdiamond{ a }{ Z } & =  & \setof{\pi}{a_1(\pi)=a \wedge \pi^2 \in Z} 
    &\semdiamond{ g }{ v,Z } & =  & \setof{\pi}{v \models g \wedge \pi \in Z}
    \end{array}
  \end{equation}
  and where environment $\rho$ is an element of $\setrho$, the set of partial functions $\vars \rightharpoonup \parts{\twords}$,
  and $v$ a valuation in $\mathcal{V}$. %

  Let $\dom{f}$ denote the domain of definition of a partial function $f$. %
  The notation~$\sem{\phi,\rho}_v$, with $\fv{\phi} \subseteq \dom{\rho}$, denotes the set of timed words that satisfy
  $\termF$ \wrt 
  the environment $\rho$ and the valuation~$v$. %

\paragraph{Properties of the semantics.}
For any $\termF \in \terms$ , $v\in\mathcal{V}, X\in\vars$ and $\rho\in\setrho$ such that
$\fv{ \termF  } \subseteq \dom{\rho} \cup \set{ X }$ we define the \emph{semantic function} 
\begin{equation}
  \label{eq:widehat}
  \begin{array}{lll}
    \funof{\termF }{v}{X}{\rho} & : & \parts{\twords} \longrightarrow \parts{\twords}
    \\[5pt]
    \funof{\termF }{v}{X}{\rho}(Z) & =  & \semmu{\termF}{\rho \semsubst{ X }{  Z }}{v}
  \end{array}
\end{equation}
In other words $\funof{\termF}{v}{X}{\rho} =  \lambda y . \semmu{\termF}{\rho[ X \mapsto y ]}{v}$.

Since environments are functions from logical variables to complete
lattice, we order them in the standard pointwise way to
obtain a complete lattice $\poset{\vars \rightarrow
  \parts{\twords}}{\leq}$,
where 
infinitary least upper bounds and greatest lower bounds are defined as usual:
$$
\begin{array}{lll}
  (\bigvee \set{ \rho_1, \rho_2, \ldots })(X) & = & \bigcup\set{
    \rho_1(X), \rho_n(X), \ldots  }
  \\
  (\bigwedge \set{ \rho_1, \rho_2, \ldots })(X) & = & \bigcap\set{
    \rho_1(X), \rho_2(X), \ldots }
\end{array}
$$

The semantic function (\cref{eq:widehat})
is monotone \wrt the pointwise ordering of environments, therefore
Knaster-Tarski's fixpoint theorem 
ensures that the largest and least fixpoints of every function
$\funof{\termF}{v}{X}{\rho}$ exist, and are the semantics of the terms
defined via $\minfp{-}$ and $\maxfp{-}$:
\begin{equation}
  \label{eq:consequence-tarski-fixed-point}
  \begin{array}{lllll}
    \semmu{\maxfp{\termF}}{\rho}{v}
    & = & \bigcup \setof{Z}{ Z \subseteq \funof{\termF}{v}{X}{\rho}(Z)}
    & = & \nu y.\funof{\termF}{v}{X}{\rho}(y)
    \\
    \semmu{\minfp{\termF}}{\rho}{v}
    & = & \bigcap \setof{Z}{\funof{\termF}{v}{X}{\rho}(Z) \subseteq Z}
    & = & \mu y.\funof{\termF}{v}{X}{\rho}(y)
  \end{array}
\end{equation}

The semantic function is continuous
\wrt directed sets over the CPO $\poset{\parts{\twords} }{ \subseteq
}$.
An immediate consequence of this 
and of \cref{eq:consequence-tarski-fixed-point} is 
that we can reason on terms defined via~$\minfp{-}$
(resp. $\maxfp{-}$) using Kleene fixpoint theorem (\resp its formal dual):
for every term $\minfp{\termF} \in \terms$, $v\in\mathcal{V}$, and $\rho\in\setrho$ such
that $\fv{\termF} \subseteq \dom{\rho} \cup \set{X}$
  we have
  \begin{equation}
    \label{eq:inductive-proof-method-min}
    \begin{array}{lll}
    \semmu{\minfp{\termF}}{\rho}{v}
    & 
    = &\bigcup_{n \in
      \naturals}\left(\funofN{\termF}{v}{X}{\rho}{n}(\emptyset)
    \right)
    \\
    \semmu{\maxfp{\termF}}{\rho}{v}
    &= &\bigcap_{n \in \naturals}\left(\funofN{\termF}{v}{X}{\rho}{n}(\twords)
    \right)
    \end{array}
  \end{equation}
  where $\funofN{\termF}{v}{X}{\rho}{ n }$ is defined by induction on~$n$ thus,
  \begin{equation}
    \label{eq:function-iteration}
  \funofN{\termF}{v}{X}{\rho}{ n } = 
  \begin{cases}
    \mathit{id} & \text{ if } n = 0,\\
    \funof{\termF}{v}{X}{\rho} \circ \funofN{\termF}{v}{X}{\rho}{n-1} & \text{ if } n \geq 1.
  \end{cases}
  \end{equation}

  \paragraph{Formulae.}
We say that a variable $X \in \vars$ is {\em guarded} in a term~$\termF$ if all free occurrences of~$X$ in $\termF$ are 
in the scope of at least 
one modality (\ie~$\dmd{ a }{-}$, $a\in\act$). 
We say that a term~$\termF$ is {\em guarded}
if for all its subterms
$\minfp[X]{\termG_1}$ and $\maxfp[Y]{\termG_2}$,
$X$ is guarded in $\termG_1$ and $Y$ is guarded in $\termG_2$.
For example, the term $\minfp[X]{\dmd{a}x \lin \lbox{x \leq 2}X \vee \dmd{b}Y}$ is guarded,
and the term $\minfp[X]{\dmd{a}x \lin \lbox{x \leq 2}X \vee X}$ is
not, because the rightmost occurrence of $X$ is not in the scope of a
modality. 
The term $\minfp{x \lin [x < 3] X}$ is not guarded, because neither $x\lin$ nor $[x < 3]$ is
 a modality. 
We may now define $\logic$, the set of formulae in $\terms$. 

\begin{definition}
\label{def:formulae}
The set of formulae $\logic$ is defined as the set of closed and guarded terms in $\terms$: 
$\logic = \{\phi\in\terms\, \, |\, \, \phi \text{ is guarded and } \fv{\phi} = \fvTstamp{\phi} = \emptyset\}$
\end{definition}  
Since elements of $\logic$ are closed, their semantics 
does not depend on valuations and environments. 
As a consequence, we simply write $\sem{ \termF }$ instead of $\sem{ \termF, \rho }_{v}$ to denote the semantics of any~$\termF \in \logic$ in any environment $\rho\in\setrho$ and \wrt to any  $v\in\mathcal{V}$. %

\paragraph{Syntactic sugar}
We use the notation $\bar{a}$ to denote the set $\act \setminus \{a\}$.
For $S \subseteq \act$, we write $\dmd{S}\phi$ as an abbreviation for
$\bigvee_{a \in S}\dmd{a}\phi$ (if $S$ is empty, then as expected the empty disjunction, and therefore $\dmd{S}\phi$, is equivalent to $\false$).
To improve the readability of formulae and of the complementation procedure that we will develop later on, we introduce the \emph{box operators} 
$\lbox{a}\phi$ and $\lbox{g}\phi$ as
abbreviations for
$\dmd{a}\phi \vee \dmd{\bar{a}}\true$ and
$\dmd{g}\phi \vee \dmd{\bar{g}}\true$, respectively. 

\begin{remark}
\label{rk:alpha-conv}
Note that we work up to $\alpha$-conversion of bound timestamp and recursion variables.  %
For clarity, we assume these to be distinct and pick fresh ones when manipulating formulae. We also assume each recursion variable to be bound at most once, \eeg the formula $\minfp{\dmd{a}X \vee \dmd{b}X} \vee \maxfp{\dmd{c}X}$ can be expressed as $\minfp{\dmd{a}X \vee \dmd{b}X} \vee \maxfp[Y]{\dmd{c}Y}$.
\end{remark}

Let us now give two examples of the semantics of a formula $\termF \in \logic$, as to explain concretely the meaning of the new constructs $x \lin \phi$ and $\dmd{g}\phi$, relying on~\cref{fig:lhml-semantics}. %

\begin{example}
\label{ex:simple-example}
Let $\phi\in\logic$ be the following formula:
  $$\phi = x \lin \lbox{x \leq 5}\dmd{b}\true$$
  The semantics of this formula equals the set of all infinite timed words $\pi$ that satisfy the following: either $t_1(\pi) > 5$ or $t_1(\pi) \leq 5$ \emph{and} $a_1(\pi) = b$. %
  That is,  $\phi$ expresses the property: ``if the first timestamp is smaller than $5$, then the action associated with it must be a $b$''.  
\end{example}

\begin{example}
  \label{ex:not-in-mtl} 
  Let us discuss a more complex example. %
  The following untimed formula, given in~\cite[Ex. 4.2]{aceto2021operational}, expresses the property ``every other action is a $b$'':
  $$\phi_{even(b)} = \maxfp{\dmd{\act}\dmd{b}X}$$

 We want a timed version of this formula, \ie to express $\phi_{even(b),I}$, the generalisation of $\phi_{even(b)}$ to any interval $I\in\mathcal{I}$. %
  More precisely, the semantics of $\phi_{even(b),I}$ must contain every infinite timed word $\pi$ that satisfies the following: for every even position $i\in\naturals_{>1}$, \ie ${i}\bmod{2} = 0$, if $t_i(\pi) - t_1(\pi)$ is within interval $I$, then $a_i(\pi)$ must be a $b$, and $a_i(\pi)$ can be any action otherwise. %
  For this, we use the constructs $x\lin - $ and $\dmd{g} - $ to build up the following formula:
  
  $$\phi_{even(b),I} = x_s\lin \maxfp{\dmd{\act} (x_e \lin \dmd{x_e - x_s\in I}\dmd{b}X \vee y \lin  \dmd{y - x_s\notin I}\dmd{\act}X)}$$

 Let us explain through evaluating whether some infinite timed word $\pi$ belongs to the semantics of $\phi_{even(b),I}$. %
  The $x_s\lin - $ is used outside of the maximal-fixpoint operator, allowing timestamp variable $x_s$ (for $x$ ``start'') to record the first timestamp in $\pi$. %
  Then, each recursion step contains two consecutive actions (as in $\phi_{even(b)}$), but this time with two disjoint recursion ``paths'' (the terms $x_e \lin \dmd{x_e - x_s\in I}\dmd{b}X$ and $y \lin  \dmd{y - x_s\notin I}\dmd{\act}X)$. %
  The first path requires the second action to be a $b$ occurring within $I$, and the second allows the second action to be any action in $\act$ occurring outside of $I$. %
  To assess whether an action occurs \eeg in $I$, the first recursion path uses a timestamp variable $x_e$ (for $x$ ``end''), updated with the timestamp of every other action in $\pi$, and requires the difference between $x_e$ and $x_s$ to belong to $I$ ($x_e \lin \dmd{x_e - x_s\in I}$). %
  Therefore, the only condition for $\pi$ to be outside of the semantics of $\phi_{even(b),I}$ is the existence of some even position at which the action is not a $b$ \emph{and} occurs within $I$. %
  For instance, for $I=[3,5]$ and any $\pi,\pi',\pi''\in\twords$ starting respectively with the prefix $(a,3)(c,4)(a,7)(c,9)$, $(a,3)(b,4)(b,7)(b,8)(c,10)$ and $(a,2)(b,5)(c,6)(c,7)$ we have: 
  \begin{itemize}
  \item{} $\pi\in\sem{\phi_{even(b),[3,5]}}$ because there is no timed action at an even position that happens within $[3,5]$: at even position $2$ we have $4-3 \notin [3,5]$, at the next even position $4$ we have $9-3 \notin [3,5]$, and since at this position the right bound of the interval is already exceeded then actions at subsequent even positions will occur outside of $I$,
  \item{} $\pi'\in\sem{\phi_{even(b),[3,5]}}$ because the only even position with actions occurring within $I$ is position $4$ and we have $a_4(\pi') = b$, 
  \item{}  $\pi''\notin\sem{\phi_{even(b),[3,5]}}$ because at position $4$ an action occurs within $I$ ($7-2 \in [3,5]$) and such action is not a $b$.  
  \end{itemize}
  In general, we have: $$\pi\in\sem{\phi_{even(b),I}} \text{ iff } \forall i\in\naturals_{>1}: {i}\bmod{2} = 0 \wedge t_i(\pi) - t_1(\pi) \in I \implies a_i(\pi) = b$$
  which coincides with the generalisation of $\phi_{even(b)}$ to any interval $I\in\mathcal{I}$. 
  Note that when $I = [0,\infty)$, we obtain exactly the untimed version $\phi_{even(b)}$ using the fact that $x_e - x_s \in [0,\infty)$ is always true and $y - x_s \notin [0,\infty)$ is always false.   
         \end{example}

\subsection{Closure under complement}\label{sec:clos}
In order to handle the case $\neg \phi$ when translating MTL or TPTL formulae into \logic, we prove the more generic property of closure under complement of $\terms$. %
  This is done 
  by the function $\mathsf{not} : \terms \to \terms$ which maps any term $\phi$ to its dual (\ie a term whose semantics is $\twords\setminus{\sem{\phi}}$),
  defined recursively as follows (note that we make use of the box modalities for readability)
  $$
    \begin{array}{llllll}
      \tlinnot{\true} &=& \false & \tlinnot{\false} &=& \true \\ 
      \tlinnot{X} &=& X & \tlinnot{x \lin \phi} &=& x \lin \tlinnot{\phi} \\ 
      \tlinnot{\phi \wedge \psi} &=& \tlinnot{\phi} \vee \tlinnot{\psi} & \tlinnot{\phi \vee \psi} &=& \tlinnot{\phi} \wedge \tlinnot{\psi} \\ 
      \tlinnot{\dmd{a}{\phi}} &=& \lbox{a}{\tlinnot{\phi}} &  \tlinnot{\dmd{g}{\phi}} &=& \lbox{g}{\tlinnot{\phi}} \\
      \tlinnot{\maxfp{\phi}} &=& \minfp{\tlinnot{\phi}} & \tlinnot{\minfp{\phi}} &=& \maxfp{\tlinnot{\phi}} \\ 
    \end{array}
  $$

  \begin{proposition}
    \label{prop:complementation-tlin}
    For all $\phi \in \logic$, $\sem{\tlinnot{\phi}} = \compl{\sem{\phi}}$.
  \end{proposition}
  
  The proof of~\cref{prop:complementation-tlin} follows immediately from the proof of the following more general lemma. 
  
  \begin{restatable}{lemma}{closure}\label{lem:clos}
  Let $\bar{\rho}$ 
    denote the environment that maps any $X \in \dom{\rho}$ to $\compl{\rho(X)}$. For all $\phi \in \terms$, $v \in \mathcal{V}$ 
    and $\rho \in \setrho$ with $\fv{\phi} \subseteq \dom{\rho}$, the following holds: 
    $$
      \sem{\tlinnot{\phi},\bar{\rho}}_v = \compl{\sem{\phi,\rho}}_v
    $$
    \end{restatable}
    The proof of~\cref{lem:clos} is detailed in~\cref{app:clos}.

\subsection{Expressiveness}\label{sec:expressiveness}
The proofs related to this subsection are detailed in~\cref{sec:appendix-tohml}. %

\subsubsection{From MTL to \logic}\label{sec:mtllog}

We show that every formula in MTL has a counterpart with exactly the same semantics in \logic. %
We proceed through proposing the function $\tohmlSym : MTL \longrightarrow \logic$, 
defined in~\cref{fig:mitl-compiler}, that transforms any MTL formula into an equivalent \logic formula.

\begin{restatable}{proposition}{totlin}%
  \label{prop:correctness-tohml}
  \label{prop-correctness-tohml}
  For every~$\phi \in MTL$ we have~$\sem{\phi} = \sem{ \tohml{\phi} }$.
\end{restatable}

\begin{figure}[ht!]
  \hrulefill
    $$
      \begin{array}{lll}
        \tohml{\top} &=& \true
        \\[3pt]
        \tohml{\bot} &=& \false
        \\[3pt]
        \tohml{a} &=& \dmd{a}\true
        \\[3pt]
        \tohml{\neg \phi} &=& \tlinnot{\tohml{\phi}}
        \\[3pt]
        \tohml{\phi \wedge \psi} &=& \tohml{\phi} \wedge \tohml{\psi}
        \\[8pt]
        \tohml{\until{\phi}{I}{\psi}} &=&
        x_s \lin \dmd{\act} \minfp[Y]{x_e \lin (\dmd{x_e - x_s \in I}{\tohml{\psi}}) \\ 
          && \phantom{x_s \lin \dmd{\act} \minfp[Y]{x_e}} \vee (\tohml{\phi} \wedge \dmd{ \act }{Y})}
      \end{array}
      $$
  \hrulefill
  \caption{From MTL to \logic.}
  \label{fig:mitl-compiler}
  \hrulefill
\end{figure}

\begin{example}
  \label{ex:bUIC-toTlin} 
  Let $b$,$c$ $\in \act$. We explain how $\tohmlSym$ preserves the semantics of $\until{b}{I}{c}$ (\cref{ex:until}). %
  Using the rules in~\cref{fig:mitl-compiler}, we obtain the following $\logic$ formula: 
  $$\phi = \tohmlSym(\until{b}{I}{c}) = x_s \lin\dmd{\act} \minfp[Y]{((x_e \lin \dmd{x_e - x_s \in I}{\dmd{c}\true}) \vee  
    ({\dmd{b}\true} \wedge {\dmd{\act}{Y}}))}$$
  Let $\psi_1 = (x_e \lin \dmd{x_e - x_s \in I}{\dmd{c}\true})$ and $\psi_2 = {\dmd{b}\true} \wedge {\dmd{\act}{Y}}$. 
  We want to evaluate whether some infinite timed word $\pi$ belongs to the semantics of~$\phi$. %
  Similarly to~\cref{ex:not-in-mtl}, $x_s$ evaluates constantly to $t_1(\pi)$, whereas $x_e$ is updated when progressing through~$\pi$, starting from position $2$ (the action at position $1$ is ``skipped'' due to the $\dmd{\act}$ preceding the min). %
         That is, $x_e$ equals initially $t_2(\pi)$ and, at each recursion step (${\dmd{\act}{Y}}$), 
         $Y$ is replaced by $\minfp[Y]{\psi_1\vee \psi_2}$ and $x_e$ is updated with the timestamp $t_i$ of the timed event in $\pi$ triggering the recursion. %
         Within $\phi$, the semantics of $\psi_1$ contains $\pi^i$ iff the difference between $t_i$ and $t_1$ is within the interval $I$ \emph{and} $a_i(\pi)$, the action associated with $t_i$, is a $c$, whereas $\pi^i$ belongs to the semantics of $\psi_2$ iff $a_i(\pi)$ is a $b$ \emph{and} the continuation of $\pi^i$ (\ie $\pi^{i+1}$) is in the semantics of $\minfp[Y]{\psi_1\vee \psi_2}$. 
         Following the minimal fixpoint semantics, $\pi$ is therefore in $\sem{\phi}$ iff there exists a finite number of recursions through~$\pi^2$ such that $\psi_1$ or $\psi_2$ holds,  %
         \ie $\pi\in\sem{\phi}$ iff~$b$ always holds from position $2$ included  until~$c$ holds within~$I$. 
         That is, $\phi$ expresses the same property as $\until{b}{I}{c}$ (\cref{ex:until}).%
         \end{example}

\subsubsection{From TPTL to \logic}\label{sec:tptllog}
Similarly to MTL, we provide a semantic-preserving translation 
$\totlinSym$ from TPTL terms to $\terms$ (\cref{fig:tptl-to-tlin}), and require the input of $\totlinSym$ to be a TPTL formula guaranteeing the output to be a formula as well, \ie in \logic. %

\begin{restatable}{proposition}{totlintwo}%
  \label{prop:correctness-totlin}
  \label{prop-correctness-totlin}
  For every formula~$\phi \in TPTL$ we have~$\sem{\phi} = \sem{ \totlinn{\phi} }$.
\end{restatable}

\begin{figure}[!t]
  \hrulefill
  $$
    \begin{array}{lll}
    \totlinn{\top} &=& \true \\[0.1em]
    \totlinn{\bot} &=& \false \\[0.1em]
      \totlinn{a} &=& \dmd{a}\true \\[0.1em]
      \totlinn{\phi \wedge \psi} &=& \totlinn{\phi} \wedge \totlinn{\psi} \\[0.1em] 
      \totlinn{\neg \phi} &=& \tlinnot{\totlinn{\phi}} \\[0.1em]
      \totlinn{\uuntil{\phi}{\psi}} &=& \dmd{\act}{\minfp{\totlinn{\psi} \vee (\totlinn{\phi} \wedge \dmd{\act}X)}} \\[0.1em]
      \totlinn{\freeze{x}{\phi}} &=& x \lin \totlinn{\phi} \\[0.1em]
      \totlinn{x \sim c} &=& y \lin \dmd{y - x \sim c}\true \\[0.1em]
    \end{array}
  $$
  \hrulefill
  \caption{From TPTL terms to $\terms$.}
  \hrulefill
  \label{fig:tptl-to-tlin}
\end{figure}

\begin{example}
Let us consider the TPTL formula $\varphileq$ of~\cref{ex:tptl}, not expressible in MTL:

$$\varphileq = \freeze{x}{\ueventually{(b \wedge x\leq 1 \wedge (\uglobally{x\leq 1 \Rightarrow \neg b})})}$$ 
Its translation into \logic gives the following formula (recalling that $\ueventually{\phi}\equiv\uuntil{\top}{\phi}$ and $\uglobally{\phi}\equiv\neg(\uuntil{\top}{\neg\phi})$):
$$
\begin{aligned}
\totlinn{\varphi} &=  x \lin \dmd{\act}  \minfp{(y \lin \dmd{y - x \leq 1}\true \wedge\dmd{b}\true \wedge \\
&\tlinnot{\dmd{\act}\minfp[Y]{z \lin \dmd{z - x \leq 1}\true\wedge\dmd{b}\true \vee (\true \wedge\dmd{\act}Y)}}\vee (\true \wedge \dmd{\act}X)}
\end{aligned}
$$

Applying the complementation function and simplifying $\true \wedge \psi$ into $\psi$, $\psi\in\{\dmd{\act}Y,\dmd{\act}X\}$ we get

$$
\begin{aligned}
&= x \lin \dmd{\act}  \minfp{(y \lin \dmd{y - x \leq 1}\true \wedge\dmd{b}\true \wedge \\
&\bigwedge_{a\in\act}\lbox{a}\maxfp[Y]{(z \lin \lbox{z - x \leq 1}\false\vee\lbox{b}\false)\bigwedge_{a\in\act}\lbox{a}Y}\vee \dmd{\act}X}
\end{aligned}
$$

The semantics of $\psi$ coincides with the set of all infinite timed words such that there exists an action $b$ at a position strictly greater than $1$ (due to the first $\dmd{\act}$), necessarily occurring (due to the minimal fixpoint operator) within $1$ time unit relative to the first timestamp (recorded in $x$ outside of the min). %
Now, the maximal fixpoint operator ensures that, starting from the action following such $b$, the only way to satisfy the formula as long as the bound $1$ relative to the first timestamp is still not exceeded, is to have an action different than $b$, otherwise both $\lbox{z - x \leq 1}\false$ and $\lbox{b}\false$ evaluate to false. %
When the bound is exceeded, $\lbox{z - x \leq 1}\false$ evaluates to true and recursion can be carried on infinitely within the max. %
Therefore, for some $\pi$ to satisfy $\psi$, there must exist a position $i>1$ such that $t_i(\pi) - t_1(\pi) \leq 1$, $a_i(\pi) = b$, and for every $j > i$, $t_i(\pi) - t_1(\pi) \leq 1$ implies $a_i(\pi) \neq b$, which corresponds to the formal definition of $\sem{\varphileq}$ given within~\cref{ex:tptl}. 
\end{example}  

The following theorem follows directly from~\cref{thm:tptl} and~\cref{prop-correctness-totlin}. 

\begin{theorem}\label{thm:tlin}
Satisfiability and validity of \logic are undecidable. 
\end{theorem}  

\subsubsection{\logic is strictly more expressive than TPTL and MTL}
\label{sec:expre}

We finally prove that \logic is strictly more expressive than both MTL and TPTL. %
Since TPTL is already known to be more expressive than MTL~\cite{bouyer2005expressiveness}, it suffices thus to prove that \logic is more expressive than the former, as states the following theorem.  

\begin{theorem}
\label{thm:expressivity}
\begin{enumerate}
\item $\forall \phi\in \text{ TPTL }\exists\psi\in\logic: \sem{\phi} = \sem{\psi}$,\label{thm:expressivity1}
\item $\exists \psi \in \logic\, \,  \forall \phi\in \text{ TPTL }: \sem{\psi} \neq \sem{\phi}$.\label{thm:expressivity2}
\end{enumerate}
\end{theorem}

The proof of~\cref{thm:expressivity1} is direct from~\cref{prop-correctness-totlin}. To prove~\cref{thm:expressivity2}, it suffices to find a formula $\psi\in\logic$ that is not expressible in TPTL. %
As a matter of fact, the formula $\phi_{even,I}$ that we used as an example for $\logic$ (\cref{ex:not-in-mtl}) is not expressible in TPTL, as states the following proposition%
:
\begin{proposition}
\label{prop:phieven}
For any $I\in\mathcal{I}$, the formula $$\phi_{even(b),I} = x_s\lin \maxfp{\dmd{\act} (x_e \lin \dmd{x_e - x_s\in I}\dmd{b}X \vee y \lin \dmd{y - x_s\notin I}\dmd{\act}X)}$$ 
is not expressible in TPTL.
\end{proposition}

The proof of~\cref{prop:phieven} follows directly from~\cite[Sect. 3.4]{alur1993real}. %
This concludes the proof of~\cref{thm:expressivity2} of~\cref{thm:expressivity}, and therefore the proof of~\cref{thm:expressivity}. %
Accordingly, \logic is strictly more expressive than TPTL, and by~\cite{bouyer2005expressiveness}, strictly more expressive than MTL as well. 
We summarise these results in~\cref{venn}, where $\sem{L} = \setof{\sem{\phi}}{\phi\in L}$, with $L\in \{MTL, TPTL, \logic\}$,  %
and the semantics of example formulae is denoted by thick dots for simplicity. %

As we will see in~\cref{sec:monitorability}, $\phi_{even(b),I}$ is monitorable for violations in the general case (and also for satisfactions whenever $I$ is bounded), which means that \logic allows
for a larger set of monitorable formulae than TPTL and MTL. %

\begin{figure}[htbp]
    \centering
    \begin{tikzpicture}[
        set/.style={draw, thick, rounded corners=10pt},
        element/.style={circle, fill=black, inner sep=2pt},
        every node/.style={font=\normalsize}
    ]
        
        \draw[set] (0,0) rectangle (14,7);
        \node[anchor=north west] at (0.3,6.7)
            {$\sem{\logic}$};

        \draw[set] (1,0.6) rectangle (10.5,6);
        \node[anchor=north west] at (1.3,5.7)
            {$\sem{\mathrm{TPTL}}$};

        \draw[set] (2,1.2) rectangle (7.5,5.1);
        \node[anchor=north west] at (2.3,4.8)
            {$\sem{\mathrm{MTL}}$};

        \node[element] at (3,3) {};
        \node[anchor=west, align=left] at (3.25,3)
            {$\sem{\until{b}{I}{c}}$\\[-1pt]
             {\footnotesize \cref{ex:until}}};

        \node[element] at (8,3) {};
        \node[anchor=west, align=left] at (8.25,3)
            {$\sem{\varphileq}$\\[-1pt]
             {\footnotesize \cref{ex:tptl}}};

        \node[element] at (11.2,3) {};
        \node[anchor=west, align=left] at (11.45,3)
            {$\sem{\phi_{even(b),I}}$\\[-1pt]
             {\footnotesize \cref{ex:not-in-mtl}}};

    \end{tikzpicture}
    \caption{Expressiveness of \logic}
    \label{venn}
\end{figure}

\subsubsection{Ease of Expression}
We conclude this section with an informal comparison regarding the ease or \emph{flexibility} in expression between \logic and TPTL. %

\begin{example}\label{ex:alt}
Consider the following \logic formula:

$$\phi_{\mathit{alt},I} = \maxfp{x_b \lin (\dmd{x_b \in I} \dmd{b} x_c \lin (\dmd{x_c \in I}\dmd{c} X \vee \dmd{x_c \notin I}\dmd{\act}X) \vee \dmd{x_b \notin I}\dmd{\act}X}$$

The formula $\phi_{\mathit{alt},I}$, parameterised with non-empty interval $I\in \iv$, expresses the following precise property over infinite timed words: 
\begin{itemize}
\item{} For all timestamps outside the interval $I$, there is no requirement on the actions, 
\item{} The smallest timestamp belonging to $I$ must be associated with action $b$, 
\item{} Actions must alternate between $b$ and $c$ as long as they occur within $I$. 
\end{itemize}

For example, given $I = (2,7]$ and $\act = \{a,b,c\}$, any infinite timed word starting with: 

\begin{itemize}
\item{}  $(a,2)(a,7.2)$ is in the semantics of $\phi_{\mathit{alt},I}$ (there are no actions in the absolute interval $I = (2,7]$), 
\item{} $(c,2.5)(b,3.2)(c,7)(b,8)$ is \emph{not} in the semantics of $\phi_{\mathit{alt},I}$ since the first action occurring within $(2,7]$ is not a $b$ (even though the alternation between $b$ and $c$ is respected within $(2,7]$), 
\item{} $(b,6)(a,7.3)$ is in the semantics of $\phi_{\mathit{alt},I}$. 
\end{itemize}
\end{example}

The formula $\phi_{\mathit{alt},I}$ in~\cref{ex:alt} is not expressible in MTL for any $I\in\iv$ different that $[0,\infty)$%
, following the same arguments used in~\cite{bouyer2005expressiveness} regarding the formula of~\cref{ex:tptl}. 
In TPTL, however, its expressibility depends on the model of timed words. %
In brief, this formula cannot be expressed in TPTL unless we force all timed words to start with timestamp $0$. %
This is because the interval $I$ is used in its absolute sense in $\phi_{\mathit{alt},I}$, \ie there is no reference to a particular timestamp to ``subtract'' within guards. %
In contrast, a guard in TPTL $x\in I$ forces an implicit subtraction of the value of $x$ when it was last frozen from its current value (\cref{sec:tptl}), making it necessary to have a particular timestamp as a reference (a simpler example is provided below). 

Now, this is not an expressiveness result \emph{per se}, as extending actions with a ``special'' symbole in order to have all timed words start with timestamp $0$ is a common and even necessary practice to delineate factual from artificial expressiveness results (see \eeg~\cite{bouyer2005expressiveness,DBLP:conf/time/Foughali26}). %
Under such assumption, $\phi_{\mathit{alt},I}$ can be expressed in TPTL as shows the following example. 

\begin{example}\label{ex:altt}
Over the set of infinite timed words starting with timestamp $0$, \ie over $\setof{\pi\in\twords}{t_1(\pi) = 0}$, the following TPTL formula is equivalent to $\phi_{\mathit{alt},I}$ from~\cref{ex:alt}: 

$$\psi_{\mathit{alt},I} = x.(\psi_1\wedge  \psi_2 \wedge \psi_3 \wedge \psi_4)$$

Where $\psi_1$ (\resp $\psi_2$) ensures the first action to be a $b$ if $I$ is left-closed at $0$ (\resp otherwise): 

$$\psi_1 = x\in I \Rightarrow b$$

$$\psi_2 = \uwglobally ((x\preccI \rb{\lcomplement{I}} \wedge \unxt{x\in I}) \Rightarrow \unxt{b})$$

And $\psi_3$ and $\psi_4$ guarantee alternation between $b$ and $c$ inside $I$: 

$$\psi_3 = \uwglobally ((x\in I \wedge b \wedge \unxt{x\in I}) \Rightarrow \unxt{c})$$

$$\psi_4 = \uwglobally ((x\in I \wedge c \wedge \unxt{x\in I}) \Rightarrow \unxt{b})$$

\end{example}

The above examples show however that, in contrast to its lengthy formulae, \logic comes with a noticeable flexibility thanks to the mechanism of its timestamp variables, decoupling recording a timestamp from subtracting two values, as opposed to TPTL's clock-like variables with implicit, hardcoded subtraction. %
Actually, very simple formulae such as $x \lin \dmd{x\leq 3}\true$ turn out to be inexpressible in TPTL unless forcing all timed words to start with timestamp $0$. 
Indeed, the above formula requires that the first timestamp is less than $3$, yet in TPTL one needs first to freeze a reference time at some position. 
The formula $\freeze{x}{x\leq 3}$, the closest to $x \lin \dmd{x\leq 3}\true$, will then freeze $x$ at the first timestamp and compare instead the difference between such timestamp and $x$, therefore giving always $0$. %
If all timed words are forced to start with timestamp $0$, $x \lin \dmd{x\leq 3}\true$ becomes equivalent to $\true$ and therefore trivially expressible in TPTL. %

\section{Monitoring a System}%
\label{sec:monitoring}

In this section, we formally define \emph{timed monitors} (\cref{sec:monitors}), hereafter referred to simply as monitors, and the special
instrumentation operator allowing to compose an arbitrary timed system of interest, hereafter denoted a \emph{system under monitoring} (SUM), with a monitor (\cref{sec:instrumentation}). %
The proofs related to~\cref{sec:monitors} and~\cref{sec:instrumentation} can be found in~\cite[Appendix D]{DBLP:conf/ecoop/Amara0FF25}. %
  
Moreover, we define a new \emph{reactive} semantics for monitors allowing to abstract away the instrumentation operator without loss of generality, and to generalise accordingly the maximality results of~\cite{DBLP:conf/ecoop/Amara0FF25}, requiring monitors to be reactive by design, to all monitors. %

\subsection{Monitors Syntax and Semantics}
\label{sec:monitors}
\begin{figure}[!h]
  \hrulefill
  $$
  \begin{array}{lllllllll}
    M, N & ::= & \sver & {\BNFsep}a.M & {\BNFsep}g.M
    & {\BNFsep}M + N & {\BNFsep}\rec{M} & {\BNFsep}X & {\BNFsep}\letin{x}{M}\\
    \sver & ::= & \mend & {\BNFsep}\no & {\BNFsep}\yes
  \end{array}
  $$
  \hrulefill
  \caption{Syntax of m-terms in $\tmon$.}
  \label{fig:syntax-monitors}
  \hrulefill
\end{figure}

We first define $\tmon$, the set of monitor terms, hereafter abbreviated m-terms. %
$\tmon$ is given by the grammar in
\cref{fig:syntax-monitors} where $x \in \mathcal{D}$, $a \in \act$, $g \in \guards{\mathcal{D}}$ and $X \in \vars$.
$\yes ,\no, \mend$ are special m-terms called \emph{verdicts}, corresponding respectively 
to acceptance, rejection and termination (\ie an inconclusive outcome).  %
By abuse of notation, we write $E$ to denote $L(E)$, the set of m-terms generated by the grammar of \cref{fig:syntax-monitors} starting from 
the nonterminal symbol $E$ \cite{Knuth1968SemanticsOC}. %
 
Notions of guardedness and free variables are defined similarly as for $\terms$ (\cref{sec:ltrt-HML}). 
In an m-term, $\rec[X]{M}$ (\resp $\letin{x}{M}$) is a binder  of the recursion variable $X$ (\resp timestamp variable $x$). %
The free recursion (\resp timestamp) variables of an m-term $M$ are given by the function 
$\mathit{mfv} : \tmon \longrightarrow \parts{\vars}$ (\resp $\mathit{mfv_{\mathit{time}}} : \tmon \longrightarrow \parts{\mathcal{D}}$), 
defined in the expected way. For example, $\mfv{a.\rec[X]{(b.Y + (x \leq 2).X)}} = \set{Y}$, 
$\mfvtime{(x - y \leq 2).X} = \set{x,y}$ and $\mfvtime{\letin{x}{\letin{y}{(x - y \leq 2).X}}} = \emptyset$. %
A recursion variable $X$ is guarded in an m-term $M \in \tmon$ if every occurrence of $X$ in $M$ 
is in the scope of at least one modality $(a.)$, and 
$M \in \tmon$ is guarded if for all its subterms $\rec[X]{N}$, $X$ is guarded in $N$. %
For example the m-term $\rec[X](a.(\letin{x}g.X \extc Y))$ is guarded, and the m-terms 
$\rec[X](a.X \extc X)$, $\rec[X]{\letin{x}X}$ and $\rec[X]{g.X}$ are not guarded. %
We may now define the set of monitors $\mon$. 
\begin{definition}
\label{def:mon}
The set of monitors $\mon$ is defined as the guarded m-terms in $\tmon$ closed vis-à-vis recursion variables: 
$$\mon = \{M\in\tmon\, \, | \, \, \text{ M is guarded and } \mfv{M} = \emptyset\}$$
\end{definition}
For example, $\rec[X](a.\letin{x}g.X \extc Y) \notin\mon$ because it is not closed \wrt recursion variable $Y$, whereas $\rec[X]{\letin{y}{(x - y \leq 2)}.a.X}\in\crmon$.

\begin{remark}
Note that, compared to that of formulae (\cref{def:formulae}), \cref{def:mon} is more relaxed as monitors are allowed to have free timestamp variables. %
The practical reason behind this is illustrated later in this subsection (\cref{ex:imon}). %
\end{remark}  

\begin{remark}
When manipulating monitors, we assume the same conventions as in~\cref{rk:alpha-conv}; in particular that fresh variables are picked at each recursion step.
\end{remark} 

The semantics of \emph{all} monitors in $\mon$ is given by the TLTS\,  $\mathbb{M} =  \langle S_\mathbb{M} , \labb, \longrightarrow_\mathbb{M} \rangle$. %
A state $m\in S_\mathbb{M}$ is of the form $(M,v_c,v_t)$ where $M$ is a monitor (in $\mon$), $v_c\in\rplus$ the value of \emph{monitor time} (to record the ``global'' time) and $v_t\in\mathcal{V}$ a valuation (to update timestamp variables and evaluate guards). %
The rules of the transition relation $\longrightarrow_\mathbb{M}$ are given by \cref{fig:sos-monitors},
where $a \in \act$ and $d \in \rplus$. %
In~\cref{fig:sos-monitors} and hereafter, we drop the subscript $_{\mathbb{M}}$ from (weak or strong) transition symbols whenever clear from the context. 
The rule \textsc{mVer} states that a verdict state, \ie a state of the form $(M,-,-)$, with $M \in E$ can 
perform a transition with any visible action and reach a state preserving its first component, \ie verdicts are {\em persistent}. %
Rule \textsc{mTime} corresponds to a delay transition. %
The remaining rules determine when a state with a (non-verdict) first component can perform an action $a$ depending on its syntax. %
These are better explained through examples. %
Hereafter, we let $\mathbf{0}$ denote the valuation that assigns $0$ to every $x\in\mathcal{D}$. 

\begin{remark}
In practice, in order to monitor a formula $\phi\in\logic$, we start with a \emph{closed} monitor $M$ (\ie $M\in\mon$ \emph{and} $\mfvtime{M} = \emptyset$) at initial state $m_0 = (M,0,\mathbf{0})$. %
As the results in this subsection are more generic and do not rely on these assumptions, 
the latter will be introduced, justified and enforced in~\cref{sec:monitorability}. %
Without loss of generality, we still use initial states of the form  $(M,0,\mathbf{0})$ for illustration purposes. %
\end{remark}

\begin{figure}[!tb]
  \hrulefill
  $$
  \begin{array}{rl@{\hspace{1em}}rl}
    \scriptstyle \rname{mAct}
    &
    \begin{prooftree}
      \justifies
      (a.M,v_c,v_t) \stt{ a } (M,v_c,v_t)
    \end{prooftree}
    &
    \scriptstyle \rname{mVer}
    &
    \begin{prooftree}
      M \in E
      \justifies
      (M,v_c,v_t) \stt{ a } (M,v_c,v_t)
    \end{prooftree}
    \\[3em]
    \scriptstyle \rname{mLet}
    &
    \begin{prooftree}
      (M,v_c,v_t\update{x}{v_c}) \stt{a} (M',v_c,v_t')
      \justifies
      (\letin{x}{M},v_c,v_t) \stt{a}  (M',v_c,v_t')
    \end{prooftree}
    &
    \scriptstyle \rname{mTime}
    &
    \begin{prooftree}
      \justifies
      (M,v_c,v_t) \stt{d} (M,v_c+d,v_t)
    \end{prooftree}
    \\[3em]
    \scriptstyle \rname{mSelL}
    &
    \begin{prooftree}
      (M,v_c,v_t) \stt{a} (M',v_c,v_t')
      \justifies
      (M \extc N,v_c,v_t) \stt{ a } (M',v_c,v_t')
    \end{prooftree}
    &
    \scriptstyle \rname{mRec}\quad 
    &
    \begin{prooftree}
      (M\synsub{\rec{M}}{X},v_c,v_t) \stt{a} (M',v_c,v_t')
      \justifies
      (\rec{M},v_c,v_t) \stt{a}  (M',v_c,v_t')
    \end{prooftree}
    \\[3em]
    \scriptstyle \rname{mSelR}
    &
    \begin{prooftree}
      (N,v_c,v_t) \stt{ a } (N',v_c,v_t')
      \justifies
      (M \extc N, v_c,v_t) \stt{ a } (N',v_c,v_t')
    \end{prooftree}
    &
    \scriptstyle \rname{mGuard}\quad
    &
    \begin{prooftree}
      (M,v_c,v_t) \stt{a} (M',v_c,v_t') \wedge v_t \models g
      \justifies
      (g.M,v_c,v_t) \stt{a} (M',v_c,v_t')
    \end{prooftree}
  \end{array}
  $$
  \hrulefill
  \caption{SOS rules for monitors.}
  \label{fig:sos-monitors}
  \hrulefill
\end{figure}


\begin{example}
  \label{ex:simple-monitor}
   Let $\act = \{b,c\}$, $M = \letin{x}{((x > 5).\yes + (x \leq 5).(b.\yes + c.\no))}$, and
  let  $m_0 = (M,0, \vto)$ be a state of $\mathbb{M}$ (the TLTS of all monitors, see above). %
  Say $M$ performs, from $m_0$, the finite run $\theta = 6\, c$. This will give the following (unique) trace:
  $(M,0,\mathbf{0})\stt{6}(M,6,\mathbf{0})\stt{c}(\yes,6,[x=6])$, 
  where, from left to right: 
  \begin{itemize} \item the
  delay transition $d=6$ is allowed by rule \textsc{mTime}, 
  \item the action transition $c$ is allowed by rule \textsc{mLet}. %
  This is because the state $((x > 5).\yes + (x \leq 5).(a.\yes +
  b.\no),6,[x=6])$ can perform any action in $\act$ and reach $(\yes,6,[x=6])$ following, respectively, rules \textsc{mSelL},
  \textsc{mGuard} and \textsc{MVer} (\cref{fig:sos-monitors}).
  \end{itemize} 
\end{example}

\begin{example}
\label{ex:imon}
This example illustrates further the evolution of a monitor according to the rules of $\longrightarrow_\mathbb{M}$, but also explains why monitors are allowed to have free timestamp variables, following~\cref{def:mon}. Let $\act = \{b,c\}$ and the monitor 
$$M = \letin{x}(\rec[X]{(\letin{y}((y-x\leq 5).(b.X + c.\no) + (y-x > 5).\yes)))}$$
and let $$N = \rec[X]{(\letin{y}((y-x\leq 5).(b.X + c.\no) + (y-x > 5).\yes))}$$
\ie $N$ is the subterm of $M$ without the binder of timestamp variable $x$. %
Intuitively, the syntax of $M$ indicates that it accepts a run (becomes $\yes$) iff every action seen within $5$ time units, relative to the global time at which the first action occurs, is a $b$ (otherwise it rejects). %
Let us apply the rules of $\longrightarrow_\mathbb{M}$, starting at $(M,0,\mathbf{0})$,  to see whether $M$ accepts any run $\theta$ starting with $2\, b\, 6\, c$ as it should: 
$(M,0,\mathbf{0}) \stt{2} (M,2,\mathbf{0}) \stt{b} (M',2,[x=2,y=2]) \stt{6} (M',8,[x=2,y=2])\stt{c} (\yes,8,[x=2,y=8])$ where $$M' = \letin{y}((y-x\leq 5).(b.N + c.\no) + (y-x > 5).\yes)$$
  applying, respectively (from left to right) rule \textsc{mTime}, \textsc{mLet}, \textsc{mTime} and \textsc{mLet}. %
Notice now that, in $M'$, there is no longer the binder of timestamp variable $x$, and therefore $x$ is free in $M'$. %
This is because $\letin{x}$ in $M$, outside of the recursion, is intended to evaluate constantly to the value of global time corresponding to the first action. %
This simplification is handy to deal with monitors with both recursion and guards, and justifies the relaxed nature of~\cref{def:mon}, allowing monitors to have free timestamp variables.  
\end{example}

\paragraph{Parallel monitors.}

\begin{figure}[!h]
\hrulefill
$$
\begin{array}{lllll}
 M, N & ::=  \sver & \BNFsep a.M & \BNFsep g.M
           &\BNFsep M \extc N\\
           &&\BNFsep \rec{M} &\BNFsep X  & \BNFsep \letin{x}{M}\\
           &&\BNFsep & M \otimes N & \text{(par. conj.)}\\
           &&\BNFsep & M \oplus N & \text{(par. disj.)}\\
E       & ::= \syes & \BNFsep \sno & \BNFsep \smend &\\
\syes & ::= \yes & \BNFsep \syes\oplus M &\BNFsep \syes \otimes \syes\\
\sno & ::= \no & \BNFsep \sno\otimes M &\BNFsep \sno \oplus \sno\\
\smend & ::= \mend & \BNFsep \smend\odot \smend &&\\
\end{array}
$$
\hrulefill
\caption{Syntax of $\tmon$ (extended with parallel composition).}
\label{fig:syntax-parallel-monitors}
\hrulefill
\end{figure}

In practice, a formula is often monitored through composing monitors in parallel~\cite{DBLP:journals/pacmpl/AcetoAFIL19}. %
We first define the syntax of parallel composition over m-terms. %
The syntax of $\tmon$ is extended accordingly with two parallel composition operators: $\otimes$ for \emph{parallel conjunction} and  $\oplus$ for  \emph{parallel disjunction}, adapted from~\cite{DBLP:journals/pacmpl/AcetoAFIL19} to the (linear-time) timed setting. %
The extended syntax thereof is given in \cref{fig:syntax-parallel-monitors}, with $\odot\in\{\otimes,\oplus\}$. %
Accordingly, we extend the set of verdicts by grouping acceptance, rejection and termination verdicts in the set $\syes$, $\sno$ and $\smend$ respectively. %
By abuse of notation, we write $E\in\{\syes,\sno,\smend\}$ to denote $L(E)$, the set of m-terms generated by the grammar of \cref{fig:syntax-parallel-monitors} starting from 
the nonterminal symbol $E$. %
Each of the latter is defined recursively using the corresponding verdict from a \emph{regular} (\ie non-parallel) m-term. %
For example, a verdict $\syes$ (an acceptance verdict) is reached in a parallel disjunction (\resp conjunction) of two regular m-terms whenever a $\yes$ verdict is reached by one of them (\resp by both of them). %
The set of monitors $\crmon$ is therefore extended in the expected way (by applying~\cref{def:mon} to $\tmon$, as extended to parallel composition). 
Thanks to lifting verdict composition rules to the syntactic level, the extension of the semantics of monitors in~\cref{fig:sos-monitors} to parallel monitors becomes rather simple; we only need one additional rule, stating that monitors in a composition perform actions together:

\newcommand{\unionv}[3]{\ensuremath{#1 \overset{#2}{\sqcup} #3}}

$$
\begin{array}{l}
\scriptstyle \rname{mPar}\quad
    \begin{prooftree}
      M \odot N \not\in E \quad (M,v_c,v_t) \stt{a} (M',v_c,v_t') \quad (N,v_c,v_t) \stt{a} (N',v_c,v_t'')
      \justifies
      (M \odot N,v_c,v_t) \stt{a} (M' \odot N',v_c,\unionv{v_t'}{v_t}{v_t''}) 
    \end{prooftree}
\end{array}
$$
where 
$$
    \unionv{v_t'}{v_t}{v_t''} (x) = \begin{cases} 
    v_t'(x) \text{ if } v_t'(x) \neq v_t(x) \\
    v_t''(x) \text{ if } v_t''(x) \neq v_t(x) \\
    v_t(x) \text{ otherwise}\\
  \end{cases}
$$
Recall that bound timestamp variables in $M$ and $N$ are disjoint, and therefore the first two cases are disjoint. %
$\mathbb{M} =  \langle S_\mathbb{M} , \labb, \longrightarrow_\mathbb{M} \rangle$, the TLTS describing the semantics of all monitors in $\crmon$ is therefore extended to parallel monitors, %
with the transition relation $\longrightarrow_\mathbb{M}$ defined using the rules in~\cref{fig:sos-monitors} and the rule $\textsc{mPar}$ above\footnote{We prove in~\cite[Appendix D]{DBLP:conf/ecoop/Amara0FF25} that these rules are well defined over $S_\mathbb{M} \times \labb \times S_\mathbb{M}$}. %

\begin{example}
  \label{ex:parallel-monitor}
  Let $\act = \{b,c,c'\}$ and recall the monitor from \cref{ex:simple-monitor},
  $$
    M = \letin{x}{(x > 5).\yes + (x \leq 5).(b.\yes + c.\no)}
  $$
  Consider the monitor $N = M \otimes c'.\yes$ and let 
  $n_0 = (N,0,\mathbf{0})$.   
  We show how, from $n_0$, an acceptance verdict is reached after performing any run that starts with 
  $6\, c'$. 
  First, using rule \textsc{mTime} to derive $(M \otimes a.\yes,0,\mathbf{0}) \stt{6} (M \otimes a.\yes,6,\mathbf{0})$,
  then $(M \otimes c'.\yes,6,\mathbf{0}) \stt{c'} (\yes \otimes \yes,6,[x = 6])$ is given by the following derivation.

  $$
    \begin{prooftree}
      N \not\in E \quad 
      \[
        \cref{ex:simple-monitor}
        \justifies
        (M,6,\mathbf{0}) \stt{c'} (\yes,6,[x = 6])
      \] 
      \quad 
      \[
        \justifies
        (c'.\yes,6,\mathbf{0}) \stt{c'} (\yes,6,\mathbf{0})
        \using{\textsc{mAct}}
      \]
      \justifies
        (M \otimes c'.\yes,6,\mathbf{0}) \stt{c'} (\yes \otimes \yes,6,[x = 6])
      \using \textsc{mPar}
    \end{prooftree}
  $$
  However, this monitor cannot perform any run of the form $d\, b$ or $d\, c$ with $d \leq 5$, because the righthand side of the parallel disjunction, \ie $c'.\yes$ can perform neither a $b$ nor a $c$, the only possible actions by $M$ when $x \leq 5$. 
  This partly motivates the notion of  \emph{reactivity}, discussed in~\cref{sec:abstr-inst}.
\end{example}

In the sequel, we use the shorthand notation $N\, \,  +\, \,  A.M$ to denote the selection between monitor $N$ and $a.M$ for all actions in $A\subseteq \act$, with the convention $N\, \,   +\, \,   A.M = N$ if $A$ is empty. %
Moreover, we write $\overline{A}$, $A\subseteq \act$, to denote the set $\act\setminus{A}$ and similarly $\overline{a}$, $a\in\act$ to denote $\act\setminus{\{a\}}$. %
Using these compact notations, we can write \eeg for $\act = \{a_1,a_2,a_3,a_4\}$, $a_2.M + \overline{a_2}.\no$ instead of $a_2.M + a_1.\no + a_3.\no + a_4.\no$. %

\paragraph{Verdict persistence.}

As mentioned earlier, all monitors in $\mon$ are verdict-persistent. 

\begin{lemma}[Verdict presistence]
  \label{lem:verdict-persistence}
  For all $(M,v_c,v_t), (N,v_c',v_t')\in S_\mathbb{M}$ and finite weak run $\theta\in \labb^*$, if $M \in E$ then:
  $$
    (M,v_c,v_t) \wt{\theta} (N,v_c',v_t') \text{ implies } N = M
  $$
\end{lemma}
\begin{proof}
  By induction on the length $n$ of $\theta$. In the case $n=0$, the result is obtained 
  by reflexivity of $\wt{}$. In the inductive case, we have $(M,v_c,v_t) \stt{\alpha} (N',v_c'',v_t'') \wt{\theta'} (N,v_c',v_t')$
  for some $\alpha \in \labb$, $N' \in \mon$, $v_c'' \in \rplus$, $v_t'' \in \mathcal{V}$.
  We proceed by case analysis on the rule used to derive $(M,v_c,v_t,r) \stt{\alpha} (N',v_c'',v_t'')$.
  The only possible two cases are:
  \begin{itemize}
    \item \textsc{mVer}: Then we have $(M,v_c,v_t) \stt{ a } (M,v_c,v_t) \wt{\theta'} (N,v_c',v_t')$.
      Since $\theta'$ is smaller than $\theta$, we apply the induction hypothesis to conclude $N = M$.
    \item \textsc{mTime}: Then we have $(M,v_c,v_t) \stt{d} (M,v_c+d,v_t) \wt{\theta'} (N,v_c',v_t')$.
      Since $\theta'$ is smaller than $\theta$, we apply the induction hypothesis to conclude $N = M$.
  \end{itemize} 
\end{proof}

\paragraph{Verdict states.} 
Thanks to verdict persistence, we know that once a monitor reaches a \emph{verdict state}, \ie a state whose first component is a verdict, the information within such state are irrelevant at the exception of the verdict itself (\ie acceptance, rejection or termination). %
That is, \eeg for a verdict state $(M\in\syes, v_c, v_t)$, we know that regardless of what $M$ performs in the future, it will remain unchanged. %
Therefore, we may denote by $\yess$, $\nos$, $\mends$ any verdict state in $(M,v_c,v_t) \in S_\mathbb{M}$ such that $M\in \syes$, $M\in \sno$, $M\in \smend$, respectively. 
Strictly speaking, $\yess$, $\nos$, $\mends$ may be viewed as equivalence classes grouping states with acceptance, rejection and termination verdicts, respectively. %
In order to simplify the writing, we abuse notation by defining an equality instead. For all $(M,v_c,v_t),(N,v_c',v_t')\in S_\mathbb{M}$: $$ (M,v_c,v_t) = (N,v_c',v_t') = \yess \text{ iff } M\in \syes \text{ and } N\in \syes$$
$$ (M,v_c,v_t) = (N,v_c',v_t') = \nos \text{ iff } M\in \sno \text{ and } N\in \sno$$
$$ (M,v_c,v_t) = (N,v_c',v_t') = \mends \text{ iff } M\in \smend \text{ and } N\in \smend$$

For example, we write $m \wt{ \theta } \yess$ if there exists a state $(N,v_c,v_t)$ with $N\in\syes$ reachable from $m$ along $\theta$, and similarly $m\wti{ \theta } \yess$ if all states $(N,v_c,v_t)$ reachable along $\theta$ from $m$ satisfy $N\in\syes$. %

\subsection{Instrumentation}
\label{sec:instrumentation}

The notion of a monitor ``performing an action'' is in fact misleading, since a monitor is supposed to be a \emph{passive entity} that \emph{follows} an SUM. %
We need an operator to compose a monitor with an SUM accordingly; this is known as \emph{instrumentation}. %
In the following, we define the rules of the instrumentation operator $\instr$, adapted to the timed settings from~\cite{DBLP:journals/pacmpl/AcetoAFIL19}. %
Hereafter, recall $
\mathbb{M} =  \langle S_\mathbb{M} , \labb, \longrightarrow_\mathbb{M} \rangle$ the TLTS describing the behaviour of all monitors, and let $P = \langle S_P, \lab, \longrightarrow_P \rangle$ be an arbitrary TLTS describing the behaviour of an SUM. %
   That is, we do not assume any particular form of states in $S_P$ or rules in $\longrightarrow_P$, since the SUM can be \eeg a blackbox system.

\begin{definition}[Instrumentation]  
  We define the {\em
  monitored system} using the instrumentation operator $\instr$ as follows: $$\mathbb{M} \instr P = \langle S_\mathbb{M} \times S_P, \lab, 
  \stt{}_\instr \rangle$$%
  The transition relation $\stt{}_\instr$  is defined in \cref{fg:instr-SOS}, 
  where $m,m' \in S_\mathbb{M}$, $p,p' \in S_P$, $\alpha \in \labb$ and $a\in\act$. %
  To save notations, we use $m \instr p$ for $(m,p) \in S_\mathbb{M} \times S_P$.
  \label{def:instrumentation}
\end{definition}

\begin{figure}[!h]
  \hrulefill
  $$
  \begin{array}{r@{\hspace{2em}}l}
   
    \begin{prooftree}
      p \stt{ \alpha }_P p' \quad m \stt{ \alpha }_\mathbb{M} m'
      \justifies
      m \instr p \stt{ \alpha }_{\instr} m' \instr p'
      \using \textsc{iMon}
    \end{prooftree}
    &
    \begin{prooftree}
      p \stt{ \tau }_P p'
      \justifies
      m \instr p \stt{ \tau }_{\instr} m \instr p'
      \using \textsc{iAsy}
    \end{prooftree}
    \\[3em]
    \multicolumn{2}{c}{%
    \begin{prooftree}
      p \stt{ a }_P p' \quad m \Nst{ a }_\mathbb{M} \quad m = (N,v_c,v_t)
      \justifies
      m \instr p \stt{ a }_{\instr} (\mend,v_c,v_t) \instr p'
      \using \textsc{iTer}
    \end{prooftree} 
    }
    
  \end{array}
  $$
  \hrulefill
  \caption{SOS rules for Instrumentation.}
  \label{fg:instr-SOS}
  \hrulefill
\end{figure}

Rule \textsc{iMon} states that the monitor must follow any action or delay transition performed by the SUM. %
That is, if the SUM performs a delay $d\in \rplus$ or an action $a\in \act$ transition, the monitor takes the same transition with the SUM provided it is able to. %
The rule  \textsc{iAsy} corresponds to the SUM performing a $\tau$ transition. %
Finally, if the SUM may perform an action $a$ and the monitor may not, rule \textsc{iTer} allows the SUM to perform $a$ while forcing the
monitor to reach a verdict state $\mends$. %
\begin{remark}
Note that \textsc{iTer} is defined only for actions and not for time delays. %
This is because a monitor can always perform time delays (rule \textsc{mTime}, ~\cref{fig:sos-monitors}) and therefore it can always follow the SUM if it performs a delay. 
\end{remark}

In summary, the rules \textsc{iMon} and \textsc{iAsy} in~\cref{fg:instr-SOS} correspond to classical parallel composition with strong synchronisation over actions (and delays), and standard interleaving semantics for $\tau$ transitions. %
The only difference resides within the rule \textsc{iTer} that forces the monitor to reach a verdict state $\mends$ whenever it cannot follow the SUM. %

\paragraph{Preserving the SUM behaviours}
 
We prove that the instrumentation operator \emph{preserves} the behaviours of the SUM, \ie a monitor instrumented with the SUM follows the SUM when possible and never alters its behaviour. %
This is done through a proof of \emph{strong timed bisimulation}. %
Hereafter, let $\mathcal{T}$ the set of all TLTSs over $\lab$.

\begin{definition}[Timed bisimulation \cite{rs_2007}]
\label{def:bisimulation}
  A binary relation $\mathcal{R}$ over the set of states of a TLTS is a strong timed bisimulation iff whenever $s_1 \mathcal{R} s_2$, for every $\alpha \in \lab$:
  \begin{itemize}
    \item if $s_1 \stt{\alpha} s_1'$ then there is a transition $s_2 \stt{\alpha} s_2'$ and $s_1' \mathcal{R} s_2'$
    \item if $s_2 \stt{\alpha} s_2'$ then there is a transition $s_1 \stt{\alpha} s_1'$ and $s_1' \mathcal{R} s_2'$
  \end{itemize}
  Two states $s$ and $s'$ are strongly timed bisimilar, written $s \approx_s s'$, iff
  there is a strong timed bisimulation that relates them. %
\end{definition}

Using~\cref{def:bisimulation}, in order to prove that instrumenting any SUM $P$ with any monitor $M$ preserves \emph{exactly} all the behaviours of $P$, if suffices to prove the following lemma. 

\begin{restatable}{lemma}{bisimulation} 
\label{lem:instr-bisimulation}
$\forall P\in\mathcal{T}, p\in S_P, m\in S_\mathbb{M}: m\instr p \approx_s p$.
\end{restatable}   

Accordingly, given a finite run $\theta\in\lab^*$, performed by the SUM, we say that a monitor \emph{consumes} the weak run $\theta'\in\labb^*$ corresponding to $\theta$ (\cref{sec:sum}). 

\begin{example}
Let us consider the monitor $M$, starting at state $m_0 = (M,0,\vto)$ (\cref{ex:simple-monitor}), and an SUM performing a finite run $\theta = 2\, \tau\, 4\, \tau\, c$ from state $p$ to state $p'$. %
We have then $m_0\instr p \stt{\theta}\yess\instr p'$, which corresponds to $M$ consuming $\theta' = 6\, c$, the weak finite run corresponding to $\theta$. %
Informally, the monitor $M$ checks the following property at runtime: ``if the first consumed action occurs within $5$ time units, then such action must be a $b$''. 
\end{example}

\subsection{Abstracting Away the Instrumentation Operator}\label{sec:abstr-inst}\label{genmon}
In~\cite[Sect. 5.1.1]{DBLP:conf/ecoop/Amara0FF25}, we abstracted away the instrumentation operator through considering only \emph{reactive} monitors, \ie monitors that can follow any run from any state of any SUM (that is, consume any run in $\labb^*$) as states the following definition. %

\begin{definition}[Monitor Reactivity]
\label{def:reactive}
A monitor $M \in \mon$ is reactive iff $\forall\theta\in\labb^*, v_c\in\rplus,v_t\in\mathcal{V}: (M,v_c,v_t)\wt{\theta}$
\end{definition}

In a linear-time setting, the set of behaviours of an SUM equals the set of its infinite runs, thus all behaviours of all SUMs simply collapse into all infinite runs in $\lab^\omega$. %
We could write accordingly $m\wt{\theta}m'$ to denote a monitor consuming weak run $\theta$, abstracting away SUMs and the instrumentation operator: the former replaced with the sum of all infinite runs in $\lab^\omega$ and the latter collapsing into simple consumption of weak runs that is always possible thanks to the reactivity assumption. %
The downside of adopting this choice is the fact that our maximality results in~\cite{DBLP:conf/ecoop/Amara0FF25} were proven \wrt reactive monitors only, a subset of the set of all monitors $\mon$.

In this extension, we get rid of the reactivity assumption while preserving all monitorability results, including the ones on maximality. %
The standard technique to do so is to translate each monitor into 
a reactive one that is verdict-equivalent as in~\cite{DBLP:journals/pacmpl/AcetoAFIL19}.  %
However, the adoption of such technique in our timed setting would give an extremely complex procedure, if such exists, due to the high expressive power of $\mon$. %
Our approach is to avoid this translation altogether: we allow monitors to be ``syntactically'' non-reactive, and instead enforce reactivity at the semantics level. 
That is, we introduce new, more flexible semantics for monitors that we refer to as \emph{reactive semantics}. %

\begin{definition}[Reactive semantics]
  \label{def:reactive-semantics}
  The reactive semantics of monitors is given by the TLTS $\mathbb{M}_r = \langle{S_\mathbb{M},
  \labb, \longrightarrow_{\mathbb{M}_r}} \rangle$ where: 
  
  $$\longrightarrow_{\mathbb{M}_r} = \longrightarrow_\mathbb{M} \cup \longrightarrow_e$$
  with
  
  $$
    \longrightarrow_e = \setof{((M,v_c,v_t),a,(\mend,v_c,v_t))}{(M,v_c,v_t) \Nst{a}_{\mathbb{M}}}
  $$
\end{definition}

In a nutshell, the above definition extends the original semantics with a new set of transitions that allows a monitor to reach an $\mends$ state on actions it cannot perform at a given time. %
Now, the new, \emph{light} instrumentation rules (\cref{fg:instr-SOS-light}) will \emph{guide} the monitor through the possible transitions, therefore forcing it to follow the SUM, and reach an $\mends$ state if and only if and exactly when it cannot do so. %

\begin{figure}[!h]
  \hrulefill
  $$
  \begin{array}{r@{\hspace{2em}}l}
   
    \begin{prooftree}
      p \stt{ \alpha }_P p' \quad m \stt{ \alpha }_{\mathbb{M}_r} m'
      \justifies
      m \instr_r p \stt{ \alpha }_{\instr_r} m' \instr_r p'
      \using \textsc{iMon}
    \end{prooftree}
    &
    \begin{prooftree}
      p \stt{ \tau }_P p'
      \justifies
      m \instr_r p \stt{ \tau }_{\instr_r} m \instr_r p'
      \using \textsc{iAsy}
    \end{prooftree}
  \end{array}
  $$
  \hrulefill
  \caption{SOS rules for (light) Instrumentation.}
  \label{fg:instr-SOS-light}
  \hrulefill
\end{figure}

In other words, the rules of the new set of transitions $$\longrightarrow_e = \setof{((M,v_c,v_t),a,(\mend,v_c,v_t))}{(M,v_c,v_t) \Nst{a}_{\mathbb{M}}}$$
replace the \textsc{iTer} rule in the original version of the instrumentation (\cref{fg:instr-SOS}), giving thereby exactly the same outcome when monitoring an SUM, as states the following lemma. %

\begin{restatable}{lemma}{bisimulation2} 
\label{lem:instr-bisimulation2}
$\forall P\in\mathcal{T}, p,p'\in S_P, m,m'\in S_\mathbb{M}, \theta \in \labb^*$: 

$$m\instr_r p \wt{\theta}_{\instr_r} m'\instr_r p'  \Leftrightarrow  m\instr p \wt{\theta}_{\instr} m'\instr p'$$
\end{restatable}

\begin{example} 
Let $\act = \{b,c\}$, $M = \letin{x}{((x > 5).\yes + (x \leq 4).(b.\yes + c.\no))}$, and let $m_0 = (M,0, \vto)$. %
The monitor $M$ is clearly not reactive: it does not know what to do for any $4 < x \leq 5$. %
Let us analyse the behaviour of $M$ (starting at $m_0$) instrumented with $P$, an SUM that produces an infinite run starting at state $p_1$ with $5\, b$, and with $P'$, another SUM performing an infinite run starting at state $p'_1$ with $2\, \tau\, c$. %
We denote by $p_2$, $p_3$, $\ldots$ the successors of $p_1$ and $p'_2$, $p'_3$, $\ldots$ those of $p'_1$.
In each case, we show that the outcome using the original monitors TLTS $\mathbb{M}$ with the original instrumentation relation $\instr$ coincides with that using the new  monitors TLTS $\mathbb{M}_r$ with the new  instrumentation relation $\instr_r$. %

\begin{itemize}
\item{} Instrumenting $m_0$ with $p_1$. This is the interesting case as in both versions the monitor reaches an $\mends$ state yet following different rules: 
\begin{itemize}
\item{} $m_0 \instr p_1 \stt{5}_\instr (M,5, \vto) \instr p_2 \stt{b}_\instr \mends \instr p_3 \ldots$ following the rules \textsc{iMon} and \textsc{iTer}, respectively (\cref{fg:instr-SOS}), 
\item{} $m_0 \instr_r p_1 \stt{5}_{\instr_r} (M,5, \vto) \instr_r p_2 \stt{b}_{\instr_r} \mends \instr p_3 \ldots$ following the rule \textsc{iMon} twice (\cref{fg:instr-SOS-light}). %
Here, the second application of \textsc{iMon} uses the new set of transitions $\longrightarrow_e$ since $(M,5, \vto)$ cannot perform action $b$ (there is no transition in $\longrightarrow_\mathbb{M}$ of the form $((M,5, \vto),b,m)$, $m\in S_\mathbb{M}$). 
\end{itemize}
\item{} Instrumenting $m_0$ with $p'_1$. %
In this case, the same rules are applied in $\instr$ and $\instr_r$ as the monitor does not reach a state where it receives an action it cannot perform:
\begin{itemize}
\item{} $m_0 \instr p'_1 \stt{2}_\instr (M,2, \vto) \instr p'_2 \stt{\tau}_\instr (M,2, \vto) \instr p'_3 \stt{c}_\instr \nos \instr p'_4 \ldots$ following the rules \textsc{iMon}, \textsc{iAsy} and \textsc{iMon}, respectively (\cref{fg:instr-SOS}), 
\item{} $m_0 \instr_r p'_1 \stt{2}_{\instr_r} (M,2, \vto) \instr p'_2 \stt{\tau}_{\instr_r} (M,2, \vto) \instr p'_3 \stt{c}_{\instr_r} \nos \instr p'_4 \ldots$ following the rules \textsc{iMon}, \textsc{iAsy} and \textsc{iMon}, respectively (\cref{fg:instr-SOS-light}).
\end{itemize}
\end{itemize} 
\end{example}

Now, abstracting the new instrumentation operator and SUMs follows the same reasoning of~\cite{DBLP:conf/ecoop/Amara0FF25} (explained above), but requires no more the monitors to be reactive. %
We obtain the same results on monitorability as in~\cite{DBLP:conf/ecoop/Amara0FF25}, yet without any loss of generality \wrt all monitors in $\mon$. %
The only statement that was dependent on the reactivity assumption was~\cite[Lemma D.2.]{DBLP:conf/ecoop/Amara0FF25}. %
It suffices therefore to prove here that the same lemma (\cref{lem:app-monitor-combinators} below) holds now for all monitors under the new reactive semantics. %
Hereafter, we drop the subscript $_{\mathbb{M}_r}$ from (weak or strong) transition symbols whenever clear from the context.  

\begin{lemma}
  \label{lem:app-monitor-combinators}
  For every $M, N \in \mon$, every $v_c,v_t,$ and $\theta$:
  \begin{enumerate}
  \item \label{otimes-nos} $(M_1 \otimes M_2,v_c,v_t) \wt{\theta} \nos$ iff 
    $(M_1,v_c,v_t) \wt{\theta} \nos$ or $(M_2,v_c,v_t) \wt{\theta} \nos$.
  \item \label{otimes-yess} $(M_1 \otimes M_2,v_c,v_t) \wt{\theta}\yess$ iff 
    $(M_1,v_c,v_t) \wt{\theta} \yess$ and ${(M_2,v_c,v_t) \wt{\theta} \yess}$.
  \item \label{oplus-yess} $(M_1 \oplus M_2,v_c,v_t) \wt{\theta} \yess$ iff 
    $(M_1,v_c,v_t) \wt{\theta} \yess$ or ${(M_2,v_c,v_t) \wt{\theta} \yess}$.
  \item \label{oplus-nos} $(M_1 \oplus M_2,v_c,v_t) \wt{\theta} \nos$ iff 
    $(M_1,v_c,v_t) \wt{\theta} \nos$ and ${(M_2,v_c,v_t) \wt{\theta} \nos}$.
  \end{enumerate}
\end{lemma}
\begin{proof}
  We present only the proof of \cref{otimes-nos}, as the proofs of the other cases 
  are similar. We proceed by induction on $|\theta|$. When $|\theta| = 0$, We have 
  $M \otimes N \in \sno$ iff $M \in \sno$ or $N \in \sno$ by \cref{fig:syntax-parallel-monitors}.
  When $|\theta| = k+1$, 
  \begin{itemize}
    \item $\Longrightarrow$) We have $(M_1 \otimes M_2, v_c, v_t) \stt{\alpha} m \wt{\theta'} \nos$. 
      If $M_1 \otimes M_2 \in E$, then by \cref{lem:verdict-persistence} we have 
      $M_1 \otimes M_2 \in \sno$, hence $M_1 \in \sno$ or $M_2 \in \sno$.
      We then conclude by successive applications of rules \textsc{mTime} and \textsc{mAct}.
      Otherwise, either $\alpha \in \rplus$, 
      in which case the first transition is obtained by rule \textsc{mTime}, yielding $m = (M_1 
      \otimes M_2, v_c + \alpha, v_t)$, and we conclude by the inductive hypothesis together with 
      rule \textsc{mTime}; or $\alpha \in \act$, in which case rule \textsc{mPar} applies, producing 
      $m = (M_1' \otimes M_2', v_c, \unionv{v_{t_1}}{v_t}{v_{t_2}})$ such that $(M_i, v_c, v_t) 
      \stt{\alpha} (M_i', v_c, v_{t_i})$. By the inductive hypothesis, for some $i \in \{1,2\}$ we 
      have $(M_i', v_c, \unionv{v_{t_1}}{v_t}{v_{t_2}}) \wt{\theta'} \nos$. Since the variables bound 
      in $M_1$ and $M_2$ are distinct, it follows that $(M_i', v_c, v_{t_i}) \wt{\theta'} \nos$ for the 
      same $i$. Thus, we conclude that $(M_i, v_c, v_t) \stt{\alpha} (M_i', v_c, v_{t_i}) \wt{\theta'} 
      \nos$ for some $i \in \{1,2\}$, as required.
    
    \item $\Longleftarrow$) Suppose $(M_1,v_c,v_t) \wt{\theta} \nos$. If $\alpha \in \rplus$, rule \textsc{mTime} gives $(M_1, v_c, v_t) 
      \stt{\alpha} (M_1, v_c + \alpha, v_t) \wt{\theta'} \nos$; by the inductive hypothesis 
      $(M_1 \otimes M_2, v_c + \alpha, v_t) \wt{\theta'} \nos$, and we conclude again by \textsc{mTime}. 
      If $\alpha \in \act$, then $(M_1, v_c, v_t) \stt{\alpha} (M_1', v_c, v_t') \wt{\theta'} \nos$ 
      for some $M_1', v_t'$, and by \cref{def:reactive-semantics} there exist $M_2', v_t''$ with 
      $(M_2, v_c, v_t) \stt{\alpha} (M_2', v_c, v_t'')$. Since $M_1'$ and $M_2'$ have distinct timestamp 
      variables, $(M_1', v_c, \unionv{v_t'}{v_t}{v_t''}) \wt{\theta'} \nos$, and by the inductive hypothesis 
      $(M_1' \otimes M_2', v_c, \unionv{v_t'}{v_t}{v_t''}) \wt{\theta'} \nos$. Finally, rule \textsc{mPar} 
      yields $(M_1 \otimes M_2, v_c, v_t) \stt{\alpha} (M_1' \otimes M_2', v_c, \unionv{v_t'}{v_t}{v_t''})
      \wt{\theta'} \nos$, as required.
  \end{itemize}
\end{proof}

\section{Monitorability of \logic}
\label{sec:monitorability}

In this section, we present the monitorability results on~\logic. %
First, in line with existing literature, we set the framework within
which we establish such results (\cref{sec:framework}). %
Afterwards, in~\cref{sec:sound-monitorss}, we develop a compiler from
\logic to sound monitors. %
We identify subsequently~\clogic, a complete-monitorable fragment of
\logic, \ie containing formulae that are monitorable for both
satisfactions and violations (\cref{sec:complete-monitorability}). %
Then, in~\cref{sec:partial-monitorability}, we identify $\slogic$ (\resp $\vlogic$), a fragment of \logic formulae that are 
satisfaction monitorable (\resp violation monitorable), and accordingly define $\mlogic = \slogic \cup \vlogic$ as a fragment of monitorable formulae in \logic. %
Finally, we prove that \vlogic, \slogic, \mlogic and \clogic are, respectively, the largest violation-monitorable, satisfaction-monitorable, monitorable and complete-monitorable fragments of
\logic \wrt to the monitoring language $\mon$ (\cref{sec:maximality}). The proofs not included in the body of this section may be found in~\cite[Appendix F]{DBLP:conf/ecoop/Amara0FF25}.

\subsection{Setting the framework}\label{sec:framework}

\subsubsection{Monitor Initial State}

Intuitively, in order to assess whether $M$ can monitor $\phi$, monitoring should start at state $m_0 = (M,0, \vto)$. %
   That is, monitor $M$ is \emph{initialised} at state $m_0$, at which its time (monitor time) starts. %
   Since monitor time may only increase (through consuming delays), it refers to the relative time since the monitoring of $\phi$ started. %
   Examples of a monitor $M$ starting at state $m_0$ have been already discussed within the previous section, \eeg in~\cref{ex:simple-monitor}. %

  \subsubsection{Timed Words, Weak Runs and Time Divergence} 
  First, we make an important, yet natural assumption on infinite timed words, namely \emph{time divergence}. %
  To explain this, let us consider the following infinite timed word: $\pi = (a_1,\frac{1}{2}), (a_2,\frac{3}{4}),(a_3,\frac{7}{8}), (a_4,\frac{15}{16}),\ldots$, \ie where the delay between two consecutive actions is halved each time a new action occurs. %
  This results in the timestamps of $\pi$ increasing but never beyond absolute time $1$ (since $\sum_{k=1}^{\infty}(\frac{1}{2^k}) = 1$). %
  The infinite timed word $\pi$ is \emph{time convergent}, and it is clear to see that there is no real system that can produce such timed word; for the simple reason that time always diverges in reality (see \eeg~\cite{henzinger1992sooner,alur1997modularity}\cite[Sect.2]{DBLP:journals/jsa/FoughaliHZ23}). %
  Therefore, when reasoning about whether an infinite timed word is or is not in the semantics of some formula, for monitorability purposes, we restrict ourselves to the set of time-divergent infinite timed words. %
  Time divergence can be expressed in the standard manner as follows. %
  An infinite timed word $\pi$ is time divergent iff its timestamps are unbounded, formally~\cite[Sect. 4]{henzinger1992sooner},~\cite[Def. 1]{alur1994really},~\cite[Def. 2.4]{DBLP:conf/time/Foughali26}:

\begin{align}
\label{eq:div-tw}
  \forall x\in \mathbb{R},\, \exists i\in \naturals : t_i(\pi) > x
\end{align} 
We define accordingly the set $\twordsr\subset\twords$ as all infinite timed words that satisfy~\eqref{eq:div-tw}. %

  Second, as we will see later on, central notions on monitorability, such as monitor soundness and completeness, gravitate around the ability to reach a verdict, \wrt whether an infinite timed word $\pi\in\twordsr$ is in the semantics of the monitored formula, based on consuming a finite prefix of $\pi$. %
  We need therefore to define the notion of consuming a finite timed word. %
  Since (i) we already defined the notion of consuming a weak run (\cref{sec:monitoring}), and (ii) a finite timed word necessarily corresponds to a finite weak run in $\labb^*$ (\cref{sec:tw}), we simply need to define a function to retrieve the latter from the former. %
  
  \begin{definition}[From finite timed words to finite weak runs]\label{def:twd}
  $$
    \textsc{Trn} : \twordstar \longrightarrow \labb^*
  $$
where $\wordd{\pi} = d_1a_1d_2a_2 \ldots$ with  
  $d_i = t_i - t_{i-1}$ and $t_0 = 0$, and the convention that $\wordd{\pi}$ gives the empty weak run $\epsilon$ if $\pi$ is an empty timed word. %
  \end{definition}
  In the remainder of this paper, by abuse of terminology, we often say a monitor ``consumes (the finite timed word) $\pi$'' instead of ``consumes (the finite weak run) $\wordd{\pi}$''
  
\begin{example} Consider the finite timed word $\pi = (a,1.7)(b,1.7)(b,5.7)$. %
The corresponding weak run is obtained using~\cref{def:twd}: $\wordd{\pi} = 1.7\, a\, 0\, b\, 4\, b$. %
\end{example}
\begin{remark}
Note that time divergence does not pose any particular restriction on finite timed words, \ie $\setof{\pf{\pi}}{ \pi\in\twordsr} = \twordstar$. %
In words, any finite timed word (in $\twordstar$) can be retrieved as a prefix of a time-divergent infinite timed word
, see~\cite{alur1997modularity}. 
\end{remark}
 
\subsubsection{Monitor Soundness and Completeness} We now define the central monitor properties of \emph{soundness} and \emph{completeness} %
extended from~\cite[Definition 4.1]{DBLP:journals/pacmpl/AcetoAFIL19} to our 
timed setting. %

\paragraph{Monitor soundness.}
Intuitively, a monitor $M$ is sound for a formula $\phi$ iff, for a given finite timed word $\pi$, if the monitor consumes $\pi$ and reaches a verdict state $\yess$ (\resp $\nos$), then any infinite extension of $\pi$ (\cref{sec:tw})  is in the semantics of $\phi$ (\resp not in the semantics of $\phi$). %

\begin{definition}[Monitor Soundness]%
  \label{def:soundness}%
A monitor $M$ is sound for a formula $\phi\in \logic$ if,
    for every finite timed word $\pi\in \twordstar$:
    \begin{itemize}
      \item $m_0 \wt{\wordd{\pi}} \yess$ \text{ implies } ($\forall \pi' \in \ext{\pi} : \pi' \in \sem{\phi}$)\label{def:soundness-1}
      \item $m_0 \wt{\wordd{\pi}} \nos$ \text{ implies } ($\forall \pi' \in \ext{\pi} : \pi' \not\in \sem{\phi}$)\label{def:soundness-2}
    \end{itemize}
\end{definition}

Sound monitors have the important property of \emph{consistency}. %
Intuitively, a monitor $M$ is consistent if for any finite timed word $\pi$, there exists no two ways for $M$, starting from $m_0$, to consume $\pi$ and reach $\yess$ following one and $\nos$ following the other. %

\begin{definition}[Consistent monitors]
  A monitor $M$ is consistent iff $\forall\pi\in \twordstar:$ $$m_0
  \wt{\wordd{\pi}} \yess \text{ implies } m_0 \nwt{\wordd{\pi}} \nos \text{ and } m_0
  \wt{\wordd{\pi}} \nos \text{ implies } m_0 \nwt{\wordd{\pi}} \yess$$
\end{definition}

\begin{lemma}
  \label{lem:mon-consis}
  If $M$ is sound for some $\phi \in \logic$ then $M$ is consistent.
\end{lemma}
\begin{proof}
  Let $\pi$ be a finite timed word. Assume $M$ is sound for $\phi$ and $m_0 \wt{\wordd{\pi}} \yess$. Then for all 
  $\pi'\in\ext{\pi}$ we have $\pi' \in \sem{\phi}$. Now assume that there is a way to consume $\pi$ and reach $\nos$, that is $m_0 \wt{\wordd{\pi}} \nos$. Then  
  for all $\pi'\in\ext{\pi}$ we have $\pi' \not\in \sem{\phi}$. Contradiction. 
  The proof for the second clause, assuming $M$ is sound for $\phi$ and $m_0 \wt{\wordd{\pi}} \nos$ is analogous.  
\end{proof}

\paragraph{Monitor completeness.}
A monitor $M$ is complete for a formula $\phi$ if it is both violation and satisfaction complete for $\phi$. %
Intuitively, a monitor $M$ is violation (\resp satisfaction) complete for $\phi$
iff, for any time-divergent timed word $\pi$, if~$\pi$ is in (\resp is
not in) the semantics of~$\phi$, then there exists $\pi'$, a finite prefix of $\pi$, such that the monitor inevitably reaches a verdict state 
$\yess$ (\resp $\nos$) after consuming $\pi'$. %

\begin{definition}[Monitor Completenness]%
  \label{def:str-completeness}%
  Let $M\in\mon$ be a monitor and $\phi\in\logic$ a formula. 
  \begin{enumerate}
  \item $M$ is violation complete for $\phi$ if,
    for all $\pi \in \twordsr$: $$\pi \not\in \sem{\phi} \text{ implies } \exists \pi'\in \pf{\pi}: m_0 \wti{\wordd{\pi'}} \nos$$\label{def:sv-comp}
  \item $M$ is satisfaction complete for $\phi$ if,
    for all $\pi \in \twordsr$: $$\pi \in \sem{\phi} \text{ implies } \exists \pi'\in \pf{\pi}: m_0 \wti{\wordd{\pi'}} \yess$$\label{def:ss-comp}
  \item $M$ is complete for $\phi$ if it is both violation  
      and satisfaction complete for $\phi$.
      \label{def:msc}
  \end{enumerate}
\end{definition}

\begin{remark}
The definitions of a (satisfaction,violation,complete) monitorable formula $\phi$ given in the remainder of this paper rely on the existence of a sound and satisfaction-complete/violation-complete/complete monitor $M$ for $\phi$. %
Since such monitors are sought within the framework $\mon$ (\cref{sec:monitoring}), we do not claim general results on arbitrary monitors, \eeg outside of $\mon$ with TLTS semantics. %
Given the expressiveness of $\mon$ and its straightforward connexion with \logic, it is still  important to reason on properties such as maximality (\cref{sec:maximality}) with regards to $\mon$. 
\end{remark}

\subsection{Sound monitor synthesis}\label{sec:sound-monitorss}

As stated in~\cref{thm:tlin}, there exists no algorithm to decide whether a \logic formula $\phi$ is valid or satisfiable, which is already the case for the less expressive TPTL and MTL (\cref{thm:mtl}, \cref{thm:tptl}). %
Therefore, our approach to isolate monitorable fragments of \logic relies on a fully syntactic sound monitor synthesis using an intuitive correspondance between formulae in \logic and monitors in \mon. %
We then reduce the monitorability of a formula to the (violation or satisfaction) completeness of its (synthesised) monitor. %
Of course, this syntax-driven approach does not guarantee that if the synthesised monitor is not violation or satisfaction complete, then the underlying formula is not monitorable, which is expected given the undecidability of \logic satisfiability/validity. %
However, this drawback is attenuated by our maximality results (more in~\cref{sec:maximality}). %

The synthesis procedure relies on an intuitive compiler from \logic to
$\mon$ having the following important property: for every formula $\phi \in \logic$, the 
compiler outputs a monitor that is sound for $\phi$ (\cref{prop:soundness-rthml}). %

The compiler $\compile{-} :$ \logic $\longrightarrow \mon$ is
  defined in \cref{fig:compiler}, where we include box modalities for convenience (note that monitors generated by $\compile{-}$ are necessarily closed since $\compile{-}$ is applied only to formulae, \ie elements of \logic). %
  The rules of $\compile{-}$ are rather simple and straightforward, operating on the syntax of the formula provided as an input. %
For example, if the formula is $\lbox{ a }\dmd{b}\true$, whose semantics is
``if the first action is $a$, then the second must be a $b$''
(\cref{sec:ltrt-HML}), then we want a monitor that consumes the first
action, and becomes verdict $\yes$ if such action is not $a$, and monitors
that the second action is a $b$ otherwise. %
We obtain thus the monitor $\compile{\lbox{ a }\dmd{b}\true} = a.(b.\yes + \overline{b}.\nos) + \overline{a}.\yes$, %
which corresponds to the description above (under the rules defined in~\cref{sec:monitors}). %

\begin{example} The monitor $M$ in~\cref{ex:simple-monitor} is compiled from the  formula $
    x \lin \lbox{x \leq 5}\dmd{b}\true $ (\cref{ex:simple-example}) over $\act = \{b,c\}$. %
    More complex examples will follow. %
\end{example}

\begin{figure}[tb!]
\hrulefill
  $$
  \begin{array}{lll@{\hspace{2em}}lll}
    \compile{\true} & = & \yes
    &
    \compile{\false} & = & \no
    \\[10pt]
    \compile{x \lin \phi} & = & \letin{x}{\compile{\phi}}  
    &
    \compile{X} &=& X
    \\[10pt]
    \compile{ \lbox{ a } \phi }  &=& 
        a.\compile{ \phi } + \overline{a}.\yes
    &
    \compile{ \dmd{ a } \phi }  &=& 
        a.\compile{ \phi } + \overline{a}.\no
    \\[10pt]
    \compile{ \lbox{ g } \phi }  &=& 
        g.\compile{ \phi } + \overline{g}.\yes
    &
    \compile{ \dmd{ g } \phi }  &=& 
        g.\compile{ \phi } + \overline{g}.\no
    \\[10pt]
    \compile{\phi \wedge \psi} & = & \compile{\phi} \otimes \compile{\psi}
    &
    \compile{\phi \vee \psi} & = & \compile{\phi} \oplus \compile{\psi}
    \\[10pt]
    \compile{\maxfp{\phi}} &=& \rec{\compile{\phi}}
    &
    \compile{\minfp{\phi}} & = & \rec{\compile{\phi}}
  \end{array}
  $$
  \hrulefill
  \caption{$\compile{-}$ rules from \logic to monitors, defined recursively on subterms of formulae.}
  \label{fig:compiler}
  \hrulefill
\end{figure}

\begin{restatable}{proposition}{compilesound}%
  \label{prop:soundness-rthml}
  For all $\phi \in \logic$, $\compile{\phi}$ is sound for $\phi$.
\end{restatable}

The other important property of our compiler is \emph{determinism}, \ie for any formula $\phi$ and any finite timed word $\pi$, there is only one way to consume $\wordd{\pi}$ from $(\compile{\phi},0,\vto)$, as states the following proposition. 

\begin{restatable}{proposition}{compiledet}%
  \label{prop:det-rthml}
  For all $\phi \in \logic$ and $\pi\in\twordstar$, there exists $m\in S_{\mathbb{M}}$ such that $(\compile{\phi},0,\vto)\wti{\wordd{\pi}} m$.
\end{restatable}

The above proposition ensured a deterministic implementation of our compiler and the monitoring procedure thereof~\cite{amara2025theory}. %

\subsection{Complete Monitorable Fragment}
\label{sec:complete-monitorability}

In this subsection, we identify $\clogic$, a subset of $\logic$, and prove its complete monitorability (\cref{thm:clogic}). %

\begin{definition}[Complete-monitorable formulae]\label{def:cm}
  A formula $\phi \in \logic$ is complete monitorable iff there exists
  a monitor $M$ that is sound and complete for it. %
  A fragment $\Gamma \subseteq \logic$ is complete monitrorable iff each $\phi \in \Gamma$ is 
  complete monitorable.
\end{definition}

It has been shown in~\cite{DBLP:journals/pacmpl/AcetoAFIL19} that the recursion-free fragment of the untimed linear-time $\mu$-calculus ($\mu$-HML) is its largest complete-monitorable fragment. %
We show that this result does not hold in the timed setting, \ie the maximal set of complete-monitorable formulae in $\logic$ is strictly larger than that of $\mu$-HML. %
To explain this, let us first prove a handy lemma. %
Recall that $\mathcal{D}$ %
is the set of timestamp variables and $\mathcal{V}$ all valuation functions over $\mathcal{D}$ %
(\cref{sec:ltrt-HML}).  

\begin{lemma}\label{time-div-const}
Let  $y\in \mathcal{D}$ and  $n\in\naturals$. %
For all $\pi\in\twordsr,x\in \mathcal{D}\setminus{\{y\}},v\in \mathcal{V},g\in\guards{\mathcal{D}}$:

	\begin{enumerate}
	\item\label{lem:exists-not-models-constraint}%
          $\exists i \in\naturals_{>0}$ such that 
	  $v[x:=t_i(\pi)] \not\models (x - y \prec n) \wedge g$,
	\item\label{lem:not-models-constraint}%
          $\forall i\in\naturals_{>0}$: $v[x:=t_i(\pi)] \not\models (x - y \prec n)$ implies 	
	  $\forall j \geq i$: $v[x:=t_i(\pi)] \not\models (x - y \prec
          n) \wedge g$.	
          \end{enumerate}
\end{lemma}
	
\begin{proof}
	\cref{lem:exists-not-models-constraint} is a direct consequence of time divergence (\cref{sec:framework}). Suppose for a contradiction, for all $i>0$, $v[x:=t_i(\pi)] \models (x - y \prec n) \wedge g$.
	This implies that for all $i > 0$, $t_i(\pi) \prec n + v(y)$, which contradicts condition \eqref{eq:div-tw}. 

	\cref{lem:not-models-constraint} is a direct consequence of monotonicity of timestamps (\cref{sec:preliminaries}). 
	Assume $v[x:=t_i(\pi)] \not\models (x - y \prec n)$. This implies that 
	$t_i(\pi) \not\prec n + v(y)$. Since the timestamps of $\pi$ are monotonically increasing, we have 
	for all $j \geq i$, $t_j(\pi) \not\prec n + v(y)$ wich implies $v[x:=t_j(\pi)] \not\models (x - y \prec n) \wedge g$.
\end{proof}
Let us depict the consequences of the above lemma through a few examples. %

\begin{example}
\label{ex:mon-eventually}
Consider the formula $\phi = \minfp[Y]{\dmd{b}\true \vee   
         {\dmd{\act}{Y}}}$, meaning ``action $b$ will be seen now or in the future'' (which is equivalent to the MTL formula $\uweventually{b}$). %
Clearly, there is no violation-complete monitor for $\phi$, because there is no way for a monitor to say $\nos$ based on a finite timed word not containing $b$ (since a $b$ can occur later on). %
In contrast, $\phi$ is monitorable for satisfactions: it suffices for the monitor to see a $b$ to say $\yess$. %
Now, in a (time) \emph{constrained} setting, \ie whenever we use a bounded interval $I$ (with $\rb{I}\in \naturals$, \cref{def:int-synt}), $\phi$ becomes both satisfaction and violation monitorable, and therefore complete monitorable. %
Consider the following formula: 

$$
  \phi_c = x_s \lin \minfp[Y]{(x_e \lin \dmd{x_e - x_s \in [5,8]}\dmd{b}\true) \vee \dmd{\act}Y}
$$
which means "eventually $b$ will occur within [5,8]". 
The guard $x_e - x_s \in [5,8]$ is of the form $x - y \prec n \wedge
g$ where $y = x_s$ (constant), $n = 8$ (constant), $x = x_e$, ${\prec} = {\leq}$,
$g = \overline{x_e - x_s < 5}$, and $x_e$ diverges after a finite number of recursions  
(because time diverges and $Y$ is guarded by $\dmd{\act}$). %
Therefore, by~\cref{time-div-const}, $x_e - x_s \in [5,8]$ will become false after a finite number of recursions, and will remain false indefinitely afterwards. %
It follows that the monitor can safely say $\nos$ when the guard becomes false (which means that the consumed prefix until then did not contain any action $b$ between $5$ and $8$). %
\end{example}

\begin{example}
\label{ex:mon-cons-br}
Let us now consider the more complex \emph{leads to} formula

$$\psi = \maxfp{\lbox{b}\false \vee \minfp[Y]{\dmd{c}\true \vee \dmd{\act}Y}\wedge \dmd{\act}X}$$

stipulating that ``whenever an action $b$ occurs, an action $c$ follows in the future'' (which is equivalent to the MTL formula $\uwglobally{(b\rightarrow \uweventually{c})}$). %
As explained in~\cref{sec:intro}, this property is not monitorable in any sense (neither for satisfactions nor for violations). %
Now consider the following formula, a \emph{constrained} version of $\psi$: 
$$
  \psi_c = x_s \lin \maxfp[Y]{(x_e \lin \lbox{x_e - x_s \in [1,3]}([b]\false \vee \zeta)) \wedge (\dmd{\act}Y)}
$$
with:
$$
\zeta = x'_s \lin \minfp[X]{(x'_e \lin \dmd{x'_e - x'_s \in [2,5]}\dmd{c}\true) \vee (\dmd{\act}X)}
$$
That is, the constrained bounded response formula $\psi'_c$ requires only $b$ actions within the bounded interval $[1,3]$ to be followed by a $c$. Besides, any such $b$ must have within the bounded interval $[2,5]$, relative to its occurrence, its $c$ ``response''. %
Notice how $\psi'_c$ is also complete monitorable. %
If no~$b$ occurs between~$1$ and~$3$, the monitor can safely say $\yess$ (when the right bound of $[1,3]$ is exceeded). %
Otherwise, for every $b$ occurring between $1$ and $3$, a (sub)monitor checks whether a $c$ occurs between $2$ and $5$ (relative to the occurrence of such $b$), and says $\nos$ if it is not the case. %
Since $[2,5]$ is bounded, the monitor can decide the latest after $5$ time units since the last occurrence of $b$ between $1$ and $3$.  
\end{example}

\begin{example}
\label{ex:mon-not-in-mtl}
Let us now consider the formulae given in~\cref{ex:not-in-mtl}, not expressible in TPTL/MTL. The formula $$\phi_{even(b)} = \maxfp{\dmd{\act}\dmd{b}X}$$ is violation monitorable (a monitor can say $\nos$ as soon as it sees an action at an even position different than $b$), but is not satisfaction monitorable. %
Now, following the same reasoning as in the previous examples (applying~\cref{time-div-const}), the formula $$\phi_{even(b),[3,5]} = x_s\lin \maxfp{\dmd{act} (x_e \lin \dmd{x_e - x_s\in [3,5]}\dmd{b}X \vee y \lin \dmd{y - x_s\notin [3,5]}\dmd{\act}X)}$$ 
 is complete monitorable. 
\end{example}

\paragraph{Syntactic characterisation.} 

In order to characterise a sufficiently expressive syntax for complete-monitorable formulae, we remark the following. %
Formulae $\phi_c$  (\cref{ex:mon-eventually}) %
is equivalent to the following formulae:
$$
\phi'_c =  x_s \lin \minfp[Y]{(x_e \lin \dmd{x_e-x_s \in [5,8]}\dmd{b}\true) \vee y \lin \dmd{y - x_s \leq 8} \dmd{\act}Y}
$$

because when the monitor time exceeds the right bound of $[5,8]$, then there is no need to check if $\dmd{b}\true$ is satisfied, and therefore recursion can be prevented (and the monitor can say $\nos$ at this stage). %

Formula~$\psi_c$  in~\cref{ex:mon-cons-br} can be transformed similarly to the following equivalent formula:
$$x_s \lin \maxfp[Y]{(x_e \lin \lbox{x_e - x_s \in [1,3]}([b]\false \vee \zeta')) \wedge (y \lin \lbox{y - x_s \leq 3} \dmd{\act}Y))}$$ where:
$$
\zeta' = x'_s \lin \minfp[X]{(x'_e \lin \dmd{x'_e - x'_s \in [2,5]}\dmd{c}\true) \vee (z \lin \dmd{z - x'_s \leq 5}\dmd{\act}X)}
$$ 
As for formula~$\phi_{even(b),[3,5]}$ in~\cref{ex:mon-not-in-mtl}, that can be rewritten similarly as follows: 
$$x_s\lin \maxfp{\dmd{act} ( x_e \lin \dmd{x_e - x_s\in [3,5]}\dmd{b}X \vee y \lin (\dmd{y - x_s < 3}\dmd{\act}X \vee \dmd{y - x_s > 5}\dmd{\act}\true))}$$
since there is no need to recurse after the right bound of $[3,5]$ is exceeded (in such case, the timed word is in the semantics of the formula). %
 
More generally, in a complete-monitorable formula $\phi$, every subterm $\minfp{\phi}$ or $\maxfp{\psi}$ is \emph{constrained}. %
That is, in every subterm of $\phi$, every occurrence of recursion variable $X$ is preceded by $x \lin \dmd{g}$ or $x\lin \lbox{g}$ such that $g$ becomes eventually false. 
To simplify the proofs, we use a straightforward syntactic manipulation in order to require recursion variables to be immediately preceded by $x \lin \dmd{g}$ or $x\lin \lbox{g}$ through duplicating such terms modulo timestamp-variable renaming. 
For example, the formula 
$$
\zeta' = x'_s \lin \minfp[X]{(x'_e \lin \dmd{x'_e - x'_s \in [2,5]}\dmd{c}\true) \vee (z \lin \dmd{z - x'_s \leq 5}\dmd{\act}X)}
$$ 
above is semantically equivalent to
$$
\zeta'' = x'_s \lin \minfp[X]{(x'_e \lin \dmd{x'_e - x'_s \in [2,5]}\dmd{c}\true) \vee (z \lin \dmd{z - x'_s \leq 5}\dmd{\act} y \lin \dmd{y - x'_s \leq 5}X)}
$$

In order to syntactically specify that some guard becomes unsatisfied after a finite number of recursions, we introduce $\conslvar$, an infinite subset of logical variables $\vars$, such that a variable that belongs to $\conslvar$ must appear in a particular form within the scope of a fixpoint operator. %

\begin{definition}[Complete-monitorable fragment]
  \label{def:clogic}
  \label{def:comp-mon}
  A formula $\phi\in\logic$ is in the set \consrthml $\subset$ \logic of complete-monitorable formulae iff $\phi$ is generated by the grammar in \cref{fig:syntax-consrthml}  
  (with $n \in \naturals$, $g\in\guards{\mathcal{D}}$, $x,y\in\mathcal{D}$ and $X\in\conslvar$) \emph{and} for all $\minfp[X]\psi$, $\maxfp[X]\psi$ subterms of $\phi$, $y$ is free\footnote{This is to guarantee that, within the scope of $\maxfp[Y]{\psi}$ or $\minfp{\psi}$, 
  the valuation of $y$ remains constant and so~\cref{time-div-const} can be applied.} in $\psi$.
 A subterm $\maxfp[X]{\phi}$ (\resp $\minfp[X]{\phi}$) of a formula in \clogic is called a constrained \textsf{max} (\resp constrained \textsf{min}). %
  \end{definition}

\begin{figure}[tb]
  \hrulefill
  $$
    \begin{array}{llllllll}
      \phi,\psi &::=& \true &\BNFsep \false &\BNFsep \phi \vee \psi & \BNFsep \phi \wedge \psi&\BNFsep x \lin \phi\\&& 
        \BNFsep \dmd{a}\phi &\BNFsep \lbox{a}\phi &\BNFsep \dmd{g}\phi &\BNFsep \lbox{g}\phi 
         \\
        &&\BNFsep \maxfp{\phi} & \BNFsep \minfp{\phi} & \BNFsep x \lin \lbox{g_x}X & \BNFsep x \lin \dmd{g_x}X \\
    
        g_x &::=& x \prec n & x - y \prec n &\BNFsep g_x \wedge g \\
    \end{array}
  $$
  \hrulefill
  \caption{Grammar to generate formulae (closed and guarded terms) of \consrthml.}
  \label{fig:syntax-consrthml}
  \hrulefill
\end{figure}

To prove that \clogic is complete monitorable, we show that $\compile{\phi}$ is sound and  complete for every $\phi\in\clogic$. 

\begin{proposition}%
  \label{prop:snd}
  \label{prop:soundness-consrthml}
  For all $\phi \in \consrthml$, $\compile{\phi}$ is sound for $\phi$.
\end{proposition}
\begin{proof}
  Since $\consrthml \subset \logic$, we conclude by \cref{prop:soundness-rthml}.
\end{proof}

\begin{restatable}{proposition}{ctlinsatcomp}%
\label{pop:sc-ctlin}
  For all $\phi \in \consrthml$, $\compile{\phi}$ is satisfaction complete for $\phi$.
\end{restatable}

\begin{restatable}{proposition}{ctlinvcomp}%
\label{pop:vc-ctlin}
  For all $\phi \in \consrthml$, $\compile{\phi}$ is violation complete for $\phi$.
\end{restatable}

\begin{example}
Consider the following formula: $$\phi =  \minfp{(x \lin \dmd{x \in (3,7]}\dmd{b} y\lin \dmd{y-x\in[0,2)}\dmd{c}\true) \vee z\lin \dmd{z \leq 7}\dmd{\act}X}$$ easily expressible in $\clogic$ (see above), which expresses the following property ``there exists an occurrence of $b$ within the absolute interval $(3,7]$ that is immediately followed by a $c$ within the interval $[0,2)$''; where ``immediately followed by'' means that there are no other actions occurring in between. %
In other words, the semantics of $\phi$ contains all infinite timed words $\pi$ such that there exists a position $i$ with $a_i(\pi) = b$ \emph{and} $t_i(\pi) - t_1(\pi) \in (3,7]$ \emph{and}  $a_{i+1}(\pi) = c$ \emph{and} $t_{i+1}(\pi) - t_i(\pi) \in [0,2)$. To compile $\phi$ into a monitor, we first distribute over the disjunction $\dmd{\act}$ using a distinct timestamp variable $z^a$ for each clause: 
$$\phi =  \minfp{(x \lin \dmd{x \in (3,7]}\dmd{b} y\lin \dmd{y-x\in[0,2)}\dmd{c}\true) \vee \bigvee_{a\in\act} z^a\lin \dmd{z^a \leq 7}\dmd{a}X}$$

Then apply the rules of our compiler to obtain: 

$M = \compile{\phi} = \rec[X]{((\letin{x}(((x \in (3,7]).(b.N + \overline{b}.\no)) + $ \\ $(x \notin (3,7]).\no)) \oplus (\oplus_{a\in\act}\, \,  \letin{z^a}(((z^a \leq 7).(a.X + \overline{a}.\no)) +  (z^a > 7).\no)))}$ where 
$$N = \letin{y}(((y-x\in [0,2)).(c.\yes + \overline{c}.\no)) + (y-x\notin [0,2)).\no)$$ 

Monitor $M$ is sound and complete for $\phi$. %
From $m_0$, if a verdict state $\yess$ is reached, then the monitor necessarily consumed a prefix $\pi\in\twordstar$ such that a $b$ occurred within $[3,7)$ followed immediately by a $c$ within $(0,2]$ ($\letin{x}(x \in (3,7]).b.\letin{y}(y-x\in [0,2).c.\yes$) and therefore any extension of $\pi$ is in the semantics of $\phi$ (the reasoning is dual for the cases where a verdict state $\nos$ is reached). Similarly, for any $\pi\in\twordsr$ in the semantics of $\phi$ (\resp not in the semantics of $\phi$), the monitor will inevitably reach a verdict state $\yess$ (\resp $\nos$) after consuming a finite prefix of $\pi$. %
For example, for any infinite timed word $\pi$ with $\pi_1 = (a,1)(b,7)(c,8.2) \in \pf{\pi}$ (\resp $\pi'$ with $\pi_2 = (a,2)(b,3)(c,7.6) \in \pf{\pi'}$) we have $m_0\wti{\wordd{\pi_1}}\yess$ (\resp $m_0\wti{\wordd{\pi_2}}\nos$). %
\end{example}

\begin{theorem}\label{thm:clogic}
  \consrthml\ is complete monitorable.
\end{theorem}

\begin{proof} For every $\phi\in \clogic$, $\compile{\phi}$ is sound (\cref{prop:snd}), satisfaction complete (\cref{pop:sc-ctlin}) and violation complete (\cref{pop:vc-ctlin}) for $\phi$. It follows that \clogic is complete monitorable (\cref{def:cm}). 
\end{proof}

\subsection{Monitorabile Fragment}
\label{sec:partial-monitorability}

In this subsection, we identify \mlogic, a fragment of \logic, and prove its monitorability (\cref{thm:mon-m}). %

\begin{definition}[Monitorable formulae]%
  \label{def:mf}

  \begin{itemize}
  \item %
    A formula $\phi \in \logic$ is {\em violation} (\resp  \emph{satisfaction}) {\em monitorable} iff there exists a monitor 
    $M$ that is sound and violation (\resp satisfaction) complete for $\phi$. A fragment $\Gamma \subseteq \logic$ is 
    violation (\resp satisfaction) monitorable iff each $\phi \in \Gamma$ is violation (\resp satisfaction) monitorable.
  \item %
    A formula $\phi \in \logic$ is {\em monitorable} iff it is violation \emph{or}  
    satisfaction monitorable. A fragment $\Gamma \subseteq \logic$ is 
    monitorable iff each $\phi \in \Gamma$ is monitorable.
  \end{itemize}
\end{definition}

In order to identify a monitorable fragment, let us first proceed
with examples.

\begin{example}
\label{ex:br2}
 Consider the following \logic formula: 
$$
  \phi = \maxfp[Y]{([b]\false \vee \psi) \wedge (\dmd{\act}Y)}
$$
with:
$$
\psi = x_s \lin \minfp[X]{(x_e \lin \dmd{x_e - x_s \in I}\dmd{c}\true) \vee (\dmd{\act}X)}
$$
with $I\in \mathcal{I}$ bounded ($\rb{I} \in \N$). %
Formula $\phi$ expresses the important bounded response property exemplified in~\cref{sec:intro} (equivalent to the 
MTL formula $\uwglobally{(b \Rightarrow \weventually{I} c)}$). %
Notice how every \textsf{min} subformula of $\phi$ can be rewritten as a constrained \textsf{min} (\cref{sec:complete-monitorability}), whereas the \textsf{max} subformulae are clearly unconstrained. %
Formula $\phi$ is violation monitorable: after every
occurrence of~$b$, a (sub)monitor can check whether a~$c$ will follow
within~$I$ (relative to such occurrence).  %
Since~$I$ is bounded,
such check always terminates (by \cref{time-div-const}), and a rejection verdict can be issued if no~$c$ occurred. %
However, this formula is not satisfaction monitorable. %
\end{example} 
\begin{remark}
  Bounded response is not \emph{bounded-memory monitorable} in the
  general case~\cite{maler2007synthesizing} (it is when $I$ is left-closed at $0$, see \eeg~\cite{AlurFH96}). 
  While bounded-memory
  monitorability is out of this paper's scope, realistic assumptions such as \emph{bounded variability} can be adopted to enforce it~\cite{maler2007synthesizing}.
\end{remark}

\begin{example} Consider now the following \logic formula: 
$$
  \eta = \minfp[Y]{((\dmd{b}\true \wedge \zeta) \vee \dmd{\act}Y)}
$$
with:
$$
\zeta = x_s \lin \maxfp[X]{(x_e \lin \lbox{x_e - x_s \in I}\dmd{c}\true) \wedge (\dmd{\act}X)}
$$
with $I\in \mathcal{I}$ bounded. %
Notice how every \textsf{max} subformula of $\eta$ can be rewritten as a constrained \textsf{max} (\cref{sec:complete-monitorability}), whereas the \textsf{min} subformulae are clearly not constrained. %
Formula $\eta$ is satisfaction monitorable: if a $b$ occurs, a monitor can safely say $\yess$ if every action seen within $I$ after such occurrence is a $c$ (this process terminates, again thanks to \cref{time-div-const}).  
This formula is however not violation monitorable: any $\nos$ verdict may be falsified by a future $b$ after which every action within $I$ is a $c$.
\end{example}

\paragraph{Syntactic characterisation.} 
The intuition is that both violation- and satisfaction-monitorable
formulae will share the syntax of \clogic and differ for
(unconstrained) minimal and maximal fixpoint operators.  %
To simplify the syntax, we rely on the following proposition, proven in~\cite[Appendix E]{DBLP:conf/ecoop/Amara0FF25}, stating  
that the semantics of constrained \textsf{min} and constrained \textsf{max} coincide over time-divergent timed words. %
We let $\semdiv{\phi} = \twordsr\cap\sem{\phi}$. 

\begin{proposition}
  \label{prop:minimal-logic}
  \phantom{invisible}
  \begin{enumerate}[(a)]
    \item \label{clogic-max} For all $\maxfp[X]{\phi} \in \clogic$, $\semdiv{\maxfp{\phi}} = \semdiv{\minfp{\phi}}$
    \item \label{clogic-min} For all $\minfp[X]{\phi} \in \clogic$, $\semdiv{\minfp{\phi}} = \semdiv{\maxfp{\phi}}$
  \end{enumerate}
\end{proposition}

Accordingly, for violation-monitorable formulae, we do not need constrained \textsf{min} in their syntax (\cref{fig:max-fragments}). %
This is because constrained \textsf{min} formulae can be rewritten as constrained \textsf{max} formulae, and a constrained \textsf{max} is a special case of \textsf{max}. %
For example, in the bounded response property (Example~\ref{ex:br2}), and after transforming $\psi$ into a constrained $\textsf{min}$ (\cref{sec:complete-monitorability}), the minimal fixpoint operator can be replaced with the maximal one without changing the semantics. %
Dually, we do not need constrained \textsf{max} in the syntax of satisfaction-monitorable formulae  (\cref{fig:min-fragments}). %

\begin{figure}
  \hrulefill
  $$
    \begin{array}{llllllll}
      \phi,\psi &::=& \true &\BNFsep \false &\BNFsep \phi \vee \psi & \BNFsep \phi \wedge \psi&\BNFsep x \lin \phi\\&& 
        \BNFsep \dmd{a}\phi &\BNFsep \lbox{a}\phi &\BNFsep \dmd{g}\phi &\BNFsep \lbox{g}\phi 
         \\
        &&\BNFsep \minfp[X]{\phi} &\BNFsep X \\
    
    \end{array}
  $$
  \hrulefill
  \caption{Grammar to generate formulae (closed and guarded terms) of \slogic.}

  \label{fig:min-fragments}
  \hrulefill
\end{figure}

\begin{figure}
 \hrulefill
  $$
    \begin{array}{llllllll}
      \phi,\psi &::=& \true &\BNFsep \false &\BNFsep \phi \vee \psi & \BNFsep \phi \wedge \psi&\BNFsep x \lin \phi\\&& 
        \BNFsep \dmd{a}\phi &\BNFsep \lbox{a}\phi &\BNFsep \dmd{g}\phi &\BNFsep \lbox{g}\phi 
         \\
        &&\BNFsep \maxfp[Y]{\phi} &\BNFsep Y \\    
    \end{array}
  $$
  \hrulefill
  \caption{Grammar to generate formulae (closed and guarded terms) of \vlogic.}
  \label{fig:max-fragments}
  \hrulefill
\end{figure}

\begin{definition}[Monitorable fragment]\label{def:mlogic}
  The set of monitorable formulae $\mlogic\subset \logic$ is the union of the set of violation-monitorable formulae (\cref{fig:max-fragments}) and satisfaction-monitorable formulae (\cref{fig:min-fragments}), \ie $\mlogic = \vlogic \cup \slogic$. 

\end{definition}

To prove \mlogic is monitorable, we rely on our compiler $\compile{-}$ in a similar fashion to the proof of complete monitorability of \clogic (\cref{sec:complete-monitorability}). %

\begin{proposition}%
  \label{prop:soundness-prthml}
  For all $\phi \in$ \mlogic, $\compile{\phi}$ is sound for $\phi$.
\end{proposition}
\begin{proof}
  Since $\mlogic \subset \logic$, we conclude by \cref{prop:soundness-rthml}.
\end{proof}

\begin{restatable}{proposition}{scomp}%
  \label{prop:satisfaction-completeness-minrthml}
  For all $\phi \in \slogic$, $\compile{\phi}$ is satisfaction complete for $\phi$.
\end{restatable}

\begin{theorem}%
  \label{thm:slogic-sm}
  \slogic is satisfaction monitorable. 
\end{theorem}
\begin{proof}%
  Direct from~\cref{prop:soundness-prthml}, ~\cref{prop:satisfaction-completeness-minrthml} and \cref{def:mf}. %
\end{proof}

\begin{restatable}{proposition}{vcomp}
  \label{prop:violation-completeness-maxrthml}
  For all $\phi \in \vlogic$, $\compile{\phi}$ is violation complete for $\phi$.
\end{restatable}

\begin{example} The formula $\phi_{even(b)} = \maxfp{\dmd{\act}\dmd{b}X}$ (\cref{ex:not-in-mtl}) is violation monitorable. %
Its compilation by $\compile{-}$ 
gives the monitor
$$M = \compile{\phi_{even(b)}} = \rec[X]{({\oplus_{a\in\act}} \, \, a.(b.X + \overline{b}.\no) + \overline{a}.\no)}$$ 
and it is rather straightforward to see that $M$ is sound and inevitably reaches verdict $\nos$ (from $m_0$) when given any infinite timed word outside of the semantics of $\phi_{even(b)}$. %
Indeed, for any infinite timed word $\pi$, we have $\pi\notin\sem{\phi_{even(b)}}$ iff there exists a position $i\in\naturals_{>1}$, ${i}\bmod{2} = 0$ such that $a_i(\pi) \neq b$. %
For any such $\pi$ and the smallest such $i$, we have $m_0 \wti{\wordd{^{i}\pi}} \nos$. %
\end{example}

\begin{theorem}%
  \label{thm:vlogic-vm}
\vlogic is violation monitorable. 
\end{theorem}
\begin{proof}
  Direct from~\cref{prop:soundness-prthml}, ~\cref{prop:violation-completeness-maxrthml} and \cref{def:mf}. %
\end{proof}

\begin{theorem}%
  \label{thm:mon-m}
  \mlogic is monitorable.
\end{theorem}

\begin{proof} 
Direct from~\cref{thm:slogic-sm}, \cref{thm:vlogic-vm}, \cref{def:mlogic} and \cref{def:mf}. 
\end{proof}

\subsection{Maximality}\label{sec:maximality}

In this subsection, we prove the \emph{maximality} of our monitorable fragments \emph{with regards to} $\mon$. %
That is, we show that any monitorable formula $\phi$, in the sense that there exists a monitor in $\mon$ that is sound and (violation or satisfaction) complete for $\phi$, there exists a formula $\psi\in\mlogic$ equivalent to $\phi$. %

\begin{definition}[Largest monitorable fragment]\label{def:maximality} 
A fragment $\Gamma\subset\logic$ is said:
\begin{itemize}
\item{} The largest satisfaction- (\resp violation-, complete-) monitorable fragment of \logic iff (i) $\Gamma$ is satisfaction (\resp violation, complete) monitorable and (ii) for every satisfaction- (\resp violation-, complete-) monitorable formula $\psi\in\logic$, there exists $\phi\in\Gamma$ such that $\sem{\phi} = \sem{\psi}$, %
\item{} The largest monitorable fragment of \logic iff (i) $\Gamma$ is monitorable and (ii) for every monitorable formula $\psi\in\logic$, there exists $\phi\in\Gamma$ such that $\sem{\phi} = \sem{\psi}$.
\end{itemize}
\end{definition}

\begin{figure}[t]
\hrulefill
  $$
  \begin{array}{lll@{\hspace{2em}}lll}
    \tomin{M\in\syes} & = & \true 
    &
    \tomin{M\in\sno} = \tomin{M\in\smend} &=& \false 
    \\[10pt]
    \tomin{\letin{x}{M}} & = &  x \lin \tomin{M}
    &
    \tomin{M + N} &=& \tomin{M} \vee \tomin{N}
    \\[10pt]
    \tomin{M \oplus N} &=& \tomin{M} \vee \tomin{N}
    &
    \tomin{M \otimes N} &=& \tomin{M} \wedge \tomin{N}
    \\[10pt]
    \tomin{a.M} &=& \dmd{a}\tomin{M}
    &
    \tomin{g.M} &=& \dmd{g}\tomin{M}
    \\[10pt] 
    \tomin{X} &=& X 
    &
    \tomin{\rec{M}} &=& \minfp{\tomin{M}}
  \end{array}
  $$
  \hrulefill
  \caption{Compiler from monitors to \slogic, defined recursively on subterms of monitors.}
  \label{fig:compiler-to-minrthml}
  \hrulefill
\end{figure}

\begin{figure}[t]
\hrulefill
  $$
  \begin{array}{lll@{\hspace{2em}}lll}
    \tomax{M\in\syes}=\tomax{M\in\smend} &=& \true 
    &
    \tomax{M\in\sno} & = & \false 
    \\[10pt]
    \tomax{\letin{x}{M}} & = &  x \lin \tomax{M}
    &
    \tomax{M + N} &=& \tomax{M} \wedge \tomax{N}
    \\[10pt]
    \tomax{M \oplus N} &=& \tomax{M} \vee \tomax{N}
    &
    \tomax{M \otimes N} &=& \tomax{M} \wedge \tomax{N}
    \\[10pt]
    \tomax{a.M} &=& \lbox{a}\tomax{M}
    &
    \tomax{g.M} &=& \lbox{g}\tomax{M}
    \\[10pt] 
    \tomax{X} &=& X 
    &
    \tomax{\rec{M}} &=& \maxfp{\tomax{M}}
%
  \end{array}
  $$
  \hrulefill
  \caption{Compiler from monitors to \vlogic, defined recursively on subterms of monitors.}
  \label{fig:compiler-to-maxrthml}
  \hrulefill
\end{figure}

Since the satisfiability/validity of \logic are undecidable, we follow again a syntactic approach, inspired from~\cite{DBLP:journals/pacmpl/AcetoAFIL19}. %
First, we prove that for every monitor $M$, if $M$ is satisfaction (\resp violation) complete for two formulae in \logic, then such formulae are semantically equivalent (\cref{lem:monitors-and-sem}). %
Second, we define two functions: $\tomin{-}$ (\cref{fig:compiler-to-minrthml}), that maps any monitor $M$ to 
a formula in \slogic, and $\tomax{-}$ (\cref{fig:compiler-to-maxrthml}) that maps any monitor $M$ to 
a formula in \vlogic; then prove that any $M$ is sound and satisfaction complete for its image by $\tomin{-}$ (\ie $\tomin{M}$), and sound and violation complete for $\tomax{M}$ (\cref{lem:sound-complete}). %
We conclude therefore that any formula $\phi\in\logic$ that is satisfaction (\resp violation) monitorable, through the existence of a monitor $M$ that is satisfaction (\resp violation) complete for $\phi$, is necessarily equivalent to $\tomin{M}\in\slogic$ (\resp $\tomax{M}\in\vlogic$, ~\cref{prop:maximality}). %
Accordingly, \slogic is the largest satisfaction-monitorable fragment of \logic, \vlogic the largest violation-monitorable fragment of \logic, and $\mlogic = \vlogic \cup \slogic$ the largest monitorable fragment of \logic (\cref{thm:max}, ~\cref{thm:maximality-prthml}). %
We also prove that \consrthml is the largest complete-monitorable fragment of \logic (\cref{thm:maximality-consrthml}). %

\begin{restatable}{lemma}{equalsemon}
  \label{lem:monitors-and-sem}
  For every monitor $M$ and formulae $\phi,\psi\in \logic$:
  \begin{enumerate}
    \item $M$ is sound and satisfaction complete for $\phi$ and $\psi$ implies $\sem{\phi} = \sem{\psi}$\label{lem:monitors-and-sem-s},
    \item $M$ is sound and violation complete for $\phi$ and $\psi$ implies $\sem{\phi} = \sem{\psi}$\label{lem:monitors-and-sem-v}.
  \end{enumerate} 
\end{restatable}

\begin{restatable}{lemma}{compforcomp}
\label{lem:sound-complete}
 
  For every monitor $M$, if $M$ is consistent then:
  \begin{enumerate}
  \item $M$ is sound and satisfaction complete for $\tomin{M}$  \label{lem:tomin-sound-satisfaction-complete}
  \item $M$ is sound and violation complete for $\tomax{M}$ \label{lem:tomax-sound-violation-complete}
  \end{enumerate}
\end{restatable}

\begin{proposition}
  \label{prop:maximality}
  For every formula $\phi \in \logic$: 
  \begin{enumerate}
  \item{} if $\phi$ is satisfaction monitorable, then there exists 
  $\psi \in \slogic$ such that $\sem{\phi} = \sem{\psi}$,\label{prop:maximality-minrthml}
  \item{} if $\phi$ is violation monitorable, then there exists 
  $\psi \in \vlogic$ such that $\sem{\phi} = \sem{\psi}$. \label{prop:maximality-maxrthml}
  \end{enumerate}
\end{proposition}

\begin{proof}
  Of~\cref{prop:maximality-minrthml}. Assume $\phi \in \logic$ is satisfaction monitorable. By definition, there exists a  monitor $M$ that is sound and satisfaction complete for $\phi$. 
  By \cref{lem:mon-consis},
  we have $M$ is consistent. 
  Now, by~\cref{lem:sound-complete} (\cref{lem:tomin-sound-satisfaction-complete}), we deduce 
  $M$ is sound and satisfaction complete for $\tomin{M} \in $ \slogic, and finally we conclude 
  $\sem{\phi} = \sem{\tomin{M}}$ by \cref{lem:monitors-and-sem} (\cref{lem:monitors-and-sem-s}).\\
  Of~\cref{prop:maximality-maxrthml}. Analogous. 
  \qed
\end{proof}

\begin{theorem}%
  \label{thm:max}
  \begin{enumerate}
  \item{} \slogic is the largest satisfaction-monitorable fragment of \logic,\label{thm:max-s}
  \item{} \vlogic is the largest violation-monitorable fragment of \logic.\label{thm:max-v}
  \end{enumerate}
\end{theorem}
\begin{proof}
Of~\cref{thm:max-s}. Direct from~\cref{def:maximality}, \cref{thm:slogic-sm}, and~\cref{prop:maximality} (\cref{prop:maximality-minrthml}). 
Of~\cref{thm:max-v}. Direct from~\cref{def:maximality}, \cref{thm:vlogic-vm}, and~\cref{prop:maximality} (\cref{prop:maximality-maxrthml}).
\end{proof}

\begin{theorem}%
  \label{thm:maximality-prthml}
  \mlogic is the largest monitorable fragment of \logic. %
\end{theorem}
\begin{proof}
  Direct from~\cref{def:maximality}, \cref{thm:max} and~\cref{def:mlogic}. %
\end{proof}

\begin{restatable}{theorem}{maximalityconsrthml}%
  \label{thm:maximality-consrthml}
  \clogic is the largest complete-monitorable fragment of \logic. %
\end{restatable}

\section{Monitorability of MTL}\label{sec:monmtl}

In this section, we identify useful satisfaction, violation, and complete-monitorable fragments of MTL. %
To this end, we exploit the results of \cref{sec:monitorability} to determine which MTL formulae are 
compiled into the violation, satisfaction, and complete-monitorable fragments of $\logic$. %
Let us first define the set of \emph{general Until sub-formulae} of an MTL formula. %
The proofs not enclosed in this section are given in~\cref{app:monmtl}. 

\begin{definition}\label{def:sub}
Let $\phi$ be an MTL formula (\cref{fig:mitl-syntax}). %
The set of general Until sub-formulae of $\phi$, denoted $\mathit{sub_{\mathcal{U}}}(\phi)$ is defined recursively: \\
$\mathit{sub_{\mathcal{U}}}(\top)=\mathit{sub_{\mathcal{U}}}(\bot)=\mathit{sub_{\mathcal{U}}}(a)=\emptyset$; \\
$\mathit{sub_{\mathcal{U}}}(\neg \phi)=\bigcup_{\psi\in\mathit{sub_{\mathcal{U}}}(\phi)} \{\neg \psi\}$;\\
$\mathit{sub_{\mathcal{U}}}(\phi\wedge\psi)=\mathit{sub_{\mathcal{U}}}(\phi)\cup\mathit{sub_{\mathcal{U}}}(\psi)$; \\
$\mathit{sub_{\mathcal{U}}}(\until{\phi}{I}{\psi})=\{\until{\phi}{I}{\psi}\}\cup\mathit{sub_{\mathcal{U}}}(\phi)\cup\mathit{sub_{\mathcal{U}}}(\psi)$.\\

Therefore, every element of $\mathit{sub_{\mathcal{U}}}(\phi)$ is of the form $\until{\psi_1}{I}{\psi_2}$ preceded by a finite number of negations. %
We call each such $\until{\psi_1}{I}{\psi_2}$ an Until sub-formula. %

\end{definition}

That is, $\mathit{sub_{\mathcal{U}}}(\phi)$ extracts every general Until sub-formula, \ie every Until sub-formula $\psi$ of $\phi$ preceded by the number of negations $\psi$ falls under within $\phi$.

\begin{example}\label{ex:sub}
Let $\act = \{a,b,c\}$ and $\psi = \ueventually{(b \wedge \globally{[0,1]}{c})}$, requiring the existence of a future $b$ followed by only $c$ actions within the relative interval $[0,1]$. %
Formula $\psi$ is syntactic sugar for 
$$\psi' = \uuntil{\top}{(b\wedge \neg(\until{\top}{[0,1]}{\neg c})})$$

We have $\mathit{sub_{\mathcal{U}}}(\psi') = \{\theta = \neg(\until{\top}{[0,1]}{\neg c}), \theta' =  \uuntil{\top}{(b\wedge \theta)}\}$.
\end{example}

We will hereafter use ``$\psi$ is preceded by'' and ``$\psi$ falls under'' some number of negation interchangeably. %

Now, given our compiler $\tohmlSym$ (\cref{fig:mitl-compiler}), the intuition is the following. %
Let $\phi$ be an MTL formula. %
If every Until sub-formula in $\mathit{sub_{\mathcal{U}}}(\phi)$ falls under an even number of negations, then $\tohmlSym(\phi)$ will contain no maximal fixpoint terms, \ie $\tohmlSym(\phi)\in \slogic$ and therefore $\phi$ is satisfaction monitorable. %
Dually, if every Until sub-formula is preceded by an odd number of negations, then $\tohmlSym(\phi)$ will contain no minimal fixpoint terms, \ie $\tohmlSym(\phi)\in \vlogic$ and therefore $\phi$ is violation monitorable. %

The above reasoning is however too coarse leading to missing formulae that are clearly monitorable. %

\begin{example}
Consider $\psi'$ and its set of general Until sub-formulae $\mathit{sub_{\mathcal{U}}}(\psi')$ in~\cref{ex:sub}. %
In $\mathit{sub_{\mathcal{U}}}(\psi')$, we have two general Until sub-formulae $\theta$ and $\theta'$, such that the former is an Until sub-formula preceded by an odd number of negations (one) and the latter an Until sub-formula preceded by an even number of negations (zero), and therefore we cannot conclude on whether this formula is monitorable using the above reasoning. %
Yet, this formula is clearly satisfaction monitorable due to the bounded interval in $\theta$: if there exists a $b$ followed only by $c$ actions within the relative interval $[0,1]$, one may issue a satisfaction verdict when the right bound of $[0,1]$ is exceeded. %
\end{example}

Treating Until sub-formulae with bounded intervals differently would be still insufficient, however, as we can miss trivially monitorable formulae containing Until sub-formulae of the form $\uuntil{\bot}{\phi}$, \ie encoding the Next operator. 

\begin{example}
Let $\act = \{a,b,c\}$ and  $\psi' = \uglobally{(b \Rightarrow \unxt{c})}$. %
This formula requires that every future $b$ is followed immediately (\ie at the next position) by a $c$. %
Here again, the above reasoning prevents to conclude on the monitorability of $\psi'$ since the latter is syntactic sugar for
$$\neg(\uuntil{\top}{(b \wedge \neg(\uuntil{\bot}{c})))}$$
and we have one Until sub-formula falling under an odd number of negations and the another  falling under an even number of negations, and none of them carries a bounded interval. %
Yet, the above formula is actually violation monitorable: since there is, by definition, an infinity of actions in an infinite timed word, whenever we encounter a $b$ we will see its successor in a bounded amount of time, and if it is not a $c$, we can return the violation verdict. %
\end{example} 

The first step is therefore to define formally, based on the above observations, the kind of Until sub-formulae which do not impair the  monitorability of the formula at hand, which we call \emph{constrained Until (sub)-formulae}. %

\begin{definition}[Constrained Until formula]\label{def:boun}
An Until sub-formula $\until{\phi}{I}{\psi}$ is constrained iff (1) $I$ is bounded (\cref{def:int-synt}) or (2) $\phi = \bot$. %
An Until sub-formula is unconstrained otherwise. 
\end{definition} 

A more precise reasoning is therefore to delineate constrained from unconstrained Until sub-formulae (\cref{def:boun}). On the former, no condition should be imposed, whereas we require for the latter something on the number of negations they are preceded by, depending on whether we seek to characterise satisfaction or violation monitorability. %

Now, our compiler $\tohmlSym$ given in~\cref{fig:mitl-compiler} does not distinguish between constrained and unconstrained Until sub-formulae. %
In order to capture constrained Until sub-formula of type (1) in~\cref{def:boun}, \ie $\until{\phi}{I}{\psi}$ where $I$ is bounded (\cref{def:boun}), the idea is to develop an equivalent variant of $\tohmlSym$.  In this variant, called $\tomtlinSym$ and detailed in~\cref{fig:mtl-compiler2}, the compilation of $\until{\phi'}{I}{\psi}$, where $\phi' \neq \bot$, adds a guard such that, if $I$ is bounded, the recursion will be eventually prevented, \ie when the right bound of $I$ is exceeded. %
We recall that $\precrI\, =\, <$ if $I$ is right open and $\leq$ otherwise (\cref{def:int-sem}).

\begin{figure}[ht!]
  \hrulefill
    $$
      \begin{array}{lll}
        \tomtlin{\top} &=& \true
        \\[3pt]
        \tomtlin{\bot} &=& \false
        \\[3pt]
        \tomtlin{a} &=& \dmd{a}\true
        \\[3pt]
        \tomtlin{\neg \phi} &=& \tlinnot{\tomtlin{\phi}}
        \\[3pt]
        \tomtlin{\phi \wedge \psi} &=& \tomtlin{\phi} \wedge \tohml{\psi}
        \\[8pt]
        \tomtlin{\until{\phi'}{I}{\psi}} &=& x_s \lin \dmd{\act} \minfp[Y]{x_e \lin (\dmd{x_e - x_s \in I}{\tomtlin{\psi}}) \\ 
          && \phantom{x_s \lin \dmd{\act} \minfp[Y]{}} \vee (\tomtlin{\phi'} \wedge  \dmd{ \act }{y \lin \dmd{y - x_s \precrI \rb{I}}{Y}} )}\\
          \tomtlin{\until{\bot}{I}{\psi}} &=& x_s \lin \dmd{\act}x_e \lin\dmd{x_e - x_s \in I}{\tomtlin{\phi}}\\
      \end{array}
      $$
  \hrulefill
  \caption{From MTL to \logic, where $\phi'\neq \bot$}
  \label{fig:mtl-compiler2}
  \hrulefill
\end{figure}

For constrained Until sub-formulae of type (2) in~\cref{def:boun}, \ie $\until{\bot}{I}{\phi}$, the compilation gives simply $x_s \lin \dmd{\act}x_e \lin\dmd{x_e - x_s \in I}{\tomtlin{\phi}}$ which coincides with the compilation of $\nxt{I}{\phi}$ given in~\cite{DBLP:conf/ecoop/Amara0FF25} (explanation below). 

The following proposition states that $\tomtlinSym$ is an equivalent variant of $\tohmlSym$, \ie for any MTL formula $\phi$, $\tomtlinSym{\phi}$ and $\tohmlSym{\phi}$ have the same semantics. 
\begin{restatable}{proposition}{mtltwo}\label{mtl2}
	\label{prop:correctness-compiler2}
	For all $\phi \in $ MTL, we have $\sem{\tohml{\phi}} = \sem{\tomtlin{\phi}}$. 
\end{restatable}

We give a proof sketch of~\cref{mtl2} for the case $\until{\bot}{I}{\phi}$ below. %
By the rules of $\tohmlSym$ (\cref{fig:mitl-compiler}), we have 
$$
\begin{aligned}
\tohmlSym{\until{\bot}{I}{\phi}} &=
          x_s \lin \dmd{\act} \minfp[Y]{x_e \lin (\dmd{x_e - x_s \in I}{\tomtlin{\phi}}) \vee\\
          & (\false \wedge  \dmd{ \act } {Y} )}
          \end{aligned}
          $$
          And since $\false \wedge \psi$ is equivalent to $\false$ for any $\psi$ we get: 
$$x_s \lin \dmd{\act} \minfp[Y]{x_e \lin (\dmd{x_e - x_s \in I}{\tomtlin{\phi}}))}$$
Finally, since the recursion variable $Y$ does not appear at all under the \textsf{min}, the latter can be removed altogether as the above formula is equivalent to 

$$x_s \lin \dmd{\act}x_e \lin\dmd{x_e - x_s \in I}{\tomtlin{\phi}}$$

Now, in order to reason on the monitorability of an MTL formula $\phi$, a constrained Until sub-formula of $\phi$ is exempt of any requirement: regardless of the number of negations it is preceded by, our compiler will give a constrained \textsf{min}, a constrained \textsf{max}, or no fixpoint term at all, and therefore if all Until sub-formulae are constrained, $\phi$ is complete monitorable, \ie both satisfaction and violation monitorable. %
This gives us the CMTL fragment defined below. %

\begin{definition}
  \label{def:cmtl}
  The CMTL fragment of MTL is defined as the set of MTL formulae where every Until sub-formula  is constrained (\cref{def:sub},\cref{def:boun}).  
\end{definition}

Otherwise, \ie if not all Until sub-formulae are constrained, we look only at the unconstrained ones. %
If each falls within an even number of negations, then $\tomtlinSym{(\phi)}$ does not contain any max sub-formula (except for the constrained max terms resulting from the compilation of constrained Until sub-formulae of type (1) preceded by an odd number of negations, which we rewrite as min terms, \cref{prop:minimal-logic}). %
It follows that, in this case, $\tomtlinSym{(\phi)}$ falls within $\slogic$ and therefore $\phi$ is satisfaction monitorable. %
We get accordingly the SMTL fragment.

\begin{definition}
  \label{def:smtl}
  The SMTL fragment of MTL, is defined as the set of 
  all formulae in which each unconstrained Until sub-formula falls under an even number of negations (\cref{def:sub},\cref{def:boun}).
\end{definition}

For violation monitorability, the reasoning is dual requiring every unconstrained Until sub-formula to fall within an odd number of negations. %

\begin{definition}
  \label{def:vmtl}
  The VMTL fragment of MTL is defined as the set of 
  all formulae in which each unconstrained Until sub-formula falls under an odd number of negations (\cref{def:sub},\cref{def:boun}).
\end{definition}

\begin{proposition}[Monitorability of MTL]
  \label{prop:monitorability-mtl}
  The following holds:
  \begin{enumerate}
  \item \label{pt:cmtl-com-mon} CMTL is complete monitorable.
    \item \label{pt:smtl-sat-mon} SMTL is satisfaction monitorable.
    \item \label{pt:vmtl-vio-mon} VMTL is violation monitorable.
  \end{enumerate}
\end{proposition}

\begin{proof}
  \begin{enumerate}
  \item CMTL is complete monitorable: Let $\phi \in CMTL$. 
      Since by the definition of 
      CMTL, every Until sub-formula of $\phi$ is constrained, we have 
      that in each fixpoint sub-formula of $\tomtlin{\phi}$, each recursion variable appears under the form $y\lin \dmd{y-x_s \prec n}Y$, $n\in\naturals$. %
      Since $x_s$ is free in each such sub-formula, we get that every fixpoint sub-formula of $\sem{\tomtlin{\phi}}$ is constrained and therefore $\sem{\tomtlin{\phi}}\in\clogic$ (\cref{def:clogic}).  %
      Given that every formula in $\clogic$ is complete monitorable (\cref{thm:clogic}), and 
      by \cref{prop:correctness-tohml} and 
      \cref{prop:correctness-compiler2}, we have $\sem{\phi} = \sem{\tomtlin{\phi}}$, 
      we conclude that $\phi$ is complete monitorable.

    \item SMTL is satisfaction monitorable: Let $\phi \in SMTL$.  Since by the definition of 
      SMTL, each unconstrained Until sub-formula of $\phi$ falls under an even number of negation, we have 
      that each fixpoint sub-formula of $\tomtlin{\phi}$ is a \textbf{min} or a constrained \textbf{max}.
      By \cref{prop:minimal-logic}, each $\maxfp{\psi}$ sub-formula in $\tomtlin{\phi}$ can be rewritten 
      to $\minfp{\psi}$ to get an equivalent formula \wrt divergent timed words. %
      Therefore, we get an equivalent formula of $\phi$ in $\slogic$, because every fixpoint operator is a \textsf{min} (\cref{fig:min-fragments}). %
      Given that every formula in $\slogic$ is satisfaction monitorable (\cref{thm:slogic-sm}), and by \cref{prop:correctness-tohml} and 
      \cref{prop:correctness-compiler2}, we have $\sem{\phi} = \sem{\tomtlin{\phi}}$,  
      we conclude that $\phi$ is satisfaction monitorable.
    \item VMTL is violation monitorable: Analogous.
  \end{enumerate}  
\end{proof}

Reasoning on monitorability can be easily extended to formulae using syntactic sugar. Any Next sub-formula is a constrained Until since $\nxt{I}{\phi}$ is syntactic sugar for $\until{\bot}{I}{\phi}$ (\cref{sec:mtl}). %

Non-strict Until sub-formulae follow the same conditions as the strict ones. 
On the one hand, $\wuntil{\phi}{I}{\psi}$ is syntactic sugar for $\neg(\neg\psi \wedge \neg(\phi\wedge (\until{\phi}{I}{\psi})))$ or $\phi\wedge (\until{\phi}{I}{\psi})$, depending on whether $I$ is left closed at $0$ (\cref{sec:mtl}). On the other hand, we have 
$$\mathit{sub_{\mathcal{U}}}(\phi\wedge (\until{\phi}{I}{\psi})) = \mathit{sub_{\mathcal{U}}}(\until{\phi}{I}{\psi})$$
And $\mathit{sub_{\mathcal{U}}}(\neg(\psi \wedge \neg(\phi\wedge (\until{\phi}{I}{\psi}))))$ contains the same Until sub-formulae us $\mathit{sub_{\mathcal{U}}}(\until{\phi}{I}{\psi})$ where the oddity/evenness of negations preceding each of them is preserved. %
In summary, the set of Until sub-formulae, as well as the oddity/evenness of the number of negations preceding each of them, is preserved  across the two versions of the Until operator. 

An unconstrained (strict or non-strict) Finally (\resp Globally) sub-formula is simply an unconstrained Until under the scope of an even (\resp odd) number of negations. 

\begin{example}\label{ex:notmon}

Consider the MTL formula $$\psi_{\mathit{nm}} = \uwglobally{(c\Rightarrow\weventually{[2,5]}{(b\wedge \uweventually{c})})}$$ 
Notice how $\psi_{\mathit{nm}}$ contains a constrained Until sub-formula, but also two unconstrained  sub-formulae that correspond, respectively, to a non-strict Until under an odd number of negations (non-strict Globally) and a non-strict Until under an even number of negations (the innermost non-strict Finally). %
Now, $\psi_{\mathit{nm}}$ belongs neither to SMTL (there is an unconstrained Until sub-formula under an odd number of negation) nor to VMTL (there is an unconstrained Until sub-formula under an even number of negations). %
Even though we do not claim maximality of our MTL monitorable fragments, we remark that  $\psi_{\mathit{nm}}$ is indeed not monitorable. %
For satisfactions, the reasoning is similar to bounded response (\cref{ex:br2}): take the infinite set $S = \setof{\pi\in\twords}{\forall i\in\naturals_{>0}: a_i(\pi) = b}$, \ie of all infinite timed words containing only $b$ actions. %
Then for every $\pi\in S$,  we have $\pi\in\sem{\psi_{\mathit{nm}}}$. 
Pick any $\pi'$ prefix of such $\pi$. %
One cannot issue a satisfaction verdict after observing $\pi'$ because there is at least one infinite extension of $\pi'$ violating $\psi_{\mathit{nm}}$, for example $\pi'.\pi''\in\twords$ where $\pi''$ is any infinite suffix containing only $c$ actions. %
For violations, take take the infinite set $S' = \setof{\pi\in\twords}{(c,1)(b,4)\in\pf{\pi} \text{ and } \forall i \in \mathbb{N}_{>2} : a_i(\pi) = b}$. %
Then for every $\pi\in S'$, we have $\pi\notin\sem{\psi_{\mathit{nm}}}$. Yet, there is no way to reach a violation verdict based on a finite prefix of any $\pi\in S'$, because any such prefix has an infinite extension satisfying $\psi_{\mathit{nm}}$. %

This is an interesting example showing how the existence of a constrained Until sub-formula in MTL does not guarantee monitorability neither for satisfaction nor for violations. %
It also gives an idea on the importance of verdict guarantees for at least all bad behaviours or all good behaviours \wrt to a formula according to our definition of monitorability.  

\end{example}

\begin{example} Consider the following MTL formula

$$\gamma = \weventually{[0,1]}{(\unxt{b} \wedge \globally{[1,2]}{c})}$$

The formula $\gamma$ expresses the following precise property: 

\begin{itemize}
\item There exists an action at some position $i\in \naturals$, occurring at some $t$ within the interval $[0,1]$ relative to the first action, and 
\begin{itemize}
\item The action at position $i+1$ is a $b$,
\item Every action within $[1,2]$ relative to $t$ is a $c$. 
\end{itemize}
\end{itemize}

Since all temporal operators are derived from the strict Until operator, we deduce that rewriting $\gamma$ in the standard syntax results in every Until sub-formula being constrained, either because it carries a bounded interval (the Until counterparts of the Finally and Globally sub-formulae) or because it is of the form $\until{\bot}{I}{\phi}$ (the Until counterpart of the Next sub-formula) and therefore $\gamma$ is complete monitorable. %
\end{example}

\paragraph{Independence of formulae depth.} 
In MTL, the depth of a formula $\phi$ is the maximum number of \emph{nestings} of temporal operators in a sub-formula of $\phi$. %
Notice how our monitorability fragments are completely independent from the depth of a formula: in an SMTL formula $\phi$, for instance, the depth of a sub-formula of $\phi$ can be any as long as every unconstrained Until is preceded by an even number of negations.  %

\section{Monitorability of TPTL}
\label{sec:montptl}

To isolate useful monitorable fragments of TPTL, we rely again on the results of \cref{sec:monitorability}, and use a similar reasoning as in the previous section. 
In the following, we say that a TPTL term $\uuntil{\phi}{\psi}$ is constrained iff $\phi = \bot$ or $\psi$ is of the 
for $(x \prec n) \wedge \psi'$ for some variable $x$, natural number $n$, and term $\psi'$. 
 We modify the compiler defined 
in \cref{fig:tptl-to-tlin} so that constrained Until sub-terms are mapped to the appropriate $\logic$ fragment. %
The proofs not enclosed in this section are given in~\cref{app:monmtl}.

The modified compiler from TPTL to $\logic$ is presented in \cref{fig:tptl-to-tlin2} where $\phi'\neq \bot$, and its correctness 
is established in \cref{prop:correctness-tptl-to-tlin2}. The definition of the term $\mathit{NXT}(X)$ used in 
\cref{fig:tptl-to-tlin2} depends on the form of $\psi$: if $\psi$ is of the form 
$(x \prec n) \wedge \psi'$, then $\mathit{NXT}(X) = \dmd{\act}{y \lin \dmd{y - x \prec n}{X}}$; otherwise, $\mathit{NXT}(X) = \dmd{\act}{X}$. %
Notice here also that for the constrained Until formula $\uuntil{\bot}{\phi}$, the compiler output reduces to $\dmd{\act}{\tomtlinn{\psi}}$ using similar arguments as in the previous section. %

\begin{figure}[!b]
  \hrulefill
  $$
    \begin{array}{lll}
    \tomtlinn{\top} &=& \true \\[0.1em]
    \tomtlinn{\bot} &=& \false \\[0.1em]
      \tomtlinn{a} &=& \dmd{a}\true \\[0.1em]
      \tomtlinn{\phi \vee \psi} &=& \tomtlinn{\phi} \vee \tomtlinn{\psi} \\[0.1em] 
      \tomtlinn{\neg \phi} &=& \tlinnot{\tomtlinn{\phi}} \\[0.1em]
      \tomtlinn{\uuntil{\phi'}{\psi}} &=& \dmd{\act}{\minfp{\tomtlinn{\psi} \vee (\tomtlinn{\phi'} \wedge \mathit{NXT}(X))}} \\[0.1em]
      \tomtlinn{\uuntil{\bot}{\psi}} &=& \dmd{\act}{\tomtlinn{\psi}} \\[0.1em]
      \tomtlinn{\freeze{x}{\phi}} &=& x \lin \tomtlinn{\phi} \\[0.1em]
      \tomtlinn{x \sim n} &=& y \lin \dmd{y - x \sim n}\true \\[0.1em]
    \end{array}
  $$
  \hrulefill
  \caption{From TPTL terms to $\terms$ where $\phi' \neq \bot$}
  \hrulefill
  \label{fig:tptl-to-tlin2}
\end{figure}

\begin{restatable}{proposition}{tptlttwo}
	\label{prop:correctness-tptl-to-tlin2}
	For all $\phi \in $ TPTL, we have $\sem{\totlinn{\phi}} = \sem{\tomtlinn{\phi}}$. 
\end{restatable}

The monitorable fragments then follow using the same reasoning as in the previous section, \ie if all Until sub-terms of $\phi\in $ TPTL are constrained, then $\phi$ is complete monitorable; otherwise if all unconstrained Until sub-terms fall under an even (\resp odd) number of negations then $\phi$ is satisfaction (\resp violation) monitorable. %
Obtaining general Until sub-terms to reason on Until sub-terms and the number of negations they are preceded by is done in similar fashion as for MTL, with $\mathit{sub}_U(x\sim n) = \emptyset$ and  $\mathit{sub}_U(x.\phi) = \mathit{sub}_U(\phi)$, and the technical detail that $\mathit{sub}_U$ may return (open) terms. %
Notice here also how our monitorable fragments of TPTL are completely independent of the depth of sub-formulae. 

\begin{definition}
  \label{def:cmtl}
  The fragment CTPTL is defined as the set of all TPTL formulae with all Until sub-terms constrained.
\end{definition}

\begin{definition}
  \label{def:stptl}
  The fragment STPTL is defined as the set of 
  all TPTL formulae in which each unconstrained Until sub-term falls under an even number of negations.
\end{definition}

\begin{definition}
  \label{def:vmtl}
  The fragment VTPTL is defined as the set of 
  all TPTL formulae in which each unconstrained Until sub-term falls under an odd number of negations.
\end{definition}

\begin{proposition}[Monitorability of TPTL]
  \label{prop:monitorability-tptl}
  The following holds:
  \begin{enumerate}
  \item \label{pt:ctptl-com-mon} CTPTL is complete monitorable.
    \item \label{pt:stptl-sat-mon} STPTL is satisfaction monitorable.
    \item \label{pt:vtptl-vio-mon} VTPTL is violation monitorable.
    
  \end{enumerate}
\end{proposition}

The proof of \cref{prop:monitorability-tptl} follows the same arguments used in the proof of \cref{prop:monitorability-mtl}.

\begin{example}
    Recall the formula 
    
    $$\varphileq = x.\ueventually{(b \wedge x\leq 1 \wedge (\uglobally{(x\leq 1) \Rightarrow \neg b})})$$ 

    from~\cref{ex:tptl}. Removing the syntactic sugar to fall back on the minimal syntax (\cref{fig:tptl}), we get: 
    
    $$x.\uuntil{\top}{(b \wedge x \leq 1 \wedge \neg (\uuntil{\top}{(x\leq 1\wedge b})))}$$

    We remark that every Until is constrained and therefore $\varphileq$ is complete monitorable. %
    Naturally, using the compiler in \cref{fig:tptl-to-tlin2}, we get a $\logic$ formula that is directly in the complete-monitorable fragment $\clogic$
    
$$
\begin{aligned}
\tomtlinn{\varphileq}&= x \lin \dmd{\act}  \minfp{(y \lin \dmd{y - x \leq 1}\true \wedge\dmd{b}\true \wedge \\
&\bigwedge_{a\in\act}\lbox{a}\maxfp[Y]{(z \lin \lbox{z - x \leq 1}\false\vee\lbox{b}\false)\bigwedge_{a\in\act}\lbox{a}y\lin\lbox{y-x\leq 1}Y}\vee \\
& \dmd{\act}z\lin\dmd{z-x\leq 1}X}
\end{aligned}
$$
\end{example}

As in MTL, we can extend the rules to include non-strict temporal operators, where non-strict Until sub-terms should obey the same requirements as for the strict ones. 

\begin{example} Consider the formula $\psi_{\mathit{alt},I}$ from~\cref{ex:altt}, that is  

$$\psi_{\mathit{alt},I} = x.(\psi_1 \wedge  \psi_2 \wedge \psi_3 \wedge \psi_4)$$ %
Where 

$$\psi_1 = x\in I \Rightarrow b$$

$$\psi_2 = \uwglobally ((x\preccI \rb{\lcomplement{I}}\wedge \unxt{x\in I}) \Rightarrow \unxt{b})$$

$$\psi_3 = \uwglobally ((x\in I \wedge b \wedge \unxt{x\in I}) \Rightarrow \unxt{c})$$

$$\psi_4 = \uwglobally ((x\in I \wedge c \wedge \unxt{x\in I}) \Rightarrow \unxt{b})$$

For any unbounded interval $I$, we can easily deduce that $\psi_{\mathit{alt},I}$ is violation monitorable: every Next sub-term is a constrained Until, the remaining sub-terms are (non-strict) unconstrained Globally, and therefore non-strict unconstrained Until under an odd number of negations. %

If $I$ is bounded, $\psi_{\mathit{alt},I}$ becomes also satisfaction monitorable and therefore complete monitorable. %
Let us show this going through its clauses. %
There is no Until sub-term in $\psi_1$. %
For $\psi_2$, we already know that the Next sub-terms are constrained Until, so we only rewrite the non-strict Globally as non-strict Until: 

$$\neg (\uwuntil{\top}{(x\preccI \rb{\lcomplement{I}}\wedge \unxt{x\in I} \wedge \neg \unxt{b})})$$

The above term is of the form $\neg(\uwuntil{\phi}{(x \prec n \wedge \psi')})$ with $\psi' = \unxt{x\in I} \wedge \neg \unxt{b}$ and is therefore (the negation of) a constrained Until. %

For $\psi_3$, since $I$ is bounded, we have $x\in I$ is of the form $x \prec n \wedge \psi'$ with $n = \rb{I}$ and $\psi' = \neg(x\preccI \lcomplement{I}) \wedge \unxt{x\in I} \wedge \neg \unxt{c}$ and therefore we get again the non-strict Globally sub-term corresponding to a non-strict constrained Until. %
The same reasoning carries on for $\psi_4$. %
Therefore, if $I$ is bounded, then every Until sub-term of $\psi_{\mathit{alt},I}$ is constrained. 
We conclude that $\psi_{\mathit{alt},I}$ is complete monitorable whenever $I$ is bounded.   
\end{example}

\section{Related Work}\label{sec:rw}

The state of the art on RV of timed properties mainly centres around the study of LTL extensions such as MTL~\cite{koymans1990specifying} and MITL, the fragment of MTL excluding singular intervals~\cite{AlurFH96}.  
The use of a declarative logic to describe system behaviour creates a clear delineation between what can be specified and how this can then be verified algorithmically. 
This is in line with our approach and is more conducive to the study of monitorability. 
Baldor \etal~\cite{BaldorN12} were the first to propose a transition based approach to monitoring of MITL formulae and establish complexity results for their proposed synthesis.
Ho \etal~\cite{ho2014online} split unbounded and bounded parts of a dense-time MTL formula to permit a simpler construction for the bounded parts to obtain space-bounded monitors.  
Neither of these works consider monitorability issues. %

The influential work by Bauer \etal~\cite{DBLP:journals/tosem/BauerLS11} introduces the idea of an $n$-valued semantics for a timed version of LTL called TLTL, defined over \emph{trace prefixes} by incorporating inconclusive outcome, typically denoted by $``?"$. 
This is used as a definition of the expected monitor behaviour when analysing a trace prefix online, which then guides a method of obtaining executable monitors from timed properties. %
The main drawback of the approach is that monitor synthesis disregards monitorability aspects; it generates monitors for \emph{any} timed property, without information on whether such property is monitorable or not.
The recent work on monitoring MITL by Grosen \etal~\cite{DBLP:conf/formats/GrosenKLZ22} also suffers from this drawback as a result of adopting a three-valued semantics.
Despite this, they propose new symbolic method of synthesising monitors for MITL properties via a translation into timed B\"{u}chi automata (TBA)~\cite{alur1991theory}. 
Pinisetty \etal~\cite{pinisetty2017predictive} focus on how prior knowledge can be used for sound predictive monitoring.

All of the above works take into account in one sense or another infinite executions; the infinitary nature of traces poses the main complication when considering the monitorability of linear-time properties~\cite{DBLP:journals/pacmpl/AcetoAFIL19}. 
Popular work on synthesising monitors for timed LTL properties with a finitary semantics, such as that of Thati and Rosu~\cite{thati2005monitoring} and of Basin \etal \cite{BasinKMZ15} sidestep any monitorability issues since every property becomes monitorable. 
The work by Basin \etal considers a first-order version of MTL called MFOTL  to describe properties involving data ranging over an infinite domain. 
Ulus \etal~\cite{UlusFAM16} describe a monitor algorithm for Timed Regular Expressions based on the union of two-dimensional zones for finite words. 
Raszyk \etal~\cite{raszyk2019multi} extend this work to an augmented monitoring setup that uses multiple analysis points (\ie multi-head monitors) that relates to our parallel monitoring setup.
The multi-head setup is used to lower the overheads of runtime executions, but bears no effect on the analytic power of the monitors or monitorability.
Lima \etal~\cite{lima2023explainable} augment monitoring verdicts of MFOTL with explanations of why the monitored execution led to violations/satisfactions.
This relies on the definition of a local proof-system~\cite{FrancalanzaC21}, and can be developed for \logic as it has been for the untimed version of the linear-time $\mu$-calculus in~\cite{DBLP:conf/fsttcs/DaxHL06} but is left for future work. 
Roussanaly \etal~\cite{RoussanalyF22} study a decentralised monitor synthesis procedure for timed regular expressions that is similar to our parallel monitors. 
Note that a number of works study \emph{quantitative monitoring} of STL~\cite{MalerN04}, a version of MITL interpreted over continuous signals, proposing thereby a robustness measure instead of yes/no verdicts. 
Such works sidestep monitorability problems either through \emph{approximate} measures under infinitary semantics (see \eeg~\cite{donze2010robust}) or more precise measures under finitary semantics~\cite{rino2024efficiently}. 

In summary, our approach is complementary to the above efforts. %
On the one hand, our work is the first that proposes complete monitorability results for prominent timed logics within a unified framework. %
On the other hand, works such as~\cite{ho2014online} and~\cite{DBLP:conf/formats/GrosenKLZ22} focus on efficiency of monitoring algorithms rather than monitorability issues. %

Studies on monitorability generally consider one of two definitions; the one due to Pnueli and Zaks~\cite{pnueli2006psl} or that due to Schneider~\cite{Schneider00}.
Aceto \etal~\cite{aceto2021operational} have shown that the latter definition is strictly stronger and have systematised the various variants in terms of soundness and completeness criteria similar to those we used. 
In particular, the definition of monitorability assumed in our work is the stronger variant due to Schneider. %
Very recently, and subsequently to our~\cite{DBLP:conf/ecoop/Amara0FF25}, Grosen et al.~\cite{DBLP:conf/concur/GrosenKL025} studied monitorability in the timed setting using a straightforward extension of Pnueli-Zaks. %
They defined the following verdict function over finite timed words, parameterised with a property $\phi$

$$
  \mathcal{V}_\phi(\pi) = 
   \begin{cases}
    \top \text{ if } \forall \pi'\in\ext{\pi}: \pi\in\sem{\phi} \\
    \bot \text{ if } \forall \pi'\in\ext{\pi}: \pi\notin\sem{\phi} \\
    ? \text{ otherwise }
   \end{cases}
$$ 

and deemed $\phi$ (strongly) monitorable iff for every $\pi\in\twordstar$, there exists $\pi'$, a finite extension of $\pi$, such that $\mathcal{V}_\phi(\pi')\in\{\top,\bot\}$~\cite[Definition 3, point 4]{DBLP:conf/concur/GrosenKL025}. %
Our definition of monitorability differs from Grosen et al.'s \wrt to at least two aspects. %
First, we guarantee, for any monitorable formula $\phi$, the detection of \emph{all} bad behaviours or \emph{all} good behaviours, whereas the definition above  does not. %
Take for instance the MTL formula $\psi_{\mathit{nm}}$ in~\cref{ex:notmon}. %
We have showed within the same example that $\psi_{\mathit{nm}}$ is not monitorable, and actually there is an infinite number of executions that violate $\psi_{\mathit{nm}}$ \emph{and} and an infinite number of executions that satisfy $\psi_{\mathit{nm}}$ for which no verdict can ever be reached. %
In contrast $\psi_{\mathit{nm}}$ is monitorable according to Grosen et al: for any finite timed word $\pi$, take the finite extension $\pi' = \pi.(c,t_i(|\pi|)(b,t_i(|\pi|)+10)$ and get $\mathcal{V}_{\psi_{\mathit{nm}}}(\pi') = \bot$. %
This is because Pnueli-Zaks monitorability, of which Grosen et al.'s above is the timed counterpart, ensures the \emph{possibility} to reach verdicts without guarantees to actually reach them. %
Second, as proven by Grosen et al.~\cite[Theorem 1]{DBLP:conf/concur/GrosenKL025}, the verdict function above is uncomputable for expressive formalisms such as non-deterministic TBA (NTBA), MTL, TPTL or our $\logic$. %
This is because, on the one hand, the verdict $\top$ must be reached given the empty timed word iff $\sem{\phi} = \twords$, and on the other hand, the universality problem in NTBA is undecidable~\cite{alur1991theory}, and so is the validity problem in MTL, TPTL and $\logic$ (\cref{thm:mtl}, \cref{thm:tptl}, \cref{thm:tlin}). %
This prevented reasoning on monitorability of NTBA altogether\footnote{The authors still obtained the monitorability of properties given by the less expressive deterministic timed Muller automata, for which universality is decidable.} in~\cite{DBLP:conf/concur/GrosenKL025}, and would have made such reasoning impossible in MTL, TPTL or $\logic$  . %
The main reason behind this is the fact that $\mathcal{V}_\phi$ requires verdicts to be both sound and \emph{tight}, \ie it requires implicitly verdicts to be issued on the \emph{shortest prefix} with all extensions in or outside the semantics of $\phi$. %
In contrast, our sound monitors, that can be seen as sound approximations of the verdict function, are a trade-off to reason on the inevitability of reaching verdicts without tightness requirements.

\section{Conclusion}
\label{sec:conclusion}

In this paper, we solved the monitorability problem for an expressive timed logic. %
We proposed \logic, a modal timed $\mu$-calculus with linear-time semantics and proved that it is strictly more expressive than the well-known and popular MTL and TPTL. %
Then, we precisely identified the maximal fragments, \wrt a rich monitorable fragment that we proposed, of violation-monitorable, satisfaction-monitorable, complete-monitorable, and monitorable formulae of \logic. %
Moreover, relying on automatic translations from MTL and TPTL to \logic and our characterisation of monitorable fragments of \logic, we isolated important monitorable fragments in MTL and TPTL that do not depend on the depth of formulae. %

A natural direction of future work is to study the relationship between TBA and \logic, as we currently do not handle properties specified using the former. %
Another open question is the class of timed languages characterised by \logic, in comparison with that of languages recognised by alternating TBA. %

\bibliographystyle{plain}
\bibliography{biblio-monitors}

\appendix

\clearpage
\section{Closure under complement of \logic}\label{app:clos}

We first state a handy lemma (\cref{lem:fixpoint-duality}) on the 
the duality 
  between extremal fixed points of a function and its dual 
   (see, e.g., \cite{Arnold2001RudimentsO} for further details). 
  Let $f : \parts{\twords} \to \parts{\twords}$ be a function. Its dual, denoted $\tilde{f}$, 
  is defined as $\tilde{f}(S) = \compl{f(\compl{S})}$ for all $S \subseteq \twords$.
  \begin{lemma}
    \label{lem:fixpoint-duality}
    Let $f : \parts{\twords} \to \parts{\twords}$ be monotone, then $\tilde{f}$ is 
    monotone and 
    \begin{enumerate}
      \item \label{pt:min-fixpoint-duality} $\mu x.\tilde{f}(x) = \compl{\nu x.f(x)}$, and 
      \item \label{pt:max-fixpoint-duality} $\nu x.\tilde{f}(x) = \compl{\mu x.f(x)}$.
    \end{enumerate}
  \end{lemma}

We are now ready to prove the main lemma

\closure*

  \begin{proof}
  
  The proof proceeds by structural induction on $\phi$, with each case concluded by unfolding
  the definition of the semantics (\cref{fig:lhml-semantics}). For the case of fixed-point
  formulae, the proof relies on \cref{eq:consequence-tarski-fixed-point} and~\cref{lem:fixpoint-duality}. %
  
       The base cases are 
    \begin{itemize}
      \item $\phi = \true$: $\sem{\tlinnot{\true},\bar{\rho}}_v = \emptyset = \compl{\twords}
            = \compl{\sem{\true,\rho}_v}$.
      \item $\phi = \false$: $\sem{\tlinnot{\false},\bar{\rho}}_v = \twords = \compl{\emptyset}
            = \compl{\sem{\false,\rho}_v}$.
      \item $\phi = X$: $\sem{\tlinnot{X},\bar{\rho}}_v = 
                                  \sem{X,\bar{\rho}}_v =
                                  \bar{\rho}(X) = 
                                  \compl{\rho(X)} = 
                                  \compl{\sem{X,\rho}_v}$.
    \end{itemize}
    The inductive cases are:
    \begin{itemize}
      \item $\phi = \phi_1 \wedge \phi_2$:
        $$
          \begin{array}{lll}
            \sem{\tlinnot{\phi_1 \wedge \phi_2},\bar{\rho}}_v 
            &=& \sem{\tlinnot{\phi_1} \vee \tlinnot{\phi_2},\bar{\rho}}_v\\ 
            &=& \sem{\tlinnot{\phi_1},\bar{\rho}}_v \cup \sem{\tlinnot{\phi_2},\bar{\rho}}_v \\
            &=& \compl{\sem{\phi_1,\rho}_v} \cup \compl{\sem{\phi_2,\rho}_v}\ \ (I.H.) \\
            &=& \compl{\sem{\phi_1,\rho}_v \cap \sem{\phi_2,\rho}_v} \\
            &=& \compl{\sem{\phi_1 \wedge \phi_2,\rho}_v} \\
          \end{array}
        $$
      \item $\phi = \phi_1 \vee \phi_2$:
        $$
          \begin{array}{lll}
            \sem{\tlinnot{\phi_1 \vee \phi_2},\bar{\rho}}_v 
            &=& \sem{\tlinnot{\phi_1} \wedge \tlinnot{\phi_2},\bar{\rho}}_v\\ 
            &=& \sem{\tlinnot{\phi_1},\bar{\rho}}_v \cap \sem{\tlinnot{\phi_2},\bar{\rho}}_v \\
            &=& \compl{\sem{\phi_1,\rho}_v} \cap \compl{\sem{\phi_2,\rho}_v}\ \ (I.H.) \\
            &=& \compl{\sem{\phi_1,\rho}_v \cup \sem{\phi_2,\rho}_v} \\
            &=& \compl{\sem{\phi_1 \vee \phi_2,\rho}_v} \\
          \end{array}
        $$
      \item $\phi = x \lin \phi_1$: 
        $$
          \begin{array}{lll}
            \sem{\tlinnot{x \lin \phi_1},\bar{\rho}}_v 
            &=& \sem{x \lin \tlinnot{\phi_1},\bar{\rho}}_v \\
            &=& \setof{\pi}{\pi \in \sem{\tlinnot{\phi_1},\bar{\rho}}_{v\update{x}{t_1(\pi)}}}\\
            &=& \setof{\pi}{\pi \in \compl{\sem{\phi_1,\rho}_{v\update{x}{t_1(\pi)}}}}\ \ (I.H.)\\
            &=& \compl{\setof{\pi}{\pi \in \sem{\phi_1,\rho}_{v\update{x}{t_1(\pi)}}}}\\
            &=& \compl{\sem{x \lin \phi_1,\rho}_v}
          \end{array} 
        $$
      \item $\phi = \dmd{a}{\phi}$: 
        $$
          \begin{array}{lll}
            \sem{\tlinnot{\dmd{a}{\phi_1}},\bar{\rho}}_v 
            &=& \sem{\lbox{a}{\tlinnot{\phi_1}},\bar{\rho}}_v \\
            &=& \sem{\dmd{a}\tlinnot{\phi_1},\bar{\rho}}_v \cup \setof{\pi}{a_1(\pi) \neq a} \\
            &=& \semdiamond{a}{\sem{\tlinnot{\phi_1},\bar{\rho}}_v} \cup \setof{\pi}{a_1(\pi) \neq a} \\
            &=& \semdiamond{a}{\compl{\sem{\phi_1,\rho}_v}} \cup \setof{\pi}{a_1(\pi) \neq a}\ \ (I.H.) \\
            &=& \setof{\pi}{a_1(\pi) \neq a \text{ or } \pi^2 \in \compl{\sem{\phi_1,\rho}_v}} \\
            &=& \compl{\sem{\dmd{a}{\phi_1},\rho}_v} \\
          \end{array} 
        $$
      \item $\phi = \dmd{g}{\phi}$:
        $$
          \begin{array}{lll}
            \sem{\tlinnot{\dmd{g}{\phi_1}},\bar{\rho}}_v 
            &=& \sem{\lbox{g}{\tlinnot{\phi_1}},\bar{\rho}}_v \\
            &=& \sem{\dmd{g}\tlinnot{\phi_1},\bar{\rho}}_v \cup \setof{\pi}{v \not\models g} \\
            &=& \semdiamond{g}{v,\sem{\tlinnot{\phi_1},\bar{\rho}}_v} \cup \setof{\pi}{v \not\models g} \\
            &=& \semdiamond{g}{v,\compl{\sem{\phi_1,\rho}_v}} \cup \setof{\pi}{v \not\models g}\ \ (I.H.) \\
            &=& \setof{\pi}{v \not\models g \text{ or } \pi \in \compl{\sem{\phi_1,\rho}_v}} \\
            &=& \compl{\sem{\dmd{g}{\phi_1},\rho}_v} \\
          \end{array} 
        $$
      \item $\phi = \maxfp{\phi_1}$: 
        $$
          \begin{array}{lll}
            \sem{\tlinnot{\maxfp{\phi_1}},\bar{\rho}}_v 
            &=& \sem{\minfp{\tlinnot{\phi_1}},\bar{\rho}}_v\\ 
            &=& \mu S. \sem{\tlinnot{\phi_1},\bar{\rho}\update{X}{S}}_v \\
            &=& \mu S. \sem{\tlinnot{\phi_1},\compl{\rho\update{X}{\compl{S}}}} \\
            &=& \mu S. \compl{\sem{\phi_1,\rho\update{X}{\compl{S}}_v}}\ \ (I.H.) \\
            &=& \compl{\nu S. \sem{\phi_1,\rho\update{X}{S}}_v}\ \ (\cref{lem:fixpoint-duality}(\ref{pt:min-fixpoint-duality})) \\
            &=& \compl{\sem{\maxfp{\phi_1},\rho}_v}
          \end{array}
        $$
      \item $\phi = \minfp{\phi_1}$:
        $$
          \begin{array}{lll}
            \sem{\tlinnot{\minfp{\phi_1}},\bar{\rho}}_v 
            &=& \sem{\maxfp{\tlinnot{\phi_1}},\bar{\rho}}_v\\ 
            &=& \nu S. \sem{\tlinnot{\phi_1},\bar{\rho}\update{X}{S}}_v \\
            &=& \nu S. \sem{\tlinnot{\phi_1},\compl{\rho\update{X}{\compl{S}}}} \\
            &=& \nu S. \compl{\sem{\phi_1,\rho\update{X}{\compl{S}}_v}}\ \ (I.H.) \\
            &=& \compl{\mu S. \sem{\phi_1,\rho\update{X}{S}}_v}\ \ (\cref{lem:fixpoint-duality}(\ref{pt:max-fixpoint-duality})) \\
            &=& \compl{\sem{\minfp{\phi_1},\rho}_v}
          \end{array}
        $$
    \end{itemize}
\end{proof}

\clearpage
\section{Proofs about $\tohmlSym$ and $\totlinSym$}
\label{sec:appendix-tohml}

\subsection{Correctness of $\tohmlSym$}
Our aim now is to prove the following
\totlin*

To avoid ambiguity, we use the symbol~$\semmtl{-}$ when reasoning on the semantics of MTL formulae (i.e. \cref{fig:mitl-semantics}).

\cref{prop:correctness-tohml} follows from \cref{lem:soundness-completeness-tohml},
where we prove the soundness and completeness of $\tohml{-}$ simultaneously.
This lemma is stated at the end of this subsection.
Rather unsurprisingly, the proof of \cref{prop:correctness-tohml} relies on continuity
and induction to reason about {\em Until} formulae.

We now review minor technicalities used in this appendix.
Recall that~$\vto$ denotes the valuation that maps any $x \in \mathcal{D}$ to $0$. Whenever
convenient we use the following fact,
\begin{equation}
\label{eq:sem-tlin-formulae}
\forall \phi \in \logic, \rho\in\setrho, v\in\mathcal{V} \suchthat \sem{\phi,\rho}_v = \sem{\phi}
\end{equation}
\noindent
Where $\sem{\phi}$ is a notation for $\sem{\phi,\emptyset}_\vto$.
This equality holds because the semantic of formulae in~\logic
(i.e. closed and guarded terms in~$\terms$) does not depend on
valuations and environments.

{\bfseries Notation:}
We use~$\alpha,\beta,\ldots$ to range over~$\act$, while $a,b,c$ are
elements of $\act$. 
Recall that~$a_i(\pi)$ (resp.~$t_i(\pi)$) is
the action (\resp the timestamp) of the timed event at position~$i$ in~$\pi$. %
Given a timed word~$\pi$ and an $i\in\naturals_{>0}$ we let
\begin{equation}
  \label{eq:properties-projection-pi}
\pi^i = 
\begin{cases}
  \pi & \mathit{if}\ i = 1,\\
  (a_i(\pi), t_i(\pi)) . \pi^{i+1} & \mathit{if}\ i > 1
\end{cases}
\end{equation}
where the operation $\cdot$ denotes concatenation.
A more formal definition on paper can be given by coinduction.
An implementation of $\pi^i$ is the function {\tt Str\_nth\_tl}
of the standard Coq library for
streams.\footnote{\url{https://coq.inria.fr/doc/v8.10/stdlib/Coq.Lists.Streams.html}
The implementation of {\tt Str\_nth\_tl} is defined {\em inductively}, and
relies on an auxiliary function {\tt tl} to compute the tail of a
stream.}

Whenever we apply the functions $\funof{\termF}{v}{Y}{\rho}$,
$\semdiamond{ \alpha }{-}$ and $\sembox{ \alpha }{-}$ to a singleton set,
for instance $\set{ \pi }$,
we shorten the notation to $\funof{\termF}{v}{Y}{\rho}( \pi )$, 
$\semdiamond{ \alpha }{ \pi }$, $\sembox{ \alpha }{ \pi }$.

\paragraph{An auxiliary function and its properties}
\label{sec:appendix-auxfun}
Consider the following term:
$$
\bigdisjunction
$$
It is syntactically defined thanks to the assumption that the set~$\act$ is finite
(see \cref{sec:preliminaries}), and it plays a crucial role in the
definition of $\tohmlSym$. In fact, we introduce a function from 
variables to the above term,
$$
\auxfun = \bigdisjunction
$$
The term~$\auxfun$ plays a crucial in the main proofs of this section,
so in this section we gather its key properties, that are given
in the next three lemmas.

\begin{lemma}
  \label{lem:property-auxfun}
  For every term~$\termF \in \terms$, %
  every~$v$,
  every~$Y$,
  every~$\rho$  
  such that $\fv{\termF} \subseteq \dom{\rho} \cup \set{Y}$, %
  and every $S \in \parts{\twords}$,
  it holds that 
  $$
  \funof{\auxfun}{v}{Y}{\rho}( S )
  =
  \bigcup_{\alpha \in \act}
  \setof{ \pi }{a_1( \pi ) = \alpha \text{ and }
    \pi^2 \in S  }.
  $$
\end{lemma}
\begin{proof}
  By unfolding the definitions:
        $$
        \begin{array}{lll}
          \funof{\auxfun}{v}{Y}{\rho}( S ) & = &
          \bigcup_{ \alpha \in \act}( \funof{\dmd{ \alpha
            }{Y}}{v}{Y}{\rho}( S ) )\\
          & = &
          \bigcup_{\alpha\in\act}(\semdiamond{\alpha}{\funof{Y}{v}{Y}{\rho}( S )}) \\
          & = &
          \bigcup_{\alpha\in\act}(\semdiamond{\alpha}{ S }) \\
          & = &
          \bigcup_{\alpha\in\act}\setof{ \pi }{a_1(\hat{\pi}) = \alpha \text{
              and } \pi^2 \in \set{ S } }
        \end{array}
        $$
\end{proof}

We now explain the next two lemmas. Our aim is to establish a relation between
the semantics of $MTL$-formulae and of $\logic$ ones.

On one hand the semantics of $MTL$-formulae explicitly uses {\em
  positions} in timed words.
For instance, a timed word $\pi$ is in the semantics of a formula
$\until{\phi_1}{I}{\phi_2}$ if there exists
a position (i.e. a natural)~$i$ such that the suffix $\pi^i$ is in the
semantics of $\phi_2$ and for all position $j$ strictly smaller than
$i$, the suffix $\pi^j$ is in the semantics of $\phi_1$ and the time elapsed
i.e. $t_i(\pi) - t_1(\pi)$ is in~$I$. 
On the other hand, the semantics of \logic-formulae uses fixed points
of functions instead of positions in timed words. For instance,~$\pi$
is in the semantics of $\minfp[X]{\phi}$ if it belongs to all prefixed
points of the function $\lambda x.\sem{\phi,[X \mapsto x]}_v$ (See
\cref{ex:bUIC-toTlin}).

Thanks to the continuity of the semantics of \logic, and in particular
to \cref{eq:inductive-proof-method-min}, the approach based on fixed
points captures the properties expressed via positions.
The first steps to lay this bare are carried out in \cref{lemma:aux-until-soundness}
and \cref{lemma:aux1-compiler-mitl-comp}.
These lemmas are helpful to prove the soundness and completeness of
$\tohmlSym$ respectively, and some explanations are in order.






The \cref{lemma:aux-until-soundness} states that if 
a timed word~$\pi$ is in the semantics of the term $\minfp[Y]{\phi \vee (\psi
  \wedge \auxfun)}$ then there exists a position  
$i$ in $\pi$ such that $\pi^i$ is in the semantics of $\phi$ and for
every position $j$ strictly smaller that $i$,  
$\pi^j$ is in the semantics of $\psi$.
This is the intuition that we have of the semantics of this formula and one 
should notice the similarity with the semantics of
$\until{\phi_1}{I}{\phi_2}$.

First, an auxiliary fact.
\begin{lemma}
  \label{lem:property-suffix-construction}
    For all $\pi \in \twords$ and $k \in \naturals$ we have that
    $(\pi^2)^k = (\pi^k)^2 = (\pi)^{k+1}$.
\end{lemma}
\begin{proof}
By induction on $k$.
\end{proof}

\begin{lemma}
  \label{lemma:aux-until-soundness}
  For all terms $\termF,\termG \in \terms$ with $\fv{\termF} = \fv{\termG} = \emptyset$,
  every $v \in \mathcal{V}$, $x \in \mathcal{D}$, $Y \in \setrho$, $\pi \in \twords$ and $n \geq 0$, if
  \begin{equation}
    \label{eq:aux-until-soundness-hyp}
    \pi \in \funofN{x \lin (\termF \vee (\termG \wedge \auxfun))}{v}{Y}{\emptyset}{n}(\emptyset)
  \end{equation}
  then there exists some $i \geq 1$ such that 
  \begin{enumerate}
    \item\label{pt:aux-until-soundness-1}%
      $\pi^i \in  \sem{\termF}_{v\update{x}{t_i(\pi)}}$, and
    \item\label{pt:aux-until-soundness-2}%
      For all $1 \leq j < i: \pi^j \in \sem{\termG}_{v\update{x}{t_j(\pi)}}$.
  \end{enumerate}
\end{lemma}
\begin{proof}
  We prove (\ref{pt:aux-until-soundness-1}) and (\ref{pt:aux-until-soundness-2}) at the same time reasoning by induction on $n$:
  \begin{itemize}
    \item $n=0$: This case is impossible since 
      $$
        \funofN{x \lin (\termF \vee (\termG \wedge \auxfun))}{v}{Y}{\emptyset}{n}(\emptyset) = \emptyset
      $$
      which contradicts the hypothesis in \cref{eq:aux-until-soundness-hyp}.
    \item $n = m+1$:
      By \cref{eq:aux-until-soundness-hyp}, we have  
      \begin{equation}
        \label{eq:aux-until-soundness-hyp-ind}
        \pi \in \funof{x \lin (\termF \vee (\termG \wedge \auxfun))}{v}{Y}{\emptyset}(S)
      \end{equation}
      where $S = \funofN{x \lin (\termF \vee (\termG \wedge \auxfun))}{v}{Y}{\emptyset}{m}(\emptyset)$.
      To conclude the proof, we have to define $i \geq 1$ such that (\ref{pt:aux-until-soundness-1})
      and (\ref{pt:aux-until-soundness-2}) hold. From \cref{eq:aux-until-soundness-hyp} and 
      \cref{fig:lhml-semantics}, at least one the following holds 
      \begin{enumerate}[(a)]
        \item \label{pt:aux-until-soundness-hyp-or1} $\pi \in \sem{\phi,\semsubst{Y}{S}}_{v\update{x}{t_1(\pi)}}$, or 
        \item \label{pt:aux-until-soundness-hyp-or2}  $\pi \in \sem{\psi \wedge \auxfun,\semsubst{Y}{S}}_{v\update{x}{t_1(\pi)}}$.
      \end{enumerate}
      If (\ref{pt:aux-until-soundness-hyp-or1}) holds, then, for $i=1$,
      we conclude (\ref{pt:aux-until-soundness-1}) because $\fv{\phi}=\emptyset$, meaning
      that the environment is irrelevant; and (\ref{pt:aux-until-soundness-2}) holds by vacuity.
      The second case, where (\ref{pt:aux-until-soundness-hyp-or2}) is true, requires more work.
      The semantics of conjunction and \cref{lem:property-auxfun} ensure that
      \begin{equation}
        \label{eq:aux-until-soundness-hyp-or2-imp}
        \pi \in \sem{\termG,\semsubst{Y}{S}}_{v\update{x}{t_1(\pi)}} \text{ and } \exists \alpha\in\act\suchthat a_1(\pi) = \alpha \text{ and } \pi^{2} \in S
      \end{equation}
      Since $m < n$ the inductive hypothesis guarantees the existence of $k \geq 1$ such that 
      $$
        (\pi^{2})^{k} \in \sem{\termF}_{v\update{x}{t_k(\pi^2)}}
          \text{ and for all } 1 \leq j < k: (\pi^2)^{j} \in \sem{\termG}_{v\update{x}{t_j(\pi^2)}}
      $$
      which thanks to \cref{lem:property-suffix-construction} becomes
      $$
        (\pi)^{k+1} \in \sem{\termF}_{v\update{x}{t_{k+1}(\pi)}}
          \text{ and for all } 1 \leq j < k: (\pi)^{j+1} \in \sem{\termG}_{v\update{x}{t_{j+1}(\pi)}}
      $$
      By manipulating the naturals in the statement on the right, the
      above statement is equivalent to 
      $$
        (\pi)^{k+1} \in \sem{\termF}_{v\update{x}{t_{k+1}(\pi)}}
          \text{ and for all } 2 \leq j < k+1: (\pi)^{j} \in \sem{\termG}_{v\update{x}{t_{j}(\pi)}}
      $$
      Therefore, we conclude (\ref{pt:aux-until-soundness-1}) by letting
      $i = k+1$. Moreover, (\ref{pt:aux-until-soundness-2}) follows from the above
      statement and from
      $\pi \in \sem{\psi}_{v\update{x}{t_1(\pi)}}$, as given in
      \cref{eq:aux-until-soundness-hyp-or2-imp}, since $\fv{\termG}=\emptyset$.
  \end{itemize}
\end{proof}


We use the next lemma in the proof of the completeness of $\tohmlSym$.
\begin{lemma}
  \label{lemma:aux1-compiler-mitl-comp}
  For every term $\termF \in \terms$, %
  every $v$,
  every $Y$,
  every $\rho$
  such that $\fv{\termF} \subseteq \dom{\rho} \cup \set{Y}$, %
  for  every $i \geq 1$ and $\pi$,
  if 
  $$
  (\forall j. 1\leq j \leq i \suchthat \pi^j \in \funof{\termF}{v}{Y}{\rho}( \pi^{j+1} ))
  $$
  then
  $$
  \pi \in \funofN{\termF \wedge \auxfun }{v}{Y}{\rho}{i}(\pi^{i+1} ).
  $$
\end{lemma}
\begin{proof}
  By induction on i:
  \begin{itemize}
    \item $i = 1$: 
      Suppose
      $
      \pi^1 \in \funof{\termF}{v}{Y}{\rho}( \pi^{2} ).
      $
      Thanks to \cref{eq:properties-projection-pi} we know $\pi^1 = \pi$,
      and hence we have to prove
      $
      \pi \in \funof{\termF \wedge \auxfun }{v}{Y}{\rho}( \pi^2 ),
      $
      that is
      $$
      \pi \in \funof{\termF}{v}{Y}{\rho}( \pi^2 ) \cap \funof{\auxfun}{v}{Y}{\rho}( \pi^2 ).
      $$
      Since we already know $\pi \in \funof{\termF}{v}{Y}{\rho}(\pi^2)$ by assumption,
      it remains to prove that $\pi \in \funof{\auxfun}{v}{Y}{\rho}(\pi^2 )$.
      \cref{lem:property-auxfun} implies
      that
      $$
      \funof{\auxfun}{v}{Y}{\rho}( \pi^2 )
      =
      \bigcup_{\alpha\in\act}\setof{\pi}{a_1(\pi)=\alpha \text{ and } \pi^2 = \pi^2 }
      $$
      thus we have to prove that
      $$
      \pi 
      \in
      \bigcup_{\alpha\in\act}\setof{\pi}{a_1(\pi)=\alpha \text{ and } \pi^2 = \pi^2 }
      $$
      and this is trivially true, because 
      $a_1(\pi) \in \act$ and $\pi^2 = \pi^2$.
      
    \item $i = k+1$:
      The hypothesis states that 
        \begin{equation}
        \label{hyp-1}
        \forall j. 1\leq j \leq i\suchthat \pi^j \in \funof{\termF}{v}{Y}{\rho}( \pi^{j+1} )
        \end{equation}
        and the goal is to prove that
        $$
        \pi \in \funofN{\termF \wedge \auxfun}{v}{Y}{\rho}{i}(\pi^{i+1}).
        $$

        Let $f = \funof{\termF \wedge \auxfun}{v}{Y}{\rho}$.
        Given that $i = k + 1$, \cref{eq:function-iteration}
        implies
        $$
        \begin{array}{lll}
          f^i(\pi^{i+1})
        & = & \left( %
          f^k 
          \circ
          f
          \right)(\pi^{i+1})
        = 
        f^k(f(\pi^{i+1}))
        \end{array}
        $$
        and thus our goal becomes the proof of
        \begin{equation}
          \label{eq:goal}
          \pi \in
          f^k(f(\pi^{i+1}))
        \end{equation}
        To prove (\ref{eq:goal}) it suffices to construct a set~$B$ such that
        \begin{enumerate}
        \item\label{pt:auxiliary-completeness-goal-1} $\pi \in f^k(B)$,
        \item\label{pt:auxiliary-completeness-goal-2} $B \subseteq f(\pi^{i+1})$.
        \end{enumerate}
        The rest of the argument explains how to do this.

        The inductive hypothesis ensures the following,
        \begin{equation}
          \label{IH}
          (\forall m. 1 \leq m \leq k\suchthat \pi^{m} \in \funof{\termF}{v}{Y}{\rho}( \pi^{m+1} ))
          \text{ implies }
          \pi \in f^{k}( \pi^{k+1} )
        \end{equation}
        \cref{hyp-1} ensures that the premises of the implication
        \cref{IH} are true, hence we know that the conclusion holds:
        $
        \pi \in f^k( \pi^{k+1} ).
        $
        Since $i = k + 1$ we obtain
        \begin{equation}
          \label{hyp-2}
          \pi \in f^k( \pi^{i} )
        \end{equation}
        Now let $B = \set{ \pi^{i} }$.
        By construction (\ref{pt:auxiliary-completeness-goal-1}) is true,
        and it remains to explain why
        (\ref{pt:auxiliary-completeness-goal-2}) is true.
        
        As by definition $f = \funof{\termF \wedge
          \auxfun}{v}{Y}{\rho}$  and~$B$ contains only $\pi^i$,
        we have to prove the following
        two facts:

        \begin{equation}
          \label{final-goal-1}
          \pi^i \in \funof{\termF}{v}{Y}{\rho}(\pi^{i+1})
        \end{equation}
        and 
        \begin{equation}
          \label{final-goal-2}
          \pi^i \in \funof{\auxfun}{v}{Y}{\rho}(\pi^{i+1})
        \end{equation}
        hold.

        \cref{final-goal-1} follows from the hypothesis,
        i.e. \cref{hyp-1}, letting $j = i$.

        To prove \cref{final-goal-2}, thanks to 
        \cref{lem:property-auxfun}, it suffices to explain why
        $(\pi^i)^2 \in \set{\pi^{i+1}}$.
        This is a consequence of
        \cref{lem:property-suffix-construction},
        and so we conclude the argument.
  \end{itemize}
\end{proof}


\paragraph{The compiler $\tohmlSym$ preserves the semantics of formulae}

\newcommand{\MTL}{\mathit{MTL}}

Recall the semantics of the logics~$\MTL$ in \cref{fig:mitl-semantics},
$\logic$ in \cref{fig:lhml-semantics}, and the definition of the
function~$\tohmlSym$ in \cref{fig:mitl-compiler}.

As preliminary fact, observe that the function $\tohmlSym$
generates formulae (i.e. closed and guarded terms).

\begin{lemma}
  \label{lem:tohml-produces-formulae}
  For every $\phi \in MTL$ we have that $\tohml{ \phi } \in \logic$.
\end{lemma}
\begin{proof}
  Pick $\phi \in MTL$.
  Since $\tohml{\phi} \in \terms$ we have to show that $\tohml{\phi}$ is closed and guarded.
  We proceed by structural induction on $\phi$.
  \begin{itemize} 
    \item For the cases $\phi = \top$ and $\phi = \alpha$ we know 
      $\tohml{\phi} = \true$ and $\tohml{\phi} = \dmd{\alpha}\true$ respectively.
      These terms are both trivially closed and guarded for they contain no variables.

    \item For the case $\phi = \phi_1 \wedge \phi_2$ 
      where $\tohml{\phi} = \tohml{\phi_1} \wedge \tohml{\phi_2}$, it is immediate by the 
      inductive hypothesis applied to $\phi_1$ and $\phi_2$.
      
    \item In the case $\phi = \neg \phi_1$ by definition we have 
      $$
        \tohml{\phi} = \tlinnot{\tohml{\phi_1}}
      $$
      and $\tohml{\phi_1}$ is closed and guarded by the inductive hypothesis.
      Since $\tlinnot$ maps formulae to formulae, it follows that
      $\tohml{\phi}$ is also closed and guarded.
    \item In the case $\phi = \until{\phi_1}{I}{\phi_2}$ by definition we have
      $$
        \tohml{\phi} = x_s \lin \dmd{\act} \minfp[Y]{x_e \lin (\dmd{x_e - x_s \in I}{\tohml{\phi_2}}) \vee
        (\tohml{\phi_1} \wedge \auxfun)}
      $$
      and $\tohml{\phi_1}$ and $\tohml{\phi_2}$ are closed and guarded by the induction hypothesis.
      The variables~$x_e$, $x_s$ and $Y$ are bounded and each
      occurrence of $Y$ appear under the
      scope of the modality~$\dmd{ \alpha }{-}$ for some $\alpha\in\act$ within its binder.
      Therefore, $\tohml{\phi}$ is closed and guarded.
            
  \end{itemize}
\end{proof}


\begin{lemma}
  \label{lem:soundness-completeness-tohml}
  For all $\phi \in \MTL$ and $\pi \in \twords$:
  \begin{enumerate}
    \item \label{pt:soundness-tohml} (Soundness) $\pi \in \sem{\tohml{\phi}}$ implies $\pi \in \semmtl{\phi}$, and 
    \item \label{pt:completeness-tohml} (Completeness) $\pi \in \semmtl{\phi}$ implies $\pi \in \sem{\tohml{\phi}}$.
  \end{enumerate}
\end{lemma}
\begin{proof}
  Pick $\phi \in \MTL$, we proceed by structural induction on $\phi$ to prove 
  \ref{pt:soundness-tohml} and \ref{pt:completeness-tohml} simultaneously.
  The base cases are:
  \begin{itemize}
    \item $\phi = \top$:
      By \cref{fig:mitl-semantics}, \cref{fig:lhml-semantics}, and
      \cref{fig:mitl-compiler}, we obtain
      $\semmtl{\phi} = \sem{\tohml{\phi}} = \twords$,
      from which (\ref{pt:soundness-tohml}) and (\ref{pt:completeness-tohml}) follow.
       \item $\phi = \bot$: Analogous. 
    \item $\phi = \alpha$:
      By \cref{fig:mitl-compiler}, we have $\tohml{\phi} = \dmd{\alpha}\true$.
      To prove (\ref{pt:soundness-tohml}), suppose that
      $\pi \in \sem{\tohml{\phi}}$.
      By \cref{fig:lhml-semantics}, we have $a_1(\pi) = \alpha$, from which we
      obtain $\pi \in \semmtl{\alpha}$ by \cref{fig:mitl-semantics}.
      To prove (\ref{pt:completeness-tohml}), suppose that
      $\pi \in \semmtl{\alpha}$.
      By \cref{fig:mitl-semantics}, we have $a_1(\pi) = \alpha$, which implies
      $\pi \in \sem{\tohml{\phi}}$ by \cref{fig:lhml-semantics}.
  \end{itemize}  
  We now discuss one by one the inductive cases.
  \begin{description}
    \item[($\phi = \phi_1 \wedge \phi_2$):]
      By \cref{fig:mitl-compiler}, $\tohml{\phi} = \tohml{\phi_1} \vee \tohml{\phi_2}$. 
      \begin{itemize}
        \item Proof of (\ref{pt:soundness-tohml}):
          Suppose $\pi \in \sem{\tohml{\phi}}$, then by \cref{fig:lhml-semantics} we have  
          $$
            \pi \in \sem{\tohml{\phi_1}} \text{ and } \pi \in \sem{\tohml{\phi_2}} 
          $$
          Since $\phi_1$ and $\phi_2$ are syntactically smaller than $\phi$, the inductive hypothesis 
          (\ref{pt:soundness-tohml}) guarantees 
          $$
            \pi \in \semmtl{\phi_1} \text{ and } \pi \in \semmtl{\phi_2},
          $$
          and so \cref{fig:mitl-semantics} implies the required
          $\pi \in \semmtl{\phi_1 \vee \phi_2}$.
        \item Proof of (\ref{pt:completeness-tohml}):
          Suppose $\pi \in \semmtl{\phi}$, then by \cref{fig:mitl-semantics} we have 
          $$
            \pi \in \semmtl{\phi_1} \text{ and } \pi \in \semmtl{\phi_2}
          $$
          The inductive hypothesis (\ref{pt:completeness-tohml}) applied to $\phi_1$ and $\phi_2$ ensures than 
          $$
            \pi \in \sem{\tohml{\phi_1}} \text{ and } \pi \in \sem{\tohml{\phi_2}}
          $$
          We conclude $\pi \in \sem{\tohml{\phi}}$ by \cref{fig:lhml-semantics}.
      \end{itemize}
    \item[($\phi = \neg \phi_1$):] 
      By \cref{fig:mitl-compiler}, $\tohml{\phi} = \tlinnot{\tohml{\phi_1}}$. 
      \begin{itemize}
        \item Proof of (\ref{pt:soundness-tohml}): Suppose $\pi \in \sem{\tohml{\phi}}$, then 
          \cref{prop:complementation-tlin} implies $\pi \not\in \sem{\tohml{\phi_1}}$. The contrapositive 
          of the inductive hypothesis (\ref{pt:completeness-tohml}) ensures $\pi \not\in \semmtl{\phi_1}$ 
          which implies $\pi \in \semmtl{\phi}$ by \cref{fig:mitl-semantics}.
        \item Proof of (\ref{pt:completeness-tohml}): Suppose $\pi \in \semmtl{\phi}$ which by 
          \cref{fig:mitl-semantics} means $\pi \not\in \semmtl{\phi_1}$. The contrapositive of 
          the inductive hypothesis (\ref{pt:soundness-tohml}) ensures $\pi \not\in \sem{\tohml{\phi_1}}$,
          so we conclude $\pi \in \sem{\tohml{\phi}}$ by \cref{prop:complementation-tlin}.
      \end{itemize} 
    \item[($\phi = \until{\phi_1}{I}{\phi_2}$):]
      By definition
      $$
        \tohml{ \phi } = x_s \lin \dmd{\act}{\minfp[Y]{x_e \lin \underbrace{(\dmd{x_e - x_s \in I}{\tohml{\phi_2}})}_{\psi_1}
        \vee \underbrace{(\tohml{\phi_1} \wedge \auxfun)}_{\psi_2}}}
      $$
      \begin{itemize}  
        \item Proof of (\ref{pt:soundness-tohml}):
          Suppose $\pi \in \sem{\tohml{\phi}}$ which, by \cref{fig:lhml-semantics} means
          $$
            \pi^2 \in \sem{\minfp[Y]{x_e \lin (\dmd{x_e - x_s \in I}{\tohml{\phi_2}})
                  \vee(\tohml{\phi_1} \wedge \auxfun)}}_{\vto\update{x_s}{t_1(\pi)}}
          $$
          The continuity of the semantics (\cref{sec:logics}),
          and in particular its consequence (\cref{eq:inductive-proof-method-min}),
          guarantees that there exists~$n \in \naturals$ such that 
          \begin{equation}
            \label{eq:soundness-until-consequence-continuity}
            \pi^2 \in \funofN{x_e \lin \psi_1 \vee \psi_2}{\update{x_s}{t_1(\pi)}}{Y}{\emptyset}{n}(\emptyset)
          \end{equation}
          To prove $\pi \in \semmtl{ \phi }$, \cref{fig:mitl-semantics} requires
          us to explain why there exists $i > 1$ such that the following properties are true:
          \begin{enumerate}[(a)]
            \item\label{pt:soundness-tohml-until-aim-1}%
              $t_i(\pi) - t_1(\pi) \in I$,
            \item\label{pt:soundness-tohml-until-aim-2}%
              $\pi^i \in \semmtl{\phi_2}$, and
            \item\label{pt:soundness-tohml-until-aim-3}%
              $\forall 1 < j < i \suchthat \pi^j \in \semmtl{\phi_1}$.
          \end{enumerate}
          Thanks to \cref{eq:soundness-until-consequence-continuity}, \cref{lemma:aux-until-soundness} instantiated with
          $\termF = \psi_1$, $\termG = \tohml{\phi_1}$, $v = \vto\update{x_s}{t_1(\pi)}$, $x = x_e$, $Y$, $\pi^2$ and $n$, 
          gives us $k \geq 1$ such that
          \begin{equation}
            \label{eq:hyp1-until-sound}
            \pi^{k+1} \in \sem{\psi_1}_{\vto\update{x_s}{t_1(\pi)}{\update{x_e}{t_{k+1}(\pi)}}}
          \end{equation}       
          and
          $$
            \forall j.1 \leq j < k: \pi^{j+1} \in \sem{\tohml{\phi_1}}_{\vto\semsubst{x_s}{t_1(\pi)}\update{x_e}{t_{j+1}(\pi)}}
          $$
          which is equivalent to 
          \begin{equation}
            \label{eq:hyp2-until-sound}
            \forall j.1 < j < k+1: \pi^{j} \in \sem{\tohml{\phi_1}}_{\vto\semsubst{x_s}{t_1(\pi)}\update{x_e}{t_{j}(\pi)}}
          \end{equation}
          Let us denote $v = \vto\semsubst{x_s}{t_1(\pi)}\update{x_e}{t_{j}(\pi)}$.
          By unfolding further the definitions, \cref{eq:hyp1-until-sound} states that 
          \begin{equation}
            \label{eq:soundness-until-auxiliary-3}
            v \models x_e - x_s \in I \text{ and } \pi^{k+1} \in \sem{\tohml{\phi_2}}_{ v }     
          \end{equation}
          (\ref{pt:soundness-tohml-until-aim-1}) follows from $v \models x_e - x_s \in I$. 
          Now recall that \cref{lem:tohml-produces-formulae} ensures that both $\tohml{\phi_1}$ and 
          $\tohml{\phi_2}$ are formulae, therefore, by letting $i = k+1$, the remaining 
          (\ref{pt:soundness-tohml-until-aim-2}) and (\ref{pt:soundness-tohml-until-aim-3})
          follow from \cref{eq:soundness-until-auxiliary-3} and \cref{eq:hyp2-until-sound} and 
          the inductive hypothesis. 
        \item Proof of (\ref{pt:completeness-tohml}):
          Suppose $\pi \in \semmtl{\until{\phi_1}{I}{\phi_2}}$, which by 
          the rules in \cref{fig:mitl-semantics} imply that there exists $i > 1$ such that
          \begin{enumerate}[(a)]
            \item\label{pt:completeness-until-known-1} $t_i(\pi) - t_1(\pi) \in I$,
            \item\label{pt:completeness-until-known-2} $\pi^i \in \semmtl{\phi_2}$,
            \item\label{pt:completeness-until-known-3} $\forall 1 < j < i \suchthat \pi^j \in \semmtl{\phi_1}$.
          \end{enumerate}
          To conclude $\pi \in \sem{\tohml{\phi}}$, by \cref{eq:inductive-proof-method-min} and 
          \cref{fig:lhml-semantics}, it suffices to find $n \in \naturals$ such that 
          \begin{equation}
            \label{eq:completeness-until-goal}
            \pi^2 \in \funofN{x_e \lin \termG_1 \vee \termG_2}{\vto[x_s := t_1(\pi)]}{Y}{\emptyset}{n}(\emptyset)
          \end{equation}
          The natural~$n$ we are after is precisely~$i-1$. To prove \cref{eq:completeness-until-goal}
          with $n = i-1$, we follow the following strategy: thanks to \cref{eq:function-iteration} and 
          the fact that $i-1 \geq 1$, we know
          $$
            \begin{array}{ll}
              \funofN{x_e \lin \termG_1 \vee \termG_2}{\vto[x_s := t_1(\pi)]}{Y}{\emptyset}{i-1}(\emptyset) =
              \funofN{x_e \lin \termG_1 \vee \termG_2}{\vto[x_s := t_1(\pi)]}{Y}{\emptyset}{i-2}(S)
            \end{array}
          $$
          where
          $$
            S = \funof{x_e \lin \termG_1 \vee \termG_2}{\vto\update{x_s}{t_1(\pi)}}{Y}{\emptyset}(\emptyset),
          $$
          Then, it is enough to prove that
          \begin{equation}
            \label{completeness-until-goal2}
            \pi^2 \in \funofN{x_e \lin \termG_2}{\vto\update{x_s}{t_1(\pi)}}{Y}{\emptyset}{i-2}(S)
          \end{equation}
          Thanks the monitonicity of $\funofN{x_e \lin \termG_2}{\vto\update{x_s}{t_1(\pi)}}{Y}{\emptyset}{i-2}$ 
          , to prove \cref{completeness-until-goal2}, it suffices to show the following,
          \begin{enumerate}[(i)]
            \item \label{completeness-until-goal3} $\set{ \pi^i } \subseteq S$, and
            \item \label{completeness-until-goal4} $\pi^2 \in \funofN{x_e \lin \termG_2}{\vto\update{x_s}{t_1(\pi)}}{Y}{\emptyset}{i-2}(\pi^i)$.
          \end{enumerate}
          By definition of $S$ and \cref{fig:lhml-semantics}, to prove (\ref{completeness-until-goal3}), it suffices to show that
          $$
            \pi^i \in \sem{\psi_1, \semsubst{Y}{\emptyset}}_{\vto\update{x_s}{t_1(\pi)}\update{x_e}{t_i(\pi)}}
          $$
          which in turn is equivalent to $\vto\update{x_s}{t_1(\pi)}\update{x_e}{t_i(\pi)} \models x_e - x_s \in I$ 
          and $\pi^i \in \sem{\tohml{\phi_2}, \semsubst{Y}{\emptyset}}_{\vto\update{x_s}{t_1(\pi)}\update{x_e}{t_i(\pi)}}$. 
          The former follows from (\ref{pt:completeness-until-known-1}). The former holds by \cref{eq:sem-tlin-formulae}
          provided that $\pi^i \in \sem{\tohml{\phi_2}}$ which holds by the inductive hypothesis (\ref{pt:completeness-tohml})
          applied to $\phi_2$.

          To conclude the proof, it remains only to show (\ref{completeness-until-goal4}) which we do now.
          Thanks to (\ref{pt:completeness-until-known-3}) and the inductive hypothesis (\ref{pt:completeness-tohml}), 
          we know that
          $$
            \forall 1 < j < i \wehavethat
            \pi^j \in \sem{\tohml{\phi_1}} = \funof{\tohml{\phi_1}}{\vemptyset}{Y}{\emptyset}(\emptyset),
          $$      
          which means that
          $$
            \forall 1 < j < i \wehavethat
            \pi^j \in \funof{\tohml{\phi_1}}{\vemptyset}{Y}{\emptyset}(\pi^{j+1})
          $$
          which is equivalent to 
          $$
            \forall 1 \leq j \leq i - 2 \wehavethat
            (\pi^2)^j \in \funof{\tohml{\phi_1}}{\vemptyset}{Y}{\emptyset}((\pi^2)^{j+1})
          $$
          Using \cref{lemma:aux1-compiler-mitl-comp}, we deduce  
          $$
            \pi^2 \in \funofN{\psi_2}{\vemptyset}{Y}{\emptyset}{i-2}(\pi^i)
          $$
          Since $\psi_2$ contains no free timestamp variables, we also have
          $$
            \pi^2 \in \funofN{\psi_2}{\update{x_s}{t_1(\pi)}}{Y}{\emptyset}{i-2}(\pi^i),
          $$
          which is exactly (\ref{completeness-until-goal4}).
      \end{itemize}
  \end{description} 
\end{proof}

\subsection{Correctness of $\totlinSym$}

\totlintwo*

\begin{proof}
  We prove by structural induction that for any TPTL term $\phi$ and valuation $v$ we 
  have $\sem{\phi}_v = \sem{\totlinn{\phi}}_v$. We omit the cases other 
  than $x.\psi$ and $x \sim n$, as they follow the same reasoning as the corresponding 
  cases in the proof of \cref{prop:correctness-tohml} above. 
  \begin{itemize}
    \item $\phi = x \sim n$: By \cref{fig:tptl-to-tlin,fig:lhml-semantics,fig:tptl}, we have 
      the following equalities:
      $$
        \begin{array}{lll}
          \sem{\totlinn{\phi}}_v &=& \sem{y \lin \dmd{y - x \sim n}{\true}}_v \\
                                 &=& \setof{\pi}{t_1(\pi) - v(x) \sim n} \\
                                 &=& \sem{\phi} \\
        \end{array}
      $$
    \item $\phi = x.\psi$: By \cref{fig:tptl-to-tlin,fig:lhml-semantics,fig:tptl}, we have 
      the following equalities:
      $$
        \begin{array}{lll}
          \sem{x.\psi}_v &=& \setof{\pi}{\pi \in \sem{\psi}_{v\update{x}{t_1(\pi)}}} \\
                         &\overset{I.H.}{=}& \setof{\pi}{\pi \in \sem{\totlinn{\psi}}_{v\update{x}{t_1(\pi)}}} \\
                         &=& \sem{x \lin \totlinn{\psi}} \\
                         &=& \sem{\totlinn{\phi}} \\
        \end{array}
      $$
  \end{itemize}
\end{proof}

\clearpage
\section{Proofs about MTL and TPTL monitorability}\label{app:monmtl}
\mtltwo*

\begin{proof}
  The proof proceeds by structural induction on $\phi$. The base cases are:
  \begin{itemize}
    \item $\phi = \top$: $\tohml{\top} = \true = \tomtlin{\top}$.
    \item $\phi = \bot$: Analogous. 
    \item $\phi = a$: $\tohml{a} = \dmd{a}\true = \tomtlin{a}$.
  \end{itemize}
  The inductive cases are:
  \begin{itemize}
    \item $\phi = \neg \phi'$: This case follows immediately from the inductive hypothesis.
    \item $\phi = \phi_1 \wedge \phi_2$: This case follows immediately from the inductive hypothesis.
    \item $\phi = \until{\phi_1}{I}{\phi_2}$:
      The inclusion from right to left follows by unfolding the definition of the semantics and
      applying the inductive hypothesis. The other inclusion relies on the inductive hypothesis,
      \cref{lemma:aux-until-soundness}, and the continuity of the semantics. Let $\pi \in
      \sem{\tohml{\until{\phi_1}{I}{\phi_2}}}$. Then, by \cref{fig:mitl-compiler} and \cref{fig:lhml-semantics},
      $$
        \pi^2 \in \sem{\minfp[Y]{x_e \lin (\dmd{x_e - x_s \in I}{\tohml{\phi_2}})
          \vee (\tohml{\phi_1} \wedge \dmd{ \act }{Y})}}_{\update{x_s}{t_1(\pi)}}
      $$
      By \cref{eq:inductive-proof-method-min}, there exists $n \in \naturals$ such that
      \begin{equation}
        \label{eq:correctness-compiler2-hyp1}
        \pi^2 \in \funofN{x_e \lin (\dmd{x_e - x_s \in I}{\tohml{\phi_2}})
          \vee (\tohml{\phi_1} \wedge \dmd{ \act }{Y})}{\update{x_s}{t_1(\pi)}}{Y}{\emptyset}{n}(\emptyset)
      \end{equation}
      We prove by induction on $n$ that
      \begin{equation}
        \label{eq:correctness-compiler2-goal}
        \pi^2 \in \funofN{x_e \lin (\dmd{x_e - x_s \in I}{\tomtlin{\phi_2}})
          \vee \psi}{\update{x_s}{t_1(\pi)}}{Y}{\emptyset}{n}(\emptyset)
      \end{equation}
      where
      $$
        \psi \defeq \tomtlin{\phi_1} \wedge \dmd{ \act }{y \lin \dmd{y - x_s \precrI \rb{I}}{Y}}
      $$
      which, together with \cref{fig:lhml-semantics}, implies the required
      $\pi \in \sem{\tomtlin{\until{\phi_1}{I}{\phi_2}}}$. When $n = 0$,
      \cref{eq:correctness-compiler2-hyp1} leads to a contradiction, since its right-hand side
      is equal to the empty set. Suppose now that $n = n' + 1$. Then,
      \cref{eq:correctness-compiler2-goal} follows if we prove either
      \begin{equation}
        \label{eq:correctness-compiler2-goal1}
        t_2(\pi) - t_1(\pi) \in I \text{ and }
        \pi^2 \in \sem{\tomtlin{\phi_2},\update{Y}{S_1}}_{\update{x_s,x_e}{t_1(\pi)}}
      \end{equation}
      or
      \begin{equation}
        \label{eq:correctness-compiler2-goal2}
        \pi^2 \in \sem{\tomtlin{\phi_1},\update{Y}{S_1}}_{\update{x_s}{t_1(\pi)}\update{x_e}{t_2(\pi)}},
        t_3(\pi) - t_1(\pi) \precrI \rb{I} \text{ and }
        \pi^3 \in S_1
      \end{equation}
      where
      $$
        S_1 \defeq \funofN{x_e \lin (\dmd{x_e - x_s \in I}{\tomtlin{\phi_2}})
          \vee \psi}{\update{x_s}{t_1(\pi)}}{Y}{\emptyset}{n'}(\emptyset)
      $$

      By the hypothesis in \cref{eq:correctness-compiler2-hyp1}, we have
      $$
        \pi^2 \in \funof{x_e \lin (\dmd{x_e - x_s \in I}{\tohml{\phi_2}})
          \vee (\tohml{\phi_1} \wedge \dmd{ \act }{Y})}{\update{x_s}{t_1(\pi)}}{Y}{\emptyset}(S_2)
      $$
      where
      $$
        S_2 \defeq \funofN{x_e \lin (\dmd{x_e - x_s \in I}{\tohml{\phi_2}})
          \vee (\tohml{\phi_1} \wedge \dmd{ \act }{Y})}{\update{x_s}{t_1(\pi)}}{Y}{\emptyset}{n'}(\emptyset)
      $$
      By unfolding the definitions of the semantic functions and using
      \cref{fig:lhml-semantics}, we have two cases to consider. In the first case,
      $t_2(\pi) - t_1(\pi) \in I$ and
      $\pi^2 \in \sem{\tomtlin{\phi_2},\update{Y}{S_2}}_{\update{x_s}{t_1(\pi)}\update{x_e}{t_2(\pi)}}$.
      Since $Y$, $x_e$, and $x_s$ are not free in $\tohml{\phi_2}$, we also have
      $\pi^2 \in \sem{\tohml{\phi_2}}$. By the inductive hypothesis,
      $\pi^2 \in \sem{\tomtlin{\phi_2}}$, and we conclude \cref{eq:correctness-compiler2-goal1}
      since $Y$, $x_e$, and $x_s$ are not free in $\tomtlin{\phi_2}$. In the second case,
      we have $\pi^2 \in \sem{\tohml{\phi_1},\update{Y}{S_2}}_{\update{x_e,x_s}{t_1(\pi)}}$
      and $\pi^3 \in S_2$. Similarly to the previous case, we obtain
      $\pi^2 \in \sem{\tomtlin{\phi_2},\update{Y}{S_1}}_{\update{x_e,x_s}{t_1(\pi)}}$.
      Therefore, to conclude \cref{eq:correctness-compiler2-goal2}, it remains only to prove
      $\pi^3 \in S_1$ and
      $t_3(\pi) - t_1(\pi) \precrI \rb{I}$.
      The former follows from the inductive hypothesis on $n'$. For the latter, since we have 
      $\pi^3 \in S_2$, \cref{lemma:aux-until-soundness} ensures the existence of $i \geq 3$ such that 
      $t_i(\pi) - t_1(\pi) \in I$, therefore, we conclude $t_3(\pi) - t_1(\pi) \precrI \rb{I}$ 
      by the monitonicity of timestamps. 
  \end{itemize}
\end{proof}

\tptlttwo*

\begin{proof}
    We prove that, for every TPTL term $\phi$ and valuation $v$, $\sem{\totlinn{\phi}}_v = \sem{\tomtlinn{\phi}}_v$.
    The proof proceeds by structural induction on $\phi$. All cases except constrained Until follow directly from 
    the inductive hypothesis, since the two translations yield formulae of the same form. For the  constrained Until case with $\phi = \uuntil{\phi_1}{((x \prec n) \wedge \phi_2)}$, the proof is similar 
    to the case $\until{\phi_1}{I}{\phi_2}$ in the proof of \cref{prop:correctness-compiler2} with 
    $x_s$ replaced by $x$.
\end{proof}

\end{document}